\documentclass[fleqn,usenatbib]{mnras}

\usepackage{newtxtext,newtxmath}
\usepackage[T1]{fontenc}
\usepackage{ae,aecompl}

\usepackage{graphicx}	% Including figure files
\usepackage[export]{adjustbox}
\usepackage{amsmath}	% Advanced maths commands

\usepackage{xcolor}
\usepackage{url}

\defcitealias{2021MNRAS.507..971C}{Paper~I}
\defcitealias{2022MNRAS.513L..83C}{Paper~II}

\title[SRG/eROSITA dicovery of SNR candidate G121.1-1.9]{{ SRG/eROSITA discovery of a radio faint X-ray candidate supernova remnant SRGe~J003602.3+605421=G121.1-1.9}}

\author[Khabibullin et al.]{I.I.~Khabibullin,$^{1,2,3}$ E.M.~Churazov,$^{2,3}$ A.M.~Bykov,$^4$ N.N.~Chugai,$^5$ and R.A.~Sunyaev$^{2,3}$
\\
\\
$^1$~Universitäts-Sternwarte, Fakultät für Physik, Ludwig-Maximilians-Universität München, Scheinerstr.1, 81679 München, Germany \\
$^2$~Space Research Institute (IKI), Profsoyuznaya 84/32, Moscow 117997, Russia \\
$^3$~Max Planck Institute for Astrophysics, Karl-Schwarzschild-Str. 1, D-85741 Garching, Germany  \\
$^4$~Ioffe Institute, 26 Politekhnicheskaya str., St. Petersburg 194021, Russia \\
$^5$~Institute of Astronomy, Russian Academy of Sciences, 48 Pyatnitskaya str., Moscow 119017, Russia  \\
}

\date{Accepted XXX. Received YYY; in original form ZZZ}

\begin{document}
\label{firstpage}
\pagerange{\pageref{firstpage}--\pageref{lastpage}}
\maketitle

\begin{abstract}
{
We report the discovery of a candidate X-ray supernova remnant SRGe~J003602.3+605421=G121.1-1.9 in the course of \textit{SRG}/eROSITA all-sky survey.  The object is located at (l,b)=(121.1$^\circ$,-1.9$^\circ$), is $\approx36$ arcmin in angular size and { has a nearly circular shape}. Clear variations in spectral shape of the X-ray emission across the object are detected, with the emission from the inner (within 9') and outer (9'-18') parts dominated by iron and oxygen/neon lines, respectively. 
The non-equilibrium plasma emission model is capable of describing the spectrum of the outer part 
with the initial gas temperature 0.1 keV, final temperature 0.5 keV and the ionization age $\sim 2\times10^{10}$ cm$^{-3}$ s. The observed spectrum of the inner region is more complicated (plausibly due to the contribution of the outer shell) and requires substantial overabundance of iron for all models we have tried.
The derived X-ray absorption equals to $(4-6)\times10^{21}$ cm$^{-2}$, locating the object at the distance beyond 1.5 kpc, and implying { its age} $\sim(5-30)\times1000$ yrs. No bright radio, infrared, H$_\alpha$ or gamma-ray counterpart of this object have been found in the publicly-available archival data. A model invoking a canonical $10^{51}$ erg explosion {(either SN~Ia or core collapse)} in the hot and tenuous medium in the outer region of the Galaxy $\sim$9 kpc away might explain the bulk of the observed features.
{ This scenario can be tested} with future deep X-ray and radio observations.
}
\end{abstract}

\begin{keywords}
ISM: supernova remnants -- Interstellar Medium (ISM), Nebulae,
radiation mechanisms: thermal -- Physical Data and Processes, X-rays: general -- Resolved and unresolved sources as a function of wavelength, Galaxy: halo -- The Galaxy
\end{keywords}

%%%%%%%%%%%%%%%%%%%%%%%%%%%%%%%%%%%%%%%%%%%%%%%%%%

%%%%%%%%%%%%%%%%% BODY OF PAPER %%%%%%%%%%%%%%%%%%
%
\section{Introduction}
\label{sec:intro}
{ The supernova explosions not only expel and distribute large masses of metals in the surrounding medium, but also drive strong shock waves into the interstellar medium.  These shock waves are believed to be the primary acceleration sites responsible for the populations of cosmic ray protons and ions up to $10^{15}$ eV in our Galaxy \citep[][]{1969ocr..book.....G}. }
The direct observation of this acceleration process, however, comes from the observation of the gamma-ray radiation and radio synchrotron emission produced by the relativistic particles \citep[e.g.][]{BV04}. Indeed, the wide area surveys of the Galactic plane at radio frequencies not only allowed to discover a few hundred supernova remnants (SNR) to date \citep[][]{2019JApA...40...36G}, but also derive insightful scaling relations \citep[e.g. between the size of the remnant and its surface brightness,][]{1976MNRAS.174..267C,L81,B86,Green84,2020NatAs...4..910U}, as well as reveal intricate properties of individual objects.

{ 
{Radio emission associated with accelerated electrons turned out to be the most efficient way of finding SNR candidates.}
The expected life-time of SNRs in the radio band is $\sim$60,000~yr \citep[][]{1994ApJ...437..781F}. For a specific supernova explosions rate  calibrated with observations of other galaxies \citep[e.g.][]{2011MNRAS.412.1441L}, a few thousand SNRs should be observed in the Milky Way \citep[][]{1994ApJ...437..781F}. 
{This number is in drastic contrast to $\lesssim 400$ currently known candidates candidates  
\citep[e.g. as listed in the \texttt{SNRcat}\footnote{\url{http://snrcat.physics.umanitoba.ca}} catalogue,][]{2012AdSpR..49.1313F}.}
%}

Although new SNRs keep being discovered in the data of more sensitive and uniform radio surveys \citep[e.g.][]{2014A&A...566A..76G,2019PASA...36...45H}, the deficit of the known SNRs persists and constitutes a well known "missing SNRs problem" \citep[][]{1980PThPh..64.1587T}. Another aspect of this problem is revealed by comparing statistics of the known pulsars and SNRs, also indicating a relative deficit in the observed number of remnants \citep[e.g.][]{1977ApJ...215..885T}.} 

{
This might be connected to a number of observational issues inherent to identification of faint and extended radio sources in the Galactic plane, but might also be an indication of issues associated with extrapolation of the SNR properties as function of their type, age and environment.}

There is a growing number of the supernova remnants identified at other wavelengths and featuring very weak or non-detectable (at the current sensitivity level) radio emission. These, for instance, include supernova candidates identified in optical, X-ray and gamma-ray surveys \citep[][]{2017hsn..book.2005L}. 
Further increase in the size of radio-blind SNR samples is vital to understand whether such objects are rare examples failing to accelerate relativistic particles due to some particular reason, or there is a large population of such events missed in the previous samples because of the selection criteria.

The unprecedentedly deep and uniform maps provided by the all-sky survey by the eROSITA telescope \citep[][]{2021A&A...647A...1P} onboard the \textit{SRG} observatory \citep[][]{2021A&A...656A.132S} offer unique opportunities to look for extended objects of relatively low surface brightness, in particular in the soft X-ray band, from 0.4 to 2.3 keV.
As a result, such faint and extended X-ray structures, like eROSITA bubbles \citep[][]{2020Natur.588..227P}, Hoinga supernova remnant \citep[][]{2021A&A...648A..30B}, and the candidate Ia supernova remnant in the Galactic Halo SNR G116.6-26.1 \citep[][hereafter Paper I]{2021MNRAS.507..971C} have already been discovered. For the latter object in particular only very weak radio emission was recently discovered  \citep[][hereafter Paper II]{2022MNRAS.513L..83C} in the data of  LOFAR Two-metre Sky Survey \citep[LoTTS-DR2,][]{2022A&A...659A...1S}, possibly because this supernova exploded in the hot and tenuous medium of the Galactic Halo, reducing the shock Mach number and the efficiency of particle acceleration.

Here we report \textit{SRG}/eROSITA discovery of an X-ray bright and extremely radio-faint candidate supernova remnant located in the Galactic disc. Based on the interstellar absorption value inferred from the X-ray spectra we place it at a distance of $>1$ kpc at 90\% confidence level and $>3$ kpc at the 1$\sigma$ level,
making this object an excellent candidate for the evolved SNR located in the "normal" Galactic disc environment. No counterpart of this object is observed at other wavelengths, including radio, optical and gamma-ray bands. 

{An interesting feature of the newly found object is the dominance of iron emission lines in the X-ray spectrum of its inner half, allowing one to estimate the required amount of mass in iron ions. Potentially, the iron mass measurements alone could be used to differentiate between different SNR scenarios. Indeed, in typical type Ia and type II supernovae the iron masses differ by an order of magnitude, as expected from the theoretical models and supported by direct mass estimates from the very early (hundreds of days) direct measurements in infrared, hard X-ray and gamma-ray observations of SN1987A and SN2014J \citep[e.g.][]{1989ARA&A..27..629A,1993ApJ...419..824L,2014Natur.512..406C}.

The pronounced difference in iron mass between thermonuclear and core collapse supernovae \citep[e.g.][for a recent review]{2021NewAR..9201606I} possibly could be revealed by X-ray spectra of ejecta heated by the reverse shock for an SNR at the age of hundreds years in the galactic disk medium or significantly later, if the SNR expands in the rarefied medium \citep[e.g.][]{2003ApJ...593..358B}, and the newly discovered object might offer an example of such a situation.

}

\section{X-ray observations}
\label{sec:xrayobs}

Orbital observatory \textit{SRG} \citep[][]{2021A&A...656A.132S}, featuring two focusing X-ray telescopes, eROSITA \citep[][]{2021A&A...647A...1P} and ART-XC \citep[][4-30 keV]{2021A&A...650A..42P} was launched in July 2019 and started to perform the all-sky survey mission in December 2019. By now four complete all-sky snapshots have been obtained, resulting in the unprecedentedly deep and uniform X-ray maps of the whole sky being obtained. 

Here we use the data accumulated over 4 consecutive scans, with the total effective exposure amounting to {1225} seconds (i.e. $\approx8600$~s in equivalent exposure for one telescope module). For the imaging analysis we use the data of all seven telescope modules (TMs), while TMs 5 and 7 are excluded from the spectral analysis due to possible impact of the optical light leak on their signal \citep[e.g.][]{2021A&A...647A...1P}. Initial reduction and processing of the data were performed using standard routines of the \texttt{eSASS} software \citep{2018SPIE10699E..5GB,2021A&A...647A...1P}, while the imaging and spectral analysis were carried out with the background models, vignetting, PSF and response function calibrations built upon the standard ones via slight modifications motivated by results of calibration and performance verification observations \citep[e.g. observations of the Coma cluster,][]{2021A&A...651A..41C}. 

\subsection{X-ray imaging}
\label{ss:xrayimage}

%------------------------
\begin{figure}
\centering
\includegraphics[angle=0,width=0.99\columnwidth, bb = 55 190 545 640]{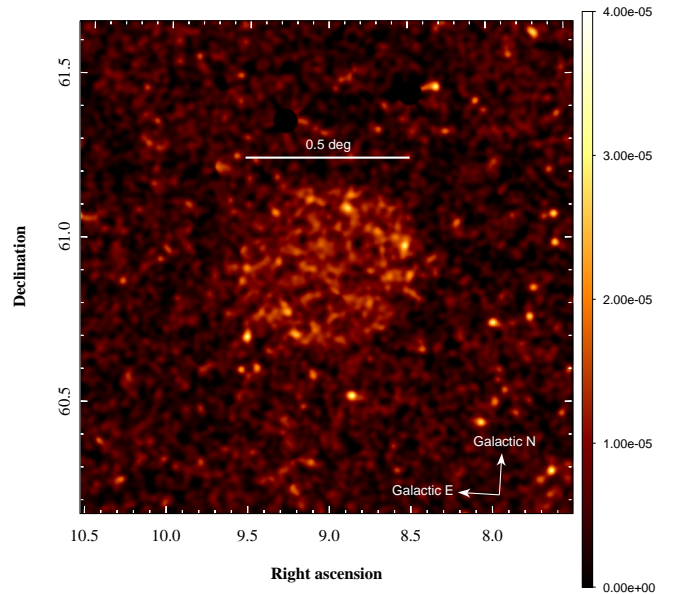}
\caption{Detector background-subtracted vignetting-corrected image of X-ray (0.4-2.3 keV) surface brightness in the 1.5$\times$1.5 deg field containing the newly discovered SNR candidate SRGe~J003602.3+605421=G121.1-1.9 obtained by \textit{SRG}/eROSITA after four surveys. The image was smoothed with a $\sigma$=30'' gaussian filter after masking point-like and mildly extended sources with the 0.5-2 keV flux above $3\times10^{-14}$ erg s$^{-1}$ cm$^{-2}$. { The image is in equatorial coordinates, RA and Dec axes here and further-on are labelled in degrees, the white compass shows direction of the Galactic coordinates.} }
\label{fig:image}
\end{figure}
%-------------------------

%------------------------
\begin{figure}
\centering
\includegraphics[angle=0, bb=30 210 590 670, width=1.\columnwidth]{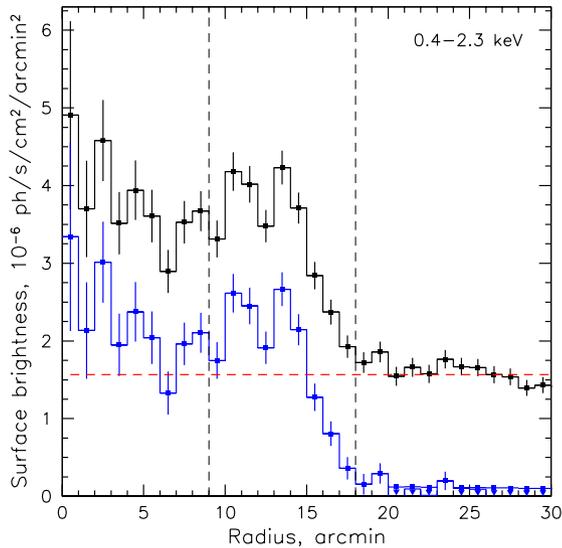}
\caption{A radial profile of the X-ray surface brightness of the diffuse emission in the 0.4 to 2.3 keV band extracted with 1 arcmin linear step without (black) and with (blue) subtraction of the estimated astrophysical background level. The error-bars show purely statistical (i.e. photon counts) 1$\sigma$ uncertainty, the red horizontal dashed line marks the estimated level of the background emission. The vertical dashed lines mark 9' and 18' radii used as boundaries of the spectral extraction regions.}
\label{fig:rprof}
\end{figure}
%------------------------

After correction for the vignetting and background subtraction, the obtained maps are routinely processed in terms of detecting point and mildly-extended sources and creating their catalogues. By means of proper modelling and subtraction of these sources, maps of the unresolved diffuse emission are obtained, consisting mainly of the unresolved cosmic X-ray background and Galactic (genuinely) diffuse emission. Already a visual inspection of these residual emission maps allows one to spot previously unknown relatively faint extended structures.

Inspection of an image in soft X-ray band (from 0.4 to 2.3 keV) centred on the position near Galactic coordinates $(l,~b)=(121.1^\circ,-1.9^\circ)$ (i.e. $\rm  (RA,Dec)\approx (9.1^\circ$, 60.9$^\circ$) in  the equatorial coordinates) revealed a faint diffuse emission of nearly circular shape half of the degree in diameter, hereafter SRGe~J003602.3+605421=G121.1-1.9. Figure~\ref{fig:image} shows the image in the 0.4-2.3 keV band, obtained by  masking point-like and mildly extended sources down to 0.5-2 keV flux\footnote{The 0.5-2 keV flux was calculated from the 0.4-2.3 keV counts using a constant factor for this purpose.} $3\times10^{-14}$ erg s$^{-1}$ cm$^{-2}$ and smoothing with a {$\sigma=30''$ gaussian window} to enhance visibility of the discovered diffuse emission. 

Figure \ref{fig:rprof} shows the corresponding radial profile of the X-ray surface brightness centred on (RA,Dec)=(9.0196$^\circ$, 60.9140$^\circ$). One can see that the excess emission extends to $\approx$18 arcmin from the geometrical center, has rather flat radial profile and varies between 1 and 2 times of the background level in amplitude. The latter is estimated to be equal to $1.5\times10^{-6}$ photons s$^{-1}$ cm$^{-2}$ arcmin$^{-2}$ from the 20'-30' annulus region around the object. 

No significant excess emission is visible at energies above 2.3 keV, suggesting that the spectrum of the object's emission is soft
{The estimated 1$\sigma$ upper limit on the 2.3-4 keV surface brightness is $\sim3\times10^{-8}$ ph s$^{-1}$ cm$^{-2}$ arcmin$^{-2}$, i.e. $\sim$10\% of the measured background level, corresponding to the total 2.3-4 keV flux $\sim10^{-13}$ erg s$^{-1}$ cm$^{-2}$.}

\begin{figure}
\centering
\includegraphics[angle=0,width=0.99\columnwidth, bb = 100 200 540 650]{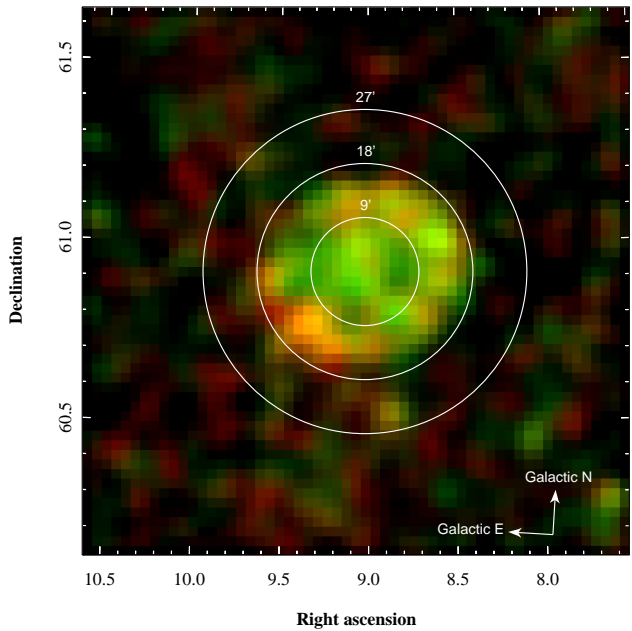}
\caption{A pseudo-{two-}colour image showing surface brightness distribution in 0.4-0.7 keV (red) and 0.7-1.3 keV (green) bands. The images were downgraded to 1.5' pixel size and smoothed with a $\sigma$=90'' gaussian filter after masking point sources in order to suppress the noise and enhance visibility of the 'colour' variations across the region of G121.1-1.9. The white circles show the 0'-9', 9'-18', 18'-27' concentric regions used for the spectra extractions.    
}
\label{fig:rgbimage}
\end{figure}

Indeed, splitting of the standard 0.4-2.3 keV into three sub-bands, 0.4-0.7, 0.7-1.3 and 1.3-2.3 keV shows no significant excess emission already in the 1.3-2.3 keV band. A pseudocolor image combining emission in 0.4-0.7 and 0.7-1.3 keV bands is shown in Figure \ref{fig:rgbimage}, where original pixel size resolution was degraded to 1.5' size and $\sigma=$90'' gaussian smoothing was applied in order to suppress low counts noise. Spectral colour variations are clearly visible on this image and in the simplest approach, we can separate harder from softer emission by r=9' radius.  

\begin{figure}
\centering
\includegraphics[angle=0,width=0.99\columnwidth, bb = 70 200 520 640]{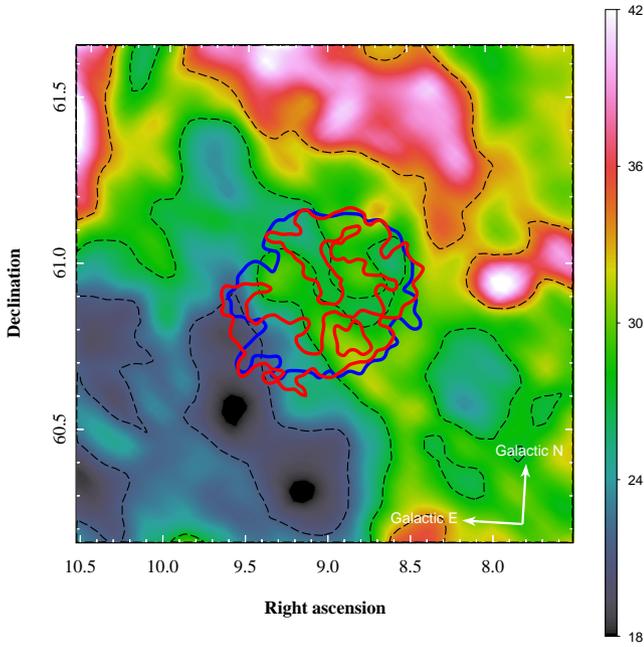}
\caption{IRIS map of emission at 100 $\mu$m in the region of G121.1-1.9 with the contours of X-ray emission in 0.4-0.7 keV (red) and 0.7-1.3 keV (blue) bands overlaid. The dashed black contours correspond to three levels of the 100 $\mu$m emission, namely the mean one within the G121.1-1.9 region and $\pm$25\% of it.}
\label{fig:iris}
\end{figure}

A possible reason for that might be the difference in the foreground absorption, causing more absorbed regions having harder spectra. In order to verify this we check the maps of the dust emission at 100$\mu m$ in the same region from the Improved Reprocessing of the IRAS Survey \citep[IRIS,][]{2005ApJS..157..302M}. Figure \ref{fig:iris} shows this map on a linear scale with the black dashed contours highlighting the mean level inside the G121.1-1.9 region, and the $\pm$25\% levels with respect to it. The red and blue contours highlight morphology of the diffuse emission in the 0.4-0.7 keV and 0.7-1.3 keV band, respectively. Although a certain degree of anti-correlation between emission in the soft band and dust emission might be suspected, in particular in the central and south-eastern part of the object, the north-western part clearly shows soft X-ray emission from the regions with the close-to-average and slightly higher amount of the dust emission. This implies that, first, the spectral variations are likely intrinsic to the source itself, and second, that the object is located further away than some of the dust features casting soft X-ray shadow on it. Next we check these suggestions via spectral analysis of the X-ray emission.

%------------------------
%-------------------------

%------------------------
%-------------------------

%-------------------------
\subsection{X-ray spectroscopy}
\label{ss:xrayspec}
%-------------------------

\begin{figure}
\centering
\includegraphics[angle=0,width=0.99\columnwidth,bb = 50 40 700 540]{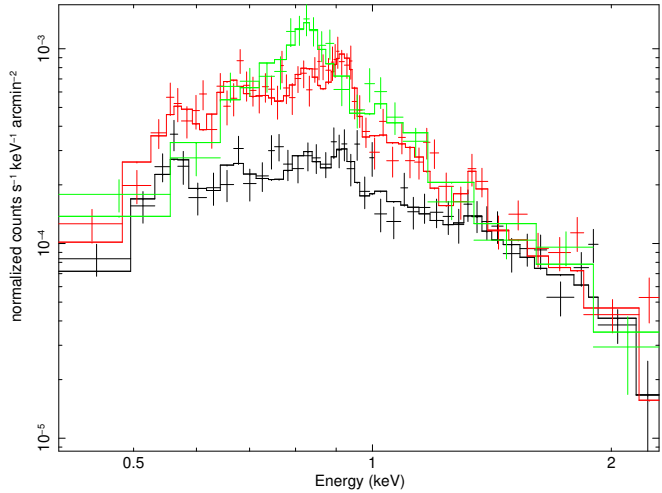}
\caption{X-ray spectra extracted from 0'-9' (green), 9'-18' (red), and 18'-27' (black) rings centred on G121.1-1.9. The latter one represents the estimated {astrophysical background/foreground} emission with the corresponding 3-component model fit (solid line). The emission model for the two G121.1-1.9 regions (also shown in solid lines) consists of the same background model plus absorbed non-equilibrium plasma emission model \texttt{rnei} with free abundance of iron in the 0'-9' region and free abundance of neon in the outer region. The resulting best fit parameters are listed in Table \ref{tab:obs}.
}
\label{fig:spec}
\end{figure}

We extract spectra, calculate corresponding response functions and estimate particle background signals for three concentric annulus regions: 0'-9, 9'-18' and 18'-27' in radius (shown as white circles in Figure \ref{fig:rgbimage}). The latter ring is used for estimation of the local astrophysical background, taking into account significant variations in the foreground absorption on a scale of $\sim1$ degree in this region close to the Galactic plane. The resulting particle background-subtracted, effective area-corrected spectra are shown in Figure \ref{fig:spec}, re-normalised to 1 arcmin$^2$ surface area. Strong excess of the emission above the background level (black points) is present only in the band from 0.5 to 1.3 keV, with the spectrum of the inner part (green points) peaking around 0.8 keV, while the spectrum from 9'-18' arcmin having relatively flat shape between 0.5 and 1 keV.

We fit the background emission with the standard three-component model \texttt{TBabs*apec+TBabs(apec+powerlaw)} in the \texttt{XSPEC} X-ray fitting package \citep[][]{1996ASPC..101...17A}. The first component stands for the Local Hot Bubble emission, and it was fitted with the fixed temperature, $kT=0.1$ keV, and metallicity, $Z=1$ (the Solar abundance set of \citealt{1989GeCoA..53..197A} is used throughout the paper).  The second component describes (in a phenomenological way) the emission of the Galactic Halo, with $kT\sim0.3$ keV, and solar metallicity. The latter component stands for the Cosmic X-ray Background radiation, with the fixed slope $\gamma=1.41$. The absorbing column density for the last two components was a free parameter and turned out to be $(6\pm0.1)\times 10^{21}$ cm$^{-2}$. 

This value is consistent with the total line-of-sight absorbing column density in this direction  estimated at the level $\approx6.1~10^{21}$~cm$^{-2}$ based on the method of \cite[][]{2013MNRAS.431..394W}, which accounts for contributions of the atomic \citep[$\approx 4.7~10^{21}$~cm$^{-2}$,][]{2005A&A...440..775K}  and molecular gas \citep[$\approx 1.4~10^{21}$~cm$^{-2}$ derived from the reddening value of $E(B-V)\approx0.91$, ][]{1998ApJ...500..525S} to it.

\begin{figure}
\centering
\includegraphics[angle=0, bb=30 200 600 660, width=1.\columnwidth]{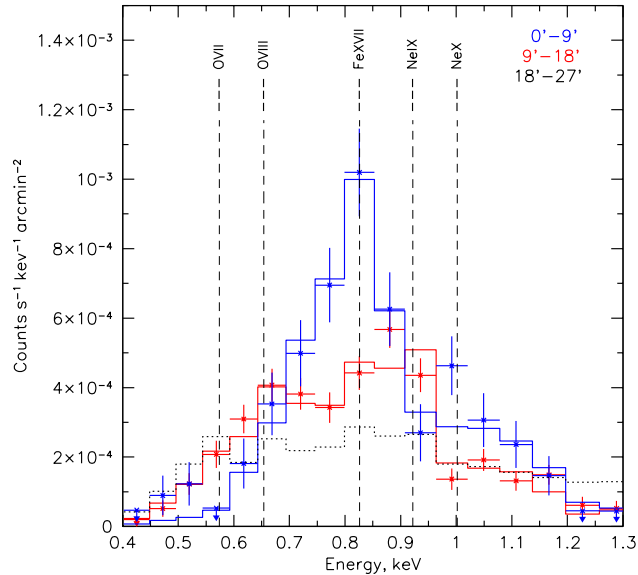}
\caption{A zoom-in in the spectra (after astrophysical background subtraction) in the 0'-9' (blue) and 9'-18'(red) rings within the 0.4-1.3 keV band. The positions of the most prominent emission lines (O~VII at 574~eV, O~VIII at 654~eV, Fe~XVII at 826~eV, Ne~IX at 922~eV and Ne~X at 1002~eV) are marked and labelled accordingly. The subtracted astrophysical background (and foreground) emission is shown by black dotted line.}
\label{fig:zoomspec}
\end{figure}

For the source regions, the emission was fitted with the same background model complemented with the additional component standing for the source emission. We have tried the collisional ionisation equilibrium plasma model \texttt{apec}, but it provided poor fits to the spectra. Instead, we used non-equilibrium plasma ionisation model with variable heavy metal abundances \texttt{vrnei} \citep[][]{2001ApJ...548..820B}.   The starting temperature was $kT_1$ was fixed at 0.1 keV for the both source regions, while the final temperature $kT_2$ was allowed to vary. {The initial gas temperature was fixed, because it was only very poorly constrained from the low side when kept as a free parameter, reflecting the fact that at moderate values of the ionization parameter $\tau\sim 10^{10}-10^{11}$ cm$^{-3}$~s there is already no strong influence of the initial temperature as far as it is high enough to ensure dominance of the helium-like ion in the ionization state of oxygen.} After the fit with the solar abundance of the all elements, some residuals were obvious in the region of the neon lines for the outer region and iron lines for the inner region.  Motivated by that, we allowed abundances of these elements to vary freely { (relative to the oxygen abundance, which was fixed at the solar value)}  during the fits and found that neon overabundance at the level of $\sim1.6$ and iron overabundance at the level of $\sim3$ can indeed improve the fit significantly. { Since the spectrum at these temperatures is dominated by line emission, normalisation and absolute value of the abundance are degenerate, so only relative abundance of the elements can be constrained.}    { The astrophysical-background subtracted spectra in the 0.4-1.3 keV band overlaid with final models are shown in Figure \ref{fig:zoomspec}, with positions of the most important lines marked and labelled correspondingly. }

The robustness of the fitting procedures was checked with an extensive survey of the parameter space over the highly non-linear parameters, namely the ionisation  parameter $\tau=n_e\times t$, where $n_e$ is the local gas number density and $t$ is the ionisation time scale \citep[see][for details]{2001ApJ...548..820B}, and the total absorbing column density $N_{H}$. Results of these surveys were used for error estimation of the model parameters. Figure \ref{fig:steppar} shows the $\chi^2$ maps on the $\tau-N_{H}$ plane for both of the source regions. The resulting parameters of the fits are listed in Table \ref{tab:obs}.

{Additionally, we have tested sensitivity of the derived parameters to the particular choice of the non-equilibrium plasma emission model. Namely, we repeated similar extensive parameter space scans for the \texttt{vnei} and \texttt{vpshock} models \citep[][]{2001ApJ...548..820B}, allowing abundance of neon and iron to vary freely for the outer and inner regions respectively. For the \texttt{vpshock}, the lower value of the ionization parameter $\tau$ was fixed at 0, and the upper value was used as the main parameter. All the derived parameters of the models turn out to be consistent with the \texttt{vrnei} fit, namely resulting in similar values for the absorbing column density $N_H$, plasma temperature $kT$, normalisations, relative Ne and Fe overabundances, and ionization parameter $\tau$. Since in the \texttt{vpshock} model a distribution over $\tau$ is assumed, it is some intermediate value that is close to the values given by the \texttt{vrnei} and \texttt{vnei} models. Although none of these models can be considered as perfectly suited for the description of the integrated spectra from such a complex medium containing shock-heated, metal-enriched and probably non-equilibrium multi-temperature plasma of the remnant, the derived quantities can be used to draw important conclusions regarding properties and nature of the object.}

First of all, no significant difference in the value of $N_{H}$ is required between the inner and the outer region, both are consistent with $N_{H}=(5\pm1)\times10^{21}$ cm$^{-2}$, which is only slightly smaller than the total absorbing column density in that direction. This implies that the source is located further away than the bulk of the absorbing gas along the line-of-sight. We exploit this information to estimate the distance to the object from the 3D absorption maps in Section \ref{ss:3dabsorption}.

Second, the best fit parameters imply that the inner and outer regions differ mostly in the value of the ionisation parameter $\tau$, and not in the final temperature $kT_2$. The latter turns out to be $\sim0.5$ keV for both regions, while the former differs by a factor of 2.5, which is not inconsistent with the assumption that the gas in the inner region was shocked earlier and had more time to equilibrate \footnote{We note in passing that we are dealing with the projected spectra, to which different radial shells can contribute, especially when the inner region is considered. It is, therefore, possible that the failure of the APEC model is caused  by the projection effects.}. On the other hand, the difference in $\tau$ might come from the higher density in the inner region. Indeed, if we combine the derived model normalizations  {$N_{\rm inner}=1.8^{+3.8}_{-0.9}\times10^{-6}$ , $N_{\rm outer}=3.1^{+7.8}_{-1.8}\times10^{-6}$} for the surface brightness within the inner and outer regions with their areas and approximate estimation of their volumes, we get

\begin{equation}
\frac{n_{\rm outer}}{n_{\rm inner}}=\sqrt{\frac{N_{\rm outer}}{N_{\rm inner}}\frac{R_{\rm outer}^2/R_{\rm inner}^2-1}{R_{\rm outer}^3/R_{\rm inner}^3-1}}\approx0.8
\end{equation}
for { $R_{\rm outer}/R_{\rm inner}=18'/9'=2$ and $N_{\rm outer}/N_{\rm inner}\approx1.6$.} Thus a large portion of the difference in $\tau$ might be attributed to the difference in the gas densities. Given the large uncertainties and degeneracies between the parameters of the fits, it is hard to draw a more quantitative conclusion on that, however. Also, a single-zone model for these regions might be an oversimplification, possibly biasing the derived parameters towards values reflecting conditions in the regions of higher emissivity. Unfortunately, low count statistics of the data does not allow us to perform more spatially-resolved spectral analysis.

Finally, the fits for the inner and outer regions require overabundance { (relative to oxygen)} in iron ($3.3\pm1.2$) and neon ($1.6\pm0.5$), respectively. Since at the temperatures of $kT\sim$0.1-0.5 keV the continuum emission is rather week, this over abundances mostly reflect the ratio of the corresponding lines to the lines of helium- and hydrogen-like oxygen at 574 eV and 654 eV  (see Figure \ref{fig:zoomspec}). The required overabundance of neon, however might be connected to the particular choice of the standard Solar abundance set (e.g. using the set of \cite{1992PhyS...46..202F} instead of \cite{1989GeCoA..53..197A} results in slightly smaller best fit value which is consistent with unity within the error bar) and imperfections of the single-zone modelling of the emission from this region. Indeed, the ratio of neon to oxygen lines might be strongly boosted in the regions of higher electron temperature due to exponential sensitivity in excitation efficiencies for the corresponding transitions \citepalias[e.g.][]{2021MNRAS.507..971C}. Although, the \texttt{vrnei} naturally accounts for this, the single-zone assumption might lead to substantial deviations from the spectrum observed in reality.

The derived overabundance of iron in the inner region cannot be fully explained by any of these effects, however. Indeed, using another set of standard abundances, e.g. \cite{1989GeCoA..53..197A}, would result in even higher values, $Z_{\rm Fe,inner}\approx5$. Also, the iron ions responsible for the observed lines correspond to significantly higher temperatures, than the lines of oxygen and neon, indicating that not only a boost in the collisional excitation but also the real change in the ionization state should have occurred.

{Given the clear signatures of radial gradients, a de-projection technique is needed in order to avoid a contamination of the SNR inner regions spectra by the emission of the outer shells. However, the statistical significance of the currently available data allows at most a two-shell approximation, which is sensitive to the choice of the boundary between the shells. We, therefore, defer reporting the results of the de-projected spectral analysis for future publications and show instead a set of radial profiles in narrow energy bands centred at the most prominent emission lines.
}

%}

\begin{figure}
\centering
\includegraphics[angle=0, bb=30 180 590 670, width=0.95\columnwidth]{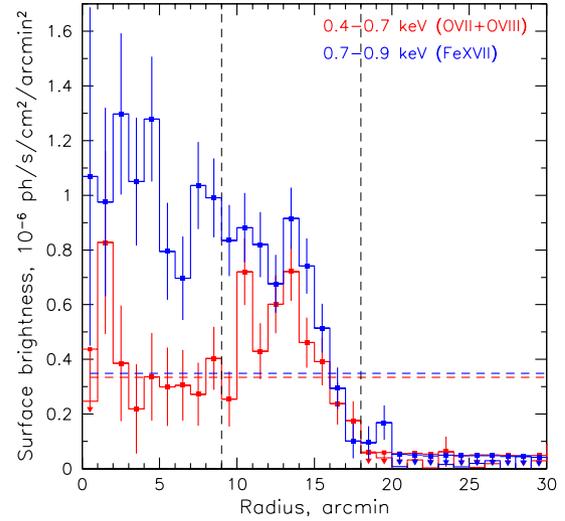}
\includegraphics[angle=0, bb=30 180 590 670, width=0.95\columnwidth]{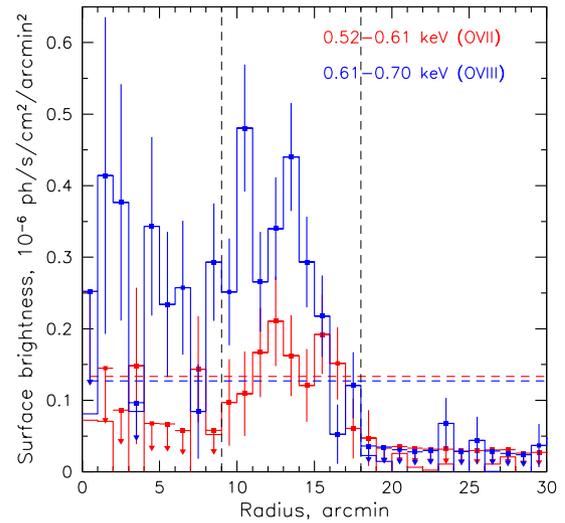}
\includegraphics[angle=0, bb=30 180 590 670, width=0.95\columnwidth]{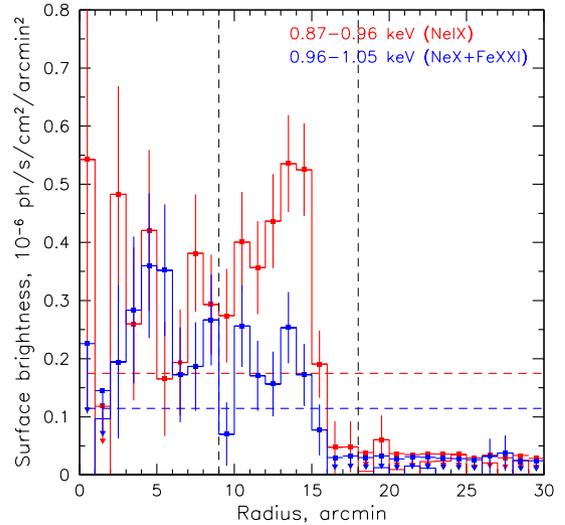}
\caption{Same as Fig. \ref{fig:rprof} but for the narrow energy bands centred on various emission lines. Top panel - 0.4-0.7 keV (O~VII and O~VIII, red) and 0.7-0.9 (Fe~XVIII, blue). Middle panel - 0.52-0.61 keV (O~VII, red) and 0.61-0.7 (O~VIII, blue). Bottom panel - 0.87-0.96 keV (Ne~IX, red) and 0.96-1.05 (Ne~X and Fe~XXI, blue) The dashed horizontal lines indicate the levels of the subtracted background emission, vertical lines mark 9' and 18' radii.}
\label{fig:rprofn}
\end{figure}

\begin{figure}
\centering
\includegraphics[angle=0, width=1.0\columnwidth]{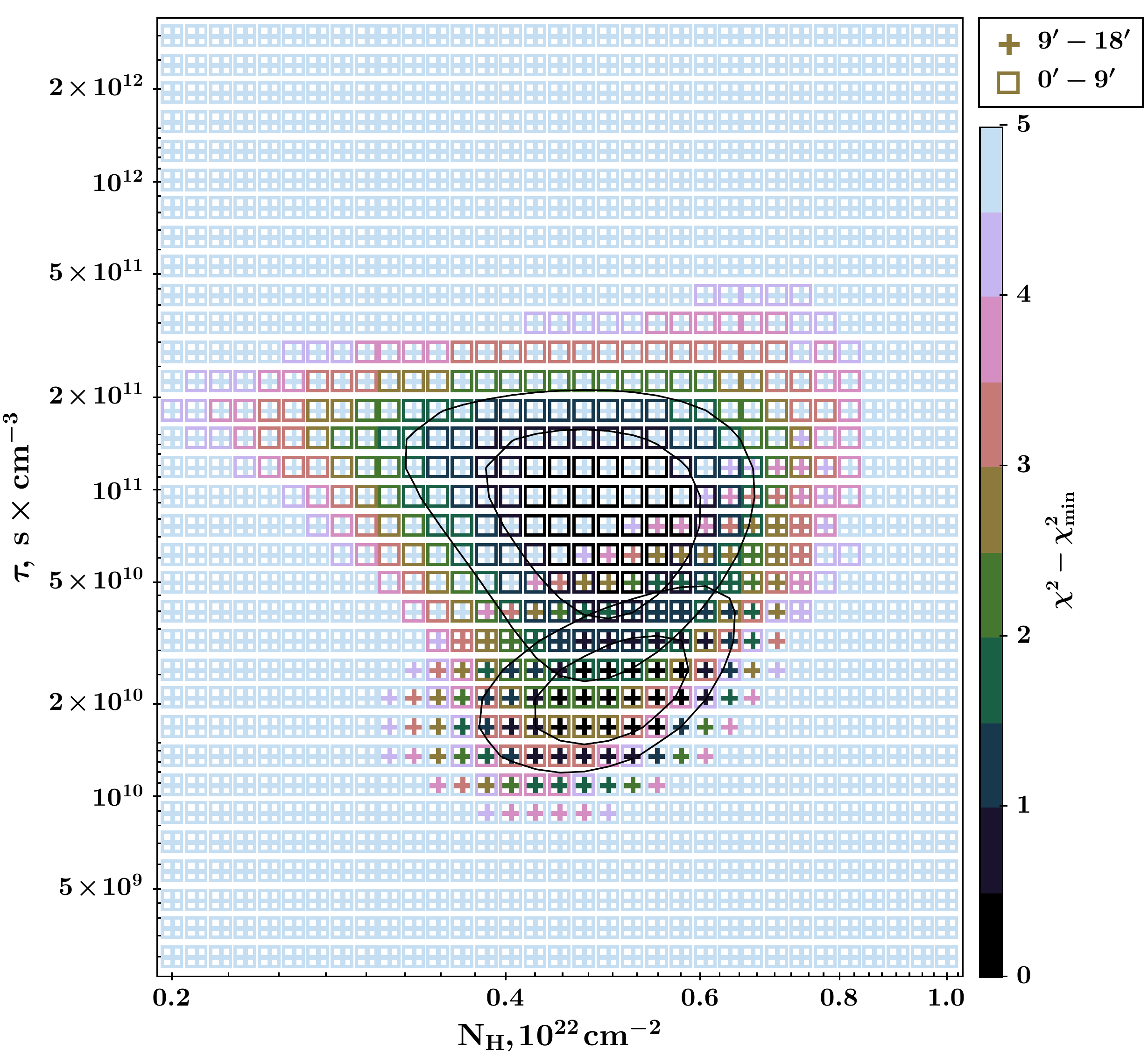}
\caption{Results of the regular parameter space mapping in $N_{H}$ and $\tau$ coordinates for the 0'-9'(squares) and 9'-18'(crosses) regions. The colour shows the difference $\Delta\chi^2$ in the $\chi^2$ value with respect to the minimum value $\chi^2_{min}$, demonstrating the shape of the confidence contours around the best fit values. The grey and black contours correspond to the $\Delta\chi^2$ values of 1 and 2 for the inner and outer regions, respectively.  }
\label{fig:steppar}
\end{figure}

{
We have built the radial profiles of the X-ray emsission in narrow energy bands, containing the oxygen lines (O~VII and O~VIII, 0.4-0.7 keV), iron lines (Fe~XVII, 0.7-0.9 keV) and neon lines (0.9-1.1 keV).  The resulting profiles (after subtraction of the constant level estimated from the ring outside the object) are shown in top panel of Figure \ref{fig:rprofn}. One can easily see the difference in the radial behaviours of the emission in these bands - for the 0.4-0.7 keV, the radial profile is edge-brightened; for the 0.7-0.9 keV, it is centrally-peaked. The major contribution to the edge-brightening in the oxygen band is provided by the O~VII line, as can be seen in the radial profiles built for even narrower energy bands, 0.52 to 0.61 and 0.61 to 0.7 keV, centred on the O~VII and O~VIII lines, respectively (middle panel in Figure \ref{fig:rprofn}). Similar behaviour is observed for the lines of helium- and hydrogen-like neon, Ne~IX and Ne~X in 0.87-0.96 keV and 0.96-1.05 keV bands (bottom panel in Figure \ref{fig:rprofn}).}

{Some part of the centrally peaked emission profile for iron might in principle be attributed to the direct contribution and gas enrichment by the iron-rich ejecta, but the observed behaviour of different lines can also be affected by the ionisation state gradient across the object, as shown by results of the spectral fitting presented above (see also the discussion in Section \ref{s:discussion}). Once again, in order to disentangle these possibilities the de-projected spectra must be considered, which requires a robust model for the radial emission measure distribution.}

{ We note that in the case of very high metal enrichment of the gas, the \texttt{vrnei} model is not applicable, and specially calculated models of metal plasma emission \citep[e.g.][]{1984ApJ...287..282H} need to be used. This could strongly affect derived emission measure of the gas but the estimated effective electron temperature is likely to stay unchanged since the constraint on it is driven primarily by the dominant ionization state of the most abundant ions (e.g. Ne-like iron and He-like oxygen). Although this might change density and mass estimates for the inner part of the object, the outer part most likely corresponds to the region of shocked circum- and interstellar medium. As a result, one might expect only moderate overabundance and not dominance of the ejecta metals.}

\begin{table}
%\centering
\caption{Main observed parameters of the X-ray emission from the newly found SNR candidate G121.1-1.9, namely equatorial and Galactic coordinates of its centre, angular size(radius), the total observed background-subtracted 0.4-2.3 keV flux {and the intrinsic 0.4-2.3 keV after correction for the interstellar absorption}. The spectral parameters are given for emission from the R=0'-9' (inner) and 9'-18' (outer) regions fitted by the absorbed \texttt{vrnei} model with the initial temperature $\rm kT_1$, current temperature $\rm kT_2$, ionisation parameter $\tau$, relative overabundances of iron $\rm Z_{\rm Fe,inner}$ (for the inner region) and neon $Z_{\rm Ne,outer}$ (for the outer region), and the absorbing column density $\rm N_H$, and the \texttt{vrnei} model normalizations per arcmin$^2$. All quoted errors correspond to 1$\sigma$ uncertainty verified by extensive parameter space mapping. 
}
\vspace{0.1cm}
\begin{center}
\begin{tabular}{rr}
\hline
 Parameter & Value  \\ 
\hline
\hline
FK5 position [deg]& (9.0, 60.9)\\
Galactic position [deg] & (121.06,-1.91) \\
Angular radius $R_a$ [arcmin]& $18$ \\
Observed 0.4-2.3 keV flux [${\rm 10^{-12} erg\,s^{-1}\,cm^{-2}}$] & $2.1\pm0.1$  \\
Intrinsic 0.4-2.3 keV flux [${\rm 10^{-12} erg\,s^{-1}\,cm^{-2}}$] & $26\pm1.5$  \\
N$_{H}$ [$10^{21} {\rm cm}^{-2}$] & $5.0\pm1.1$ \\
kT$_{1}$ [keV] & 0.1 \textit{(fixed)} \\
 kT$_{\rm 2,inner}$, kT$_{\rm 2,outer}$ [keV] & $0.50\pm0.10$, $0.44\pm0.15$\\
$ \tau_{\rm inner}$, $\tau_{\rm outer}$ [$10^{10}$ s cm$^{-3}$]& $8.9^{+8.3}_{-5.4}$, $2.0^{+1.5}_{-0.7}$\\
Z$_{\rm Fe,inner}$, Z$_{\rm Ne,outer}$& $1.6\pm0.5$, $3.3\pm1.2$   \\
N$_{\rm inner}$, N$_{\rm outer}$ [$10^{-6}$ arcmin${^{-2}}$]& $3.1^{+7.8}_{-1.8}$, $1.8^{+3.8}_{-0.9}$\\  
\hline
\hline
\end{tabular}
\end{center}
\label{tab:obs}
\end{table}

%------------------------
%-------------------------

%-------------------------
%------------------------
%------------------------

%%%
\subsection{Distance estimate from the 3D absorption profile}
\label{ss:3dabsorption}
%%%

\begin{figure}
\centering
\includegraphics[angle=0,bb=40 180 570 680,width=0.99\columnwidth]{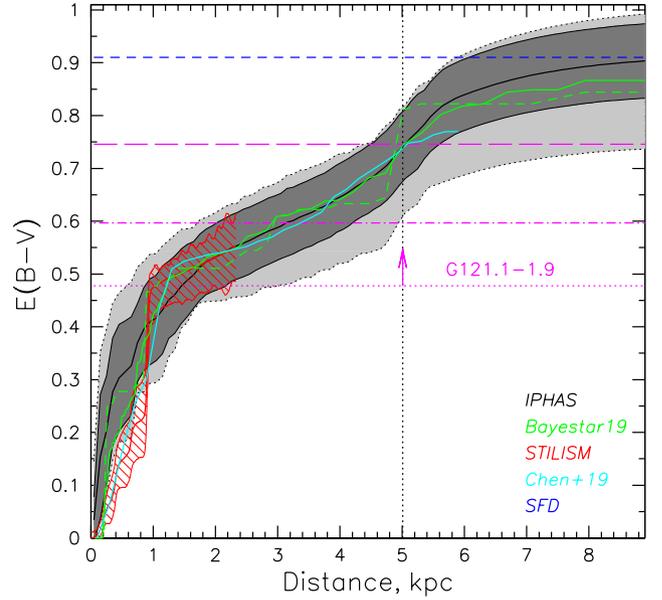}
\caption{Cumulative reddening profile in the direction of G121.1-1.9 based on the data IPHAS(black), Bayestar2019 (green), STILISM (red) and \citealt{2019MNRAS.483.4277C} (cyan). For the IPHAS profile, the black solid line shows the mean profile for the R=18' region centred on G121.1-1.9 with the dark grey region showing the uncertainty range equal to the standard deviation of the samples in each distance bin, while the light grey region depicts the range from the minimum to the maximum value. For Bayestar2019, the solid line shows the mean profile and the dashed line - best fit profile. For STILISM, the mean profile (solid line) and the uncertainty region (hatched region) are shown. For \citealt{2019MNRAS.483.4277C}, the mean profile or the R=18' region is shown. The total E(B-V) value from the \citealt{1998ApJ...500..525S} is shown as blue dashed line. All profiles have been matched at 1.5 kpc by a slight adjustment in normalizations. The dashed magenta line shows the reddening values estimated from the best fit values of the X-ray absorption, while the magenta dashed-dotted and dotted lines show the 1$\sigma$ and 90\% lower limits, correspondingly. }
\label{fig:ebvprofile}
\end{figure}

The measured value of the interstellar absorbing column density $N_H$ allows us to estimate the distance to the object based on the 3D maps of the Galactic extinction. Namely, we took advantage of the maps of \cite[][]{2014MNRAS.443.2907S} based on the photometric data of the IPHAS/IGAPS surveys, and \citep[][]{2019MNRAS.483.4277C}, STILISM \citep[][]{2017A&A...606A..65C,2019A&A...625A.135L}, and Bayestar19 \citep[][]{2019ApJ...887...93G}, all based on the \textit{Gaia} \citep[][]{2016A&A...595A...1G} parallax measurements. All these maps differ in the probed distance range and accuracy of the reconstruction, as well as in the zero-point offset and conversion to the $E(B-V)$ values. Since we are interested mostly in the shape of the cumulative distribution of the absorption, i.e. the way how it is accumulated over the line-of-sight, we start by checking that the IPHAS profile is consistent with the LOS-integrated value of absorption from \citealt{1998ApJ...500..525S} at its furthest end. After that, by slight adjustment in normalisations, we match all other profiles around 1.5 kpc, where all of them appear to be well constrained. The resulting combined profile for a $R=18'$ aperture centred on G121.1-1.9 is shown in Figure \ref{fig:ebvprofile}.    

We convert the measured X-ray absorption $N_H=(5.0\pm1.1)\times10^{21}$ cm$^{-2}$ to the $E(B-V)$ value by rescaling the total $E(B-V)$ value in that direction down to by the factor $N_{H,tot}/N_H\approx1.3\pm0.3$, resulting in $E(B-V)\approx0.75\pm0.15$, meaning that $E(B-V)\approx0.6$ is a 1$\sigma$ lower limit. Taken at face values,  $E(B-V)\approx0.75$ would locate G121.1-1.9 at the distance of 4-6 kpc, and the 1$\sigma$ lower limit at 2.5 kpc. If we consider 90\% confidence level for the measured $N_H$, we get the lower limit for the distance around 1 kpc, meaning that it should be located far enough so that the bulk of the local gas absorption is already accumulated (see the discussion in Section \ref{s:counterpart}).     
%------------------------
%-------------------------

%-------------------------
\section{Data at other wavelengths}
\label{s:counterpart}
%-------------------------

Since the newly discovered object is located in the Galactic plane with $b\approx-2$ deg, a plenty of publicly available data exist for this region, including those which have been commonly used for identification of the supernova remnants.

%-------------------------
\subsection{H$\alpha$}
\label{ss:halpha}
%-------------------------

We have used the data of the INT Galactic Plane Survey (IGAPS) in order to construct a mosaic image of the 1.4$\times$1.4 deg$^2$ region surrounding G121.1-1.9 in the narrow-band centred on H$\alpha$ emission line \citep[][]{2021A&A...655A..49G}. The resulting image, built after iterative background matching and co-adding of the individual images, performed using \texttt{MONTAGE} software\footnote{\url{http://montage.ipac.caltech.edu}},  is shown {in the atlas of the maps} in Appendix. 

The image was convolved with a 20'' filter with 3$\sigma$-clipping applied in order to suppress contribution of the numerous point point source present in the field. In addition, we have also constructed a continuum-corrected H$\alpha$ map based on the data of the Virginia Tech Spectral-Line Survey (VTSS) \citep[][]{1998PASA...15..147D}, featuring poorer angular resolution but more uniform coverage across the field. The distributions of the large-scale diffuse emission in both maps agree quite well, justifying the background-matching procedures applied to the IGAPS images.

No indication of the excess emission coming from the position of G121.1-1.9 can be noticed, except for the relatively faint and narrow filaments, which are although quite ubiquitous for a field close to the Galactic plane. 

%-------------------------
\subsection{Radio}
\label{ss:radio}
%-------------------------

Radio emission is ubiquitously detected from the supernova remnants in our Galaxy \citep[][]{2019JApA...40...36G,2019SerAJ.199...23S} and in external galaxies \citep[e.g.][]{2005A&A...435..437U}. Moreover, there is a relation connecting the physical size and surface brightness of the radio emission, i.e. the so-called $\Sigma-D$ relation \citep[][]{1976MNRAS.174..267C,L81,B86,Green84,2005A&A...435..437U,2005MNRAS.360...76A,2019SerAJ.199...23S}, which possibly reflects evolutionary track of the remnants located in some "typical" Galactic environments   \citep[e.g.][]{BV04,2006A&A...459..519A,2020NatAs...4..910U}.

Thanks to the location in the Galactic disc, the position of G121.1-1.9 falls into footprints of a number of sensitive radio surveys at various frequency bands. Namely, we constructed maps of the 1.4$\times$1.4 deg$^2$ regions surrounding the position of G121.1-1.9 based on the data of the VLA Low-Frequency Sky Survey Redux (VLSSr) at 74 MHz \citep[][]{2007AJ....134.1245C,2012RaSc...47.0K04L}, the TIFR GMRT Sky Survey (TGSS) Alternative Data Release 1 (ADR1) at 150 MHz \citep[][]{2017A&A...598A..78I}, the Westerbork Northern Sky Survey (WENSS) at 92 cm \citep[326 MHz, ][]{1997A&AS..124..259R}, the Canadian Galactic Plane Survey (CGPS) at 408 MHz and 1.4 GHz \citep[][]{2003AJ....125.3145T}, the NRAO VLA Sky Survey (NVSS) 1.4 GHz \citep[][]{1998AJ....115.1693C}, Effelsberg Radio Continuum Survey of the Galactic Plane at 11 cm \citep[2.7 GHz,][]{1984A&AS...58..197R,1990A&AS...85..691F}, the Green Bank 6cm (4.85 GHz) survey \citep[][]{1994AJ....107.1829C}, the Arcminute Microkelvin Imager Galactic Plane Survey (AMIGPS) at 15.7 GHz \citep[][]{2013MNRAS.429.3330P,2015MNRAS.453.1396P}. Some of them were generated using services provided by the the Centre d'Analyse de Données Etendues (CADE)\footnote{\url{http://cade.irap.omp.eu}}  \citep[][]{2012A&A...543A.103P}. The resulting maps are presented in the atlas in Appendix and none of them show any {obvious} signatures of the excess emission coming from the position of G121.1-1.9.

In order to estimate any possible excess surface brightness from the region of G121.1-1.9, we defined the source region as a circle with $R=18$ arcmin and selected five rectangular  regions devoid of bright point sources surrounding it for background estimation (see Appendix). In addition we used a mask for point sources constructed from the CGPS catalogue of source detected at 1.4 GHz \citep[][]{2017AJ....153..113T}. For every region, we estimate the mean surface brightness and its variance, and we also estimate {the} variance of the mean surface brightness {values} obtained from the individual background regions. For all of the inspected radio maps, the surface brightness estimate from the G121.1-1.9 was consistent with the averaged surface brightness estimates from the background regions within the statistical uncertainties. {The similar result is also obtained from the maps of polarised emission at 1.4~GHz of the CGPS survey \citep[][]{2003AJ....125.3145T}.}

From the CGPS data at 408 MHz and 1.4 GHz, we got  conservative 1$\sigma$ upper limits at the level the standard deviation between the different regions, i.e. $\sim1$ K and $\sim0.08$ K, respectively. This implies surface brightness limits of $\sim5000$ Jy/sr at 408 MHz. For the powerlaw index of 1, this means $\sim2000$ Jy/sr at 1 GHz , i.e 2$\times10^{-23}$ W m$^{-2}$ Hz$^{-1}$ sr$^{-1}$.

Taken at face value, this places G121.1-1.9 among the radio-faintest SNRs currently known, e.g. G181.1+9.5 \citep[][]{2017A&A...597A.116K}, G107.0+9.0 \citep[][]{2021A&A...655A..10R}, and G116.6-26.1 \citepalias[][]{2022MNRAS.513L..83C}. All these objects are located at relatively high latitudes out out the Galactic plane, allowing to explain faintness of their radio emission by low density of the surrounding medium. G121.1-1.9 is however located in the Galactic plane, and we explore next whether it could be associated with some low density environment.   

%-------------------------
\subsection{Global gas environment and search for cavities}
\label{ss:hi}
%-------------------------

\begin{figure}
\centering
\includegraphics[angle=0, width=1.\columnwidth]{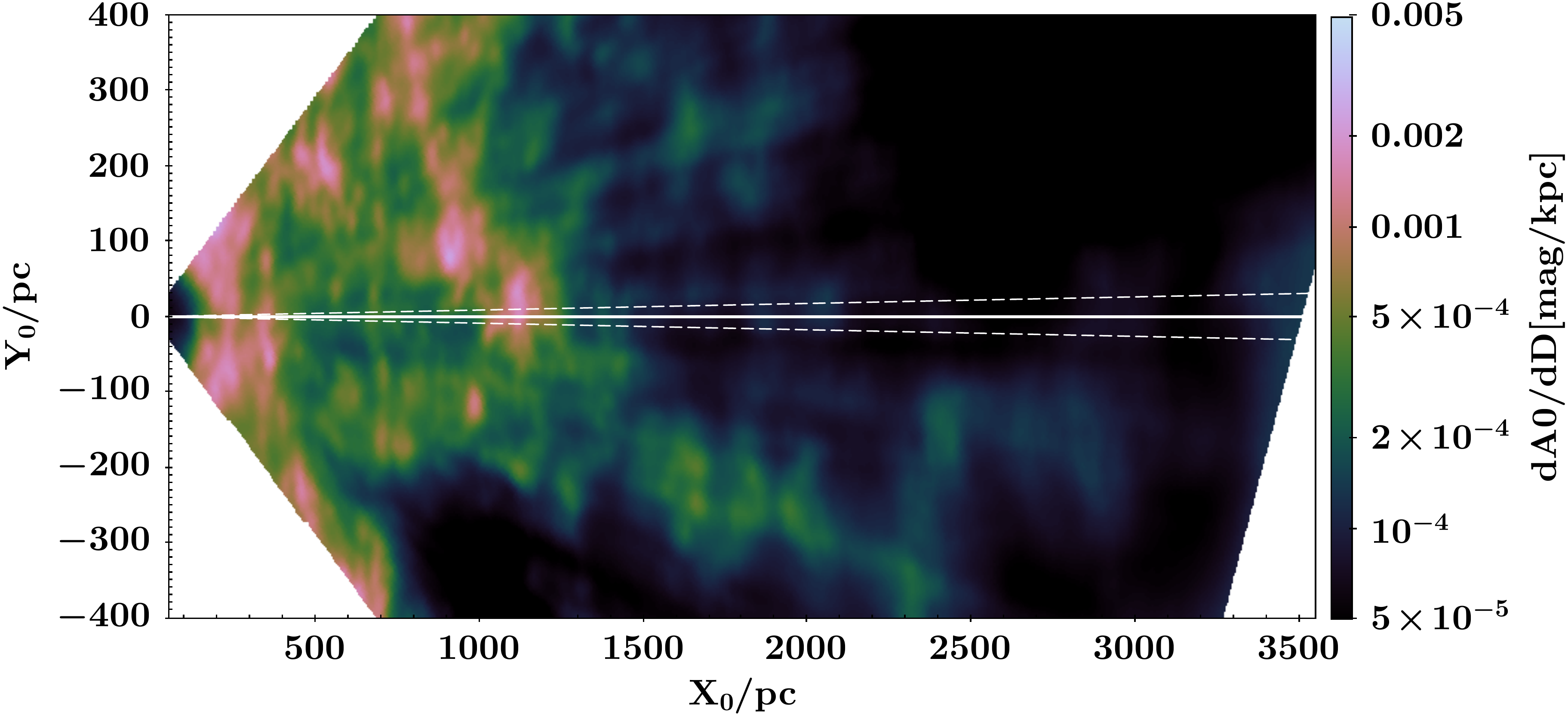}
\includegraphics[angle=0, width=1.\columnwidth]{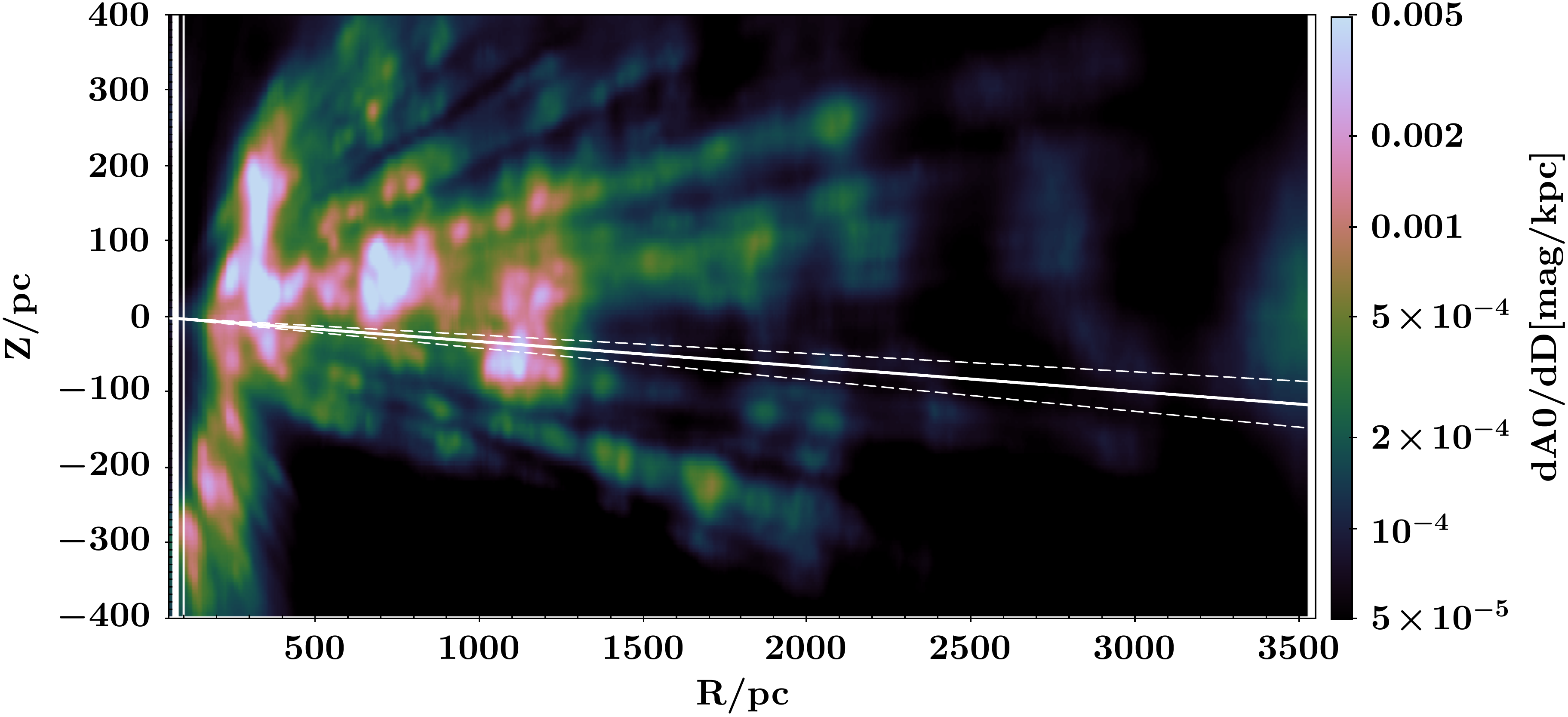}
\caption{3D maps of the dense gas distribution based on STILISM. Top panel shows projection of 200-pc thick slice in the Galactic plane, with the X axis being directed towards G121.1-1.9. Bottom panel shows gas distribution in a 1-deg wide wedge perpendicular to the Galactic plane and centred on G121.1-1.9. The white dashed lines illustrate the angular extent of the object.}
\label{fig:3dgas}
\end{figure}

Thanks to the data of \textit{Gaia} observatory \citep[][]{2016A&A...595A...1G}, the distribution of gas within a few kpc around the Sun can be now reconstructed with high accuracy \citep[e.g.][]{2019A&A...625A.135L}. In Figure \ref{fig:3dgas} we show two projections of the global gas distribution centred on the direction of G121.1-1.9 extracted from the 3D maps of the STILISM project \citep[][]{2019A&A...625A.135L}. Namely, the X-Y projection shows a view "from top" on a 200 pc-thick slice of the Galaxy, i.e. it shows distribution of the gas in plane of the Galactic disc, with the X axis being directed towards the center of G121.1-1.9. The dashed lines illustrated the angular extent of the object. Correspondingly, Z-R projection shows the projection of a 1-deg thick wedge perpendicular to the Galactic disc, with the R being a radial coordinate from G121.1-1.9.

In agreement with the radial profiles of absorption considered in Section \ref{ss:3dabsorption}, we see that the bulk of the gas in this direction is concentrated within 1.5 kpc, corresponding to the Local Arm of the Galaxy. Further gas condensations start to appear a distance of $\sim3.5$ kpc, corresponding to the Perseus Arm \citep[][]{2021A&A...645L...8X}. Location of the object beyond 1.5 kpc would indeed imply relatively low density environment, corresponding to the inter-arm region at the height of $\sim100$ pc below the Galactic plane.

On the other hand, a dark remnant can also be located in a cavity within a denser global environment. Such a cavity might have been created for instance by previous supernova explosions in its vicinity \citep[e.g.][]{2021Galax...9...13S}. Possible indications of such a circular cavity might be seen at the distance of $\sim600$ pc (see the bottom panel in Figure \ref{fig:3dgas}). At this distance, however, the absorbing column density would be significantly smaller, than the one inferred from the X-ray spectra.

Cavities in the distribution of atomic and molecular gas associated with old supernova remnants can also be searched for in the velocity-resolved data on atomic hydrogen and molecular species. In order to do so, we used the velocity-resolved HI data from the Canadian Galactic Plane Survey \citep[][]{2003AJ....125.3145T}.

 Namely, we have defined seven rectangular regions 30$\times$30 arcmin$^2$ in size spaced in a honeycomb structure centred on G121.1-1.9 and with the central one fully encompassing it (see Figure \ref{fig:hiimages}). For all these regions, we extract velocity profiles of the HI emission and look for any kinematic disturbance signatures inside the source region in comparison to the surrounding background regions.
 
\begin{figure}
\centering
\includegraphics[angle=0,bb=30 260 600 610, width=1.\columnwidth]{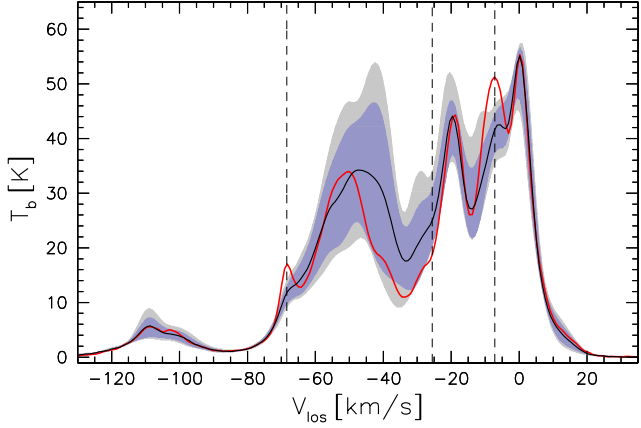}
\caption{Line-of-sight velocity profile of the HI emission from the 30$\times$30 arcmin region centred on G121.1-1.9 (red line) in comparison to the averaged profile from the six surrounding rectangular regions of the same size (black line). The blue shaded region shows the range from the averaged value minus standard deviation to the averaged value plus standard deviation, while the shaded grey region shows the span from the minimum to the maximum value among all background regions in every velocity channel.
The dashed vertical lines mark three velocity regions (around -68.2, -25.4 and -7.2 km/s), where the surface brightness within the source region falls beyond the range charted by the minimum and maximum values. }
\label{fig:hiprofile}
\end{figure}

 Of course, the extracted signal varies noticeably (and in a velocity-dependent way) from one background region to another, reflecting inhomogeneity of the atomic gas distribution on the scale of $\sim1$ deg in this field near the Galactic plane. Still, averaging over them should smear out large variations and even take into account some large scale gradients. As very simple proxies, we use the mean surface brightness value averaged over all six background regions along with its standard deviation and supplement them with the minimum and maximum values (among all six regions) for each velocity bin. After that, we compare the velocity profile, extracted from the source region, with the "averaged" profile with its deviations and min-max boundaries, as shown in Figure \ref{fig:hiprofile}.

We do not observe any obvious distortion signatures present in the source region, except for the relatively minor excesses at -68.2 and -7.2 km/s and only very marginal deficit at -25.4 km/s. In order to check for the "look-elsewhere" effect, we repeated the same procedure for all background regions used as a source regions, and confirm that such minor deviations reflect just common irregularities in the atomic gas distribution at $\sim1$~deg scale, which are more prominent at largely negative velocities corresponding to more distant structures.  At velocities above -5 km/s, corresponding to more local gas, variations are very small, and the signal within the source region is fully consistent with the those from the background regions.  The image cut-outs corresponding to velocities  -68.2 km/s, -25.4 km/s, and -7.2 km/s, are shown in Figure \ref{fig:hiimages}.

It is worth mentioning that the two main components visible on the HI profiles correspond to the gas in the Local Arm, $V_{los}\gtrsim-30$km/s, and Perseus Arm,-80 km/s $\lesssim V_{los}\lesssim-30$km/s \citep[e.g.][]{2022ApJ...925..201P}. There is also an indication of the gas in the Outer Arm at  $V_{los}\approx-110$km/s \citep[e.g.][]{2019ApJ...885..131R}. Thus, association of G121.1-1.9 with the HI distortion features at -70 km/s (and from -80 km/s to -30 km/s in general) would imply its location in the Perseus Arm, i.e. at $\sim3.5-5$ kpc from the Sun. Contrary to this, distortion features at -25 km/s and especially at -7 km/s should correspond to the gas within 1.5 kpc.

%-------------------------
\subsection{Massive stars}
\label{ss:stars}
%-------------------------

The distribution of massive star formation regions which are might be birth place for a { core collapse supernova} progenitor is closely related to the distribution of the dense gas \citep[e.g.][]{2019A&A...625A.135L}. For the direction of G121.1-1.9, the main star-formation regions \citep[e.g.][]{2019ApJ...885..131R} and massive individual O-type stars \citep[][]{2021A&A...645L...8X} are indeed concentrated within the two main arms, the Local Arm and the Perseus Arm. A progenitor star with the initial velocity of $\varv\sim10~\varv_{10}$ km/s can travel $d\approx 100~ \mathrm{pc}~\varv_{10}~t_{10}$ in 10$~t_{10}$ Myrs. This distance is in principle enough to escape denser regions of the spiral arm and enter the inter-arm medium. On the other hand, 100 pc displacement in the picture plane implies a shift by $\approx1.9 (d/3\mathrm{kpc})$ deg, so that original birth site of the progenitor star is difficult to identify.

\begin{figure}
\centering
\includegraphics[angle=0, width=1.\columnwidth]{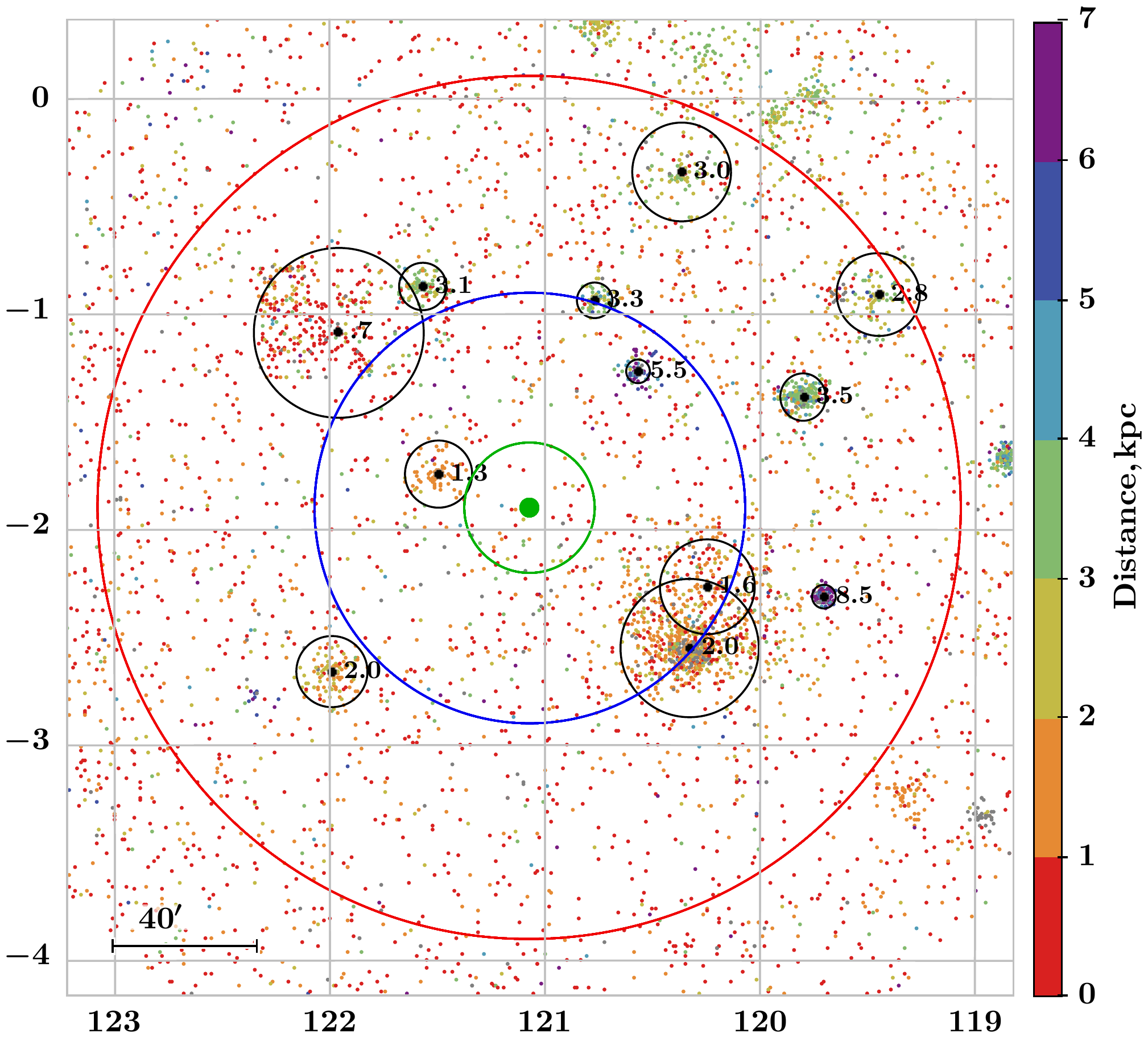}
\caption{Sky distribution (in Galactic coordinates {in degrees}) of stars and star clusters in the G121.1-1.9 field from the SIMBAD astronomical database. The green, blue and red, circles show 0.3, 1 and 2 deg radii, respectively. The stars are coloured according to their distances estimate from parallax measurements. Optical star clusters are marked with black circles, demonstrating their angular extend and labelled according the distance estimate to them. }
\label{fig:starsclusters}
\end{figure}

Indeed, we have searched SIMBAD database \citep{2000A&AS..143....9W} for all objects within $\sim3$ degrees of G121.1-1.9, and selected only stars and star clusters with the parallax measurements. The resulting sky distribution of them is shown in Figure \ref{fig:starsclusters}. The nearest star clusters, NGC~189, is located 0.45 deg away and has the distance estimate 1.3 kpc, i.e likely at the furthest edge of the Local Arm. The second one, NGC~136, is located at 5.5 kpc, instead, likely corresponding to the Perseus Arm. There is a number of other clusters being present in this field as well, with the nearest one located at 0.7 kpc and the median value for the distance being close to 3 kpc. Thus, location of the progenitor star somewhere in the inter-arm region from 1.3 kpc to 5.5 kpc away from the Sun and 50-200 pc away from the Galactic plane is not unlikely.

%-------------------------
\subsection{X-rays}
\label{ss:xray}
%-------------------------

The latter argument is further confirmed by the presence of  High Mass X-ray binary, IGR~J00370+6122, in the same field (0.46 deg to the North of G121.1-1.9 center) at the estimated distance of 3-4 kpc \citep[e.g.][]{2021PASJ...73.1389U}. Also the spectral analysis of this system shows that the likely value of the interstellar absorbing column density is $\approx6\times10^{21}$ cm$^{-2}$, consistent with the value inferred by us for G121.1-1.9 (the same is true for the inferred reddening of the optical companion, BD+60 73 = HIP 2930, \citealt{2021ApJS..254...42B}). Moreover, the measured proper motion of this star, $\mu_{\rm RA}=-1.644\pm0.034$ mas yr$^{-1}$ and $\mu_{\rm Dec}=-0.804\pm0.036$ mas yr$^{-1}$, implying in plane velocity $\approx30$ km/s at the distance of 3.5 kpc, indeed allowing to travel at least 30 pc ($\approx0.5$ deg) in 1 Myr. 	

It is also interesting to look for a possible remnant of the supernova explosion, i.e. a central compact object (CCO), among the X-ray sources observed in the field. In the case of an isolated neutron star it is expected to have very soft X-ray spectrum and lack an optical counterpart, akin CCO discovered in the Cas A \citep[][]{1999IAUC.7246....1T,2000ApJ...531L..53P} and Puppis A \citep[][]{1996ApJ...465L..43P} SNRs.

\begin{figure}
\centering
\includegraphics[angle=0,width=0.99\columnwidth, bb = 60 190 540 640]{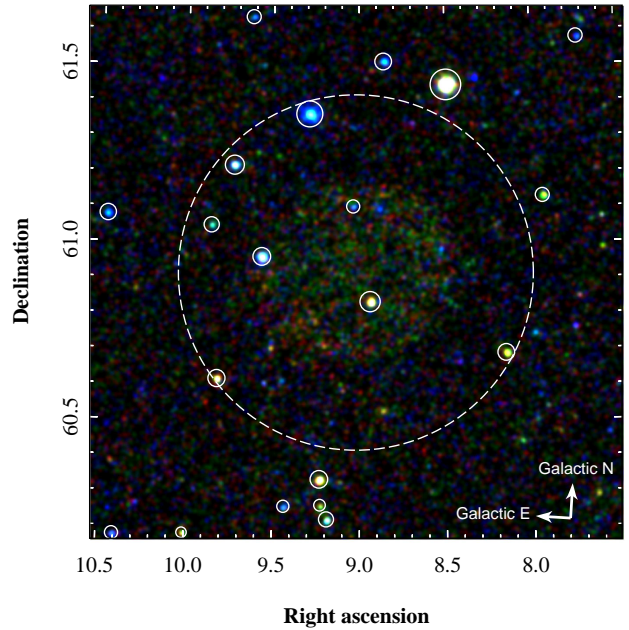}
\caption{A pseudo-colour RGB image showing surface brightness distribution in 0.4-0.7 keV (red) and 0.7-1.3 keV (green) and 1.3-2.3 keV (blue) bands without exclusion of the point sources. The sources masked in the analysis of the diffuse emission are highlighted by the white solid circles, the white dashed circle shows the 0.5 deg radius.     
}
\label{fig:rgbimagepoint}
\end{figure}

There is a number of other point sources with the 0.5-2 keV fluxes above 3$\times10^{-14}$ erg s$^{-1}$ cm$^{-2}$ within this radius, with the majority of them having relatively soft spectrum and probably associated with stars within 1 kpc (see Figure \ref{fig:rgbimagepoint}). For instance, the brightest ($F_{X,0.4-2.3}\approx2.4 10^{-13}$ erg$^{-1}$ s$^{-1}$ cm$^{-2}$) source inside the extend of G121.1-1.9, SRGe~J003545.0+604955, is probably associated with a nearby ($\sim200$ pc) star Gaia EDR3 430106477517090560 \citep{2021ApJS..254...42B}. The HMXB IGR~J00370+6122 clearly stands out in this image thanks to its relative brightness and very blue colour.

Another remarkable source is 1RXS~J003817.1+605815, which is located to the East right at the boundary of the diffuse emission. It is detected by eROSITA as SRGe~J003816+605730 with the flux of $\approx4\times10^{-13}$ erg s$^{-1}$ cm$^{-2}$. It could be associated with a \textit{Gaia} source Gaia EDR3 427131306439799296 (=PS1 181150095671982153) with the measured parallax value 0.33$\pm$0.05 and proper motions $\mu_{\rm RA}=-2.310\pm0.044$ mas yr$^{-1}$ and $\mu_{\rm Dec}=-0.111\pm0.053$ mas yr$^{-1}$ \citep[][]{2021A&A...649A...1G}. There is also a radio source CGPS~J003816+605634(= NVSS~J003816+605635) in 50'' away from it, which is however much larger then the positional accuracy for both X-ray and radio location \citep[][]{2017AJ....153..113T}. The measured proper motion implies velocity of $\sim30$ km/s, while X-ray flux corresponds to the 0.4-2.3 keV luminosity at the level of 4$\times10^{32}$ erg/s. This source however is unlikely to be a remnant, since in order to be at the SNR boundary, it has to move with the same mean velocity as the shock, i.e at least a few hundreds km/s, which is apparently not the case. 

The cross-correlation of X-ray detections down to the flux $10^{-14}$ erg~s$^{-1}$~cm$^{-2}$ with the CGPS catalogue of radio sources \citep[][]{2017AJ....153..113T} gave no probable matches within 15 arcmin from the center.
{ Also, according to the SIMBAD database, no historical nova explosions or other transients have been recorded from this direction in the past.  }

%------------------------
%-------------------------

%-------------------------
\subsection{Gamma-rays}
\label{ss:gamma}
%-------------------------

We have searched the Fermi Large Area Telescope Fourth Source Catalog \citep[4FGL, ][]{2020ApJS..247...33A} for the sources that might be associated with G121.1-1.9. The two nearest sources are 4FGL~J0041.3+6052 and 4FGL~J0035.8+6131, both being $\approx$38 arcmin from the centre of G121.1-1.9 and $\approx$20 arcmin from its boundary, which is well above the positional uncertainties of these sources,$\sim6$ arcmin. {The detected 1-100 GeV gamma rays from  4FGL~J0041.3+6052 might be produced by cosmic rays accelerated in G121.1-1.9 escaped the source and illuminating a nearby dense cloud as it is the case in some other SNRs. However, the gamma ray energy flux detected from 4FGL~J0041.3+6052 is larger than the detected thermal X-ray flux and this is not typical for many other SNRs \citep[see e.g.][]{Slane15}}. 4FGL~J0035.8+6131 is in fact a variable gamma-ray source with unclear identification so far \citep[][]{2018ApJ...862...83P}. We conclude that these sources are unlikely to be a diffuse gamma-ray counterpart of the newly discovered SNR candidate.

%-------------------------

%-------------------------
\section{Discussion}
\label{s:discussion}
%-------------------------

{
The observed X-ray properties of G121.1-1.9 listed in Table \ref{tab:obs} can be briefly summarised as follows: 1) the object has nearly circular shape and surface brightness comparable to the X-ray background level; 2) the integral X-ray spectrum is soft, with no significant emission above 1.3 keV (in excess of the background level); 3) X-ray emission in the bands dominated by O~VII and Ne~IX lines show edge-brightened morphology, while emission in the bands dominated by O~VIII, Ne~X and Fe~XVII lines is radially flat or even centrally-peaked; 4) the spectral shape of the emission from the inner and outer halves of the object can be described by non-equilibrium plasma emission models with the ionisation parameter $\tau$ in the range from $10^{10}$ to $2\times 10^{11}$ cm$^{-3}$~s, initial electron temperature $kT_1\sim0.1$ keV and final temperature $kT_2\sim$ 0.5 keV; 5) the inferred value of X-ray absorption locates object beyond the Local Arm; 6) no counterpart is visible on radio, infrared, optical or gamma-ray images.    } 

\subsection{Derived physical properties}

From the consideration presented in the previous sections, it seems reasonable to assume that G121.1-1.9 is located in the distance range from 1.5 to 6 kpc, so that we can take the fiducial value of $d=3$ kpc and allow a factor of 2 variation in it. In fact, somewhat smaller ($\sim1$~kpc) and larger ($\sim$9~kpc) distances also cannot be excluded, given the systematic uncertainties in the X-ray spectral modelling and 3D absorption profiles, so we conservatively allow a factor of 3 uncertainty in the distance instead. 

The object's apparent radius, 18 arcmin, translates into the physical size of  $R=16~(d/3\mathrm{kpc})$ pc, ranging from $\sim5$ pc at $d=1$~kpc to $\sim50$ pc at $d=9$~ kpc.

From the X-ray spectral fitting, we measured the post-shock temperature at the level of 0.5 keV, which implies the shock velocity $\varv_{sh}=\sqrt{\frac{16~kT}{3~\mu~m_{p}}}\approx$650-715 km/s for the mean molecular weight of the fully ionized gas $\mu=0.5-0.6$.

From the Sedov-Taylor solution, we can approximately estimate the age to be $t_{age}=\frac{2~R}{5~\varv_{sh}}\approx3\times10^{11}$s$~(d/3\mathrm{kpc})\approx9000$ yrs$~(d/3\mathrm{kpc})$. Combining the age estimates with the values of the $\tau$ inferred from the fit, we get estimates for the gas densities in the emitting regions: $n_{\rm e,inner,\tau}=0.3^{+0.3}_{-0.15}$ cm$^{-3}$ $(d/3\mathrm{kpc})^{-1}$ and $n_{\rm e,outer,\tau}\sim0.15^{+0.1}_{-0.05}$ cm$^{-3}$ $(d/3\mathrm{kpc})^{-1}$, where we have assumed the twice smaller age for the outer region.

On the other hand, the gas number densities can be estimated from the normalizations of the emission components obtained by the fit. From the definitions of the normalisation $N$ in the \texttt{vrnei model} \citep{2001ApJ...548..820B}, we get

\begin{equation}
    n_{e}^2\times L=\frac{4\pi~X_e~N}{10^{-14}d\Omega}=0.006~ \mathrm{cm^{-6}~pc}~ \frac{N}{10^{-6}\mathrm{arcmin}^{-2}}
\end{equation}
where $X_e=n_e/n_H$=1.21 for the fully ionised gas with Solar metallicity, $d\Omega$ is the area unit of the measured surface brightness (i.e. 1 arcmin if $N$ is measured per arcmin$^2$), and $L$ is the line-of-sight extent of the emitting region.

As the morphology and the radial profile of the X-ray emission suggest, the size of the emitting region is likely comparable to the size of the object, i.e. $L=a\times R$, with $a=0.5-2$. As a result, from the measured values of the model normlaisations $N_{\rm outer}=3.1^{+7.8}_{-1.8}\times10^{-6}$, $N_{\rm inner}=1.8^{+3.8}_{-0.9}\times10^{-6}$ we can estimate the number densities as

\begin{align}
    \mathrm{n_{\rm e,inner,N}=0.03^{+0.08}_{-0.02}~cm^{-3}~(a_{\rm inner}\times \frac{d}{3\mathrm{kpc}})^{-0.5}}\\
    \mathrm{n_{\rm e,outer,N}=0.02^{+0.04}_{-0.01}~cm^{-3}~(a_{\rm outer}\times \frac{d}{3\mathrm{kpc}})^{-0.5}}
\end{align}
    
%\end{equation}

%\begin{equation}
%\end{equation}

\begin{figure}
\centering
\includegraphics[angle=0, bb=30 200 570 660, width=0.9\columnwidth]{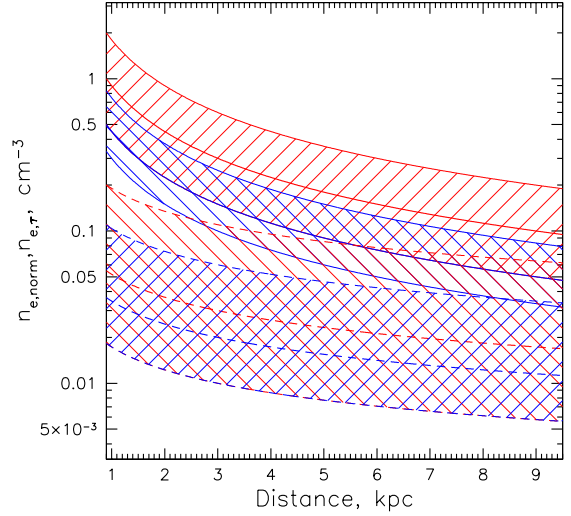}
\caption{Dependence of the $\tau$ (solid lines) and normalisation (dashed lines) based density estimates on the assumed distance to the object. The red hatched regions correspond to  1$\sigma$ uncertainties for the inner 9' arcmin region, while the blue ones - for the 9'-18' arcmin annulus.}
\label{fig:nd}
\end{figure}

Thus, there is an inconsistency between the density estimates by factor of 7-10, which is hard to reconcile by adjusting the distance by a factor of 2 or so. This point is illustrated in Figure \ref{fig:nd} where the inferred values for $n_{\rm e,norm}$ and $n_{\rm e, \tau}$ are plotted with corresponding uncertainties. In fact, the uncertainties in  $n_{\rm e,norm}$ and $n_{\rm e, \tau}$ are not independent due to degeneracies between the fitting parameters of the non-equilibrium emission model. In Appendix \ref{s:deg}, we show degeneracies between the inferred values of $\tau$, final temperature $kT_{2}$, and normalisation $N$.

The degeneracy between $\tau$ and normalisation arises from the emissivity boost $\varepsilon(\tau)$ with respect to the equilibrium emissivity at the same temperature acquired due to enhanced efficiency of collisional excitation in the overheated plasma \citepalias[][]{2021MNRAS.507..971C}. As result, the observed X-ray intensity $I$ can be written as $I\propto \varepsilon(\tau) N$, where $N$ is the usual normalisation. Thus,  $n_{\rm e,norm}\propto \sqrt{N}\propto 1/\sqrt{\varepsilon(\tau)}$ for the fixed $I$.

On the other hand, $n_{e,\tau}\propto \frac{\tau}{t_{\rm age}}\propto \tau \sqrt{kT}$, implying that we can also write down the ratio of the inferred values of densities (for the given distance) as

\begin{equation}
    \frac{n_{\rm e,norm}}{n_{e,\tau}}(\tau)\propto\frac{1}{\tau\sqrt{\varepsilon(\tau)~kT}}\propto \frac{\sqrt{N(\tau)}}{\tau\sqrt{kT(\tau)}}
\end{equation}
where we explicitly emphasised that $kT$ might be expressed as a function of $\tau$ by means of the observed degeneracy relations. An important observation is that this combination of the fitting parameters does not vary much over all area of the parameter space giving the best values of the fitting statistic (see the right panels in Figures \ref{fig:deg_r9} and \ref{fig:deg_r18}) 

Inclusion of the distance dependence then yields
\begin{equation}
    \frac{n_{\rm e,norm}}{n_{e,\tau}}(\tau,d)\propto \frac{\sqrt{N(\tau)}}{\tau\sqrt{d~kT(\tau)}}.
\end{equation}

%-------------------------

In principle, small scale clumpiness of the volume with the volume filling factor $f$ (i.e. the emitting gas occupies only the $f~V$ volume out of the apparent $V=d\Omega D^2L$) can reconcile this problem requiring $f\sim0.01$ and meaning the size of the emitting region is  $\sim\sqrt[3]{f}R\approx0.2R$. 

Another possibility might be related to the overestimation of the $\tau$ parameter in the non-equilibrium fit. The observed spectra can also be fit with very low value of $\tau\sim 10^{9}$ cm$^{-3}$~s, if the starting temperature $kT_1$ is allowed to vary freely. In this case, of course, the fitting model is not sensitive to the final temperature anymore as far as it is $\gg k T_1$. The resulting value of $kT_1$=0.2-0.4 keV, is quite hot for the normal ISM environment and would correspond more to the physical conditions in the Galactic halo \citepalias[similar to the G116.6-26.1 object discussed in][]{2021MNRAS.507..971C}.

Since the emissivity boost factor $\varepsilon(\tau)$ flattens at $\tau<3\times10^{9}$ cm$^{-3}~s$ (see Figure 7 in \citetalias{2021MNRAS.507..971C}), the fitting values for model normalisation are a factor of 4-8 smaller for $\tau$ fixed at $10^{9}$ cm$^{-3}$~s, while the $\tau$ value is smaller by a factor 20-80. As a result, for fixed values of $kT$ and $d$, one can reconcile the values of $n_{\rm e,norm}$ and $n_{\rm e,\tau}$ at $\sim0.01$ cm$^{-3}$. The estimated age would then be $t_{\rm age}\sim 3 000$ yrs, locating it at the distance of $d\sim 1$ kpc or less, which would also consistent be with lower values of the interstellar absorbing column density favoured by these models, $N_{\rm H}\lesssim 2\times10^{21}$ cm$^{-2}$.

{We note, that the scenario with large distance to the object might also imply gas metallicity lower then the Solar metallicity assumed for the X-ray analysis. In this case, since the line and recombination continuum emission from the metals dominate in the spectrum at the considered temperatures, the required emission measure of the emitting gas, and correspondingly { its} number density, would need to be increased, helping to reconcile { its} value with the estimate based on the ionization state.} 

In summary, we conclude that the number density of the emitting gas at the level of $\sim0.1$ cm$^{3}$ and a factor of 4 uncertainty can in principle explain most of the observed X-ray features of the discovered object. From the strong shock jump conditions, this would than indeed imply relatively low density environment, with $n_{e,0}=0.005-0.1$ cm$^{-3}$ consistent with the location of the object $\sim 100$ pc $(d/3 kpc)$ away from the Galactic plane in the inter-arm region.  The "hot" solution, with $kT_1\sim0.3$ keV and $\tau=10^{9}$ cm$^{-3}$~s, the ambient gas would have a factor few smaller density (as the compression ratio is probably not high in this case since relatively low Mach number).

{The observed 0.4-2.3 keV flux corrected after interstellar absorption, $F_{X,0.4-2.3}=(26\pm2)\times10^{-12}$ erg s$^{-1}$ cm$^{-2}$ (see Table \ref{tab:obs}), corresponds to the 0.4-2.3 keV luminosity $L_{X}\approx 3\times 10^{34} (d/3{\rm kpc})^2$ erg s$^{-1}$. }

Finally, we can estimate the total mass of the ambient gas then as

\begin{equation}
M=\frac{4\pi}{3} R^3 \mu_e m_p n_0\approx10 M_\odot \left(\frac{d}{3 \mathrm{kpc}}\right)^{3-\beta} \left(\frac{n_{e,0}}{0.025 \mathrm{cm^{-3}}}\right),     
\end{equation}
where $\beta=1$ for the $\tau$-based estimate, and $\beta=0.5$ for the normalisation based estimate.

The lower limit on the total energy of the explosion can then be estimated as 

\begin{equation}
E_{\rm tot}=\frac{M}{\mu mp} \kappa T\approx 2\times10^{49} \mathrm{ erg} \left(\frac{d}{3 \mathrm{kpc}}\right)^{3-\beta} \left(\frac{n_{e,0}}{0.025 \mathrm{cm^{-3}}}\right) \left( \frac{\kappa T}{0.5 \mathrm{keV}}\right).
\label{eq:etot}
\end{equation}

{For the fiducial distance of 3 kpc, the derived energy is quite small, although for the 2 times bigger distance it would be 4-6 times bigger, i.e. at the level of $\sim10^{50}$ erg, and, for the distance of $\sim9$ kpc, it reaches $(2-3)\times 10^{50}$ erg. For the "hot" solution, instead we would get $M\sim1M_{\odot}$ and $E_{\rm tot}\sim 10^{48} \mathrm{ erg}$ for the distance around 1 kpc or less. Since the temperature inferred from the X-ray fit is likely only a lower limit for the real 'mean' (i.e. averaged over ions and electrons) temperature, the estimates for the total explosion energy might also be considered as lower limits. }

{Given that the spectral fits for the inner part of the object require significant overabundance of iron, it is interesting to estimate total mass of iron in this region under certain simplifying assumptions. Namely, we can a assume uniform density distribution inside the sphere of $R_{\rm a,in}=9$ arcmin, corresponding to $R_{\rm in}=7.85$~$(d/3\mathrm{kpc})$~pc and the volume $V_{\rm in}=\frac{4\pi}{3} R_{\rm in}^3\approx2000$~$(d/3\mathrm{kpc})^{3}$~pc$^3$. The number density of iron atoms $n_{\rm Fe}$ can be estimated directly from the normalisation of the \texttt{vrnei} emission model, taking into account that it should be strongly dominated by emission lines of iron and electron and iron densities can be linked together, so that one has
}
{
\begin{equation}
\frac{N_{\rm in}}{10^{-14}} \pi R_{\rm a,in}^2~\frac{4\pi d^2}{V_{\rm in}}=n_e n_p=\frac{n_{\rm Fe}^2}{X_{\rm Fe}A_{\rm Fe}Z_{\rm Fe}}  
\end{equation}
where $X_{\rm Fe}=n_{\rm Fe}/n_e$, $A_{\rm Fe}=n_{\rm Fe}/n_p$ for the assumed Solar abundance set, and $Z_{\rm Fe}\approx3-5$ is the relative overabundance of iron required by the spectral fit. 

There are two limiting cases, that one might consider for $X_{\rm Fe}$: first, in the case of near-to-solar metal abundance of the emitting plasma,   $X_{\rm Fe}=Z_{\rm Fe}~A_{\rm Fe}/X_e\sim10^{-4}$, $X_e=n_e/n_p\approx1.21$; second, for the plasma of iron ions and electrons only, $X_{\rm Fe}\sim1/16$ if neon-like ions (responsible for Fe~XVII lines) are dominant. Now, one can combine this altogether to obtain
\begin{equation}
M_{\rm Fe,in}=m_{\rm Fe}n_{\rm Fe}V_{\rm in}=m_{\rm Fe}\left( \frac{N_{\rm in}}{10^{-14}}\frac{3\pi}{R_{\rm a,in} d}X_{\rm Fe}A_{\rm Fe}Z_{\rm Fe}\right)^{1/2}\frac{4\pi}{3} R_{\rm a,in}^3 d^3
%0.005 M_{\odot} \left(\frac{d}{\rm 3~kpc}\right)^{5/2}
\label{eq:miron}
\end{equation}
where $m_{\rm Fe}=56~m_{p}$ is the mass of an iron ion. 

For the fiducial values, we get
$M_{\rm Fe,in}=0.01 M_{\odot}\left(\frac{d}{\rm 3~kpc}\right)^{5/2}$ if $X_{\rm Fe}=10^{-4}$ or $M_{\rm Fe,in}=0.25 M_{\odot}\left(\frac{d}{\rm 3~kpc}\right)^{5/2}$ if $X_{\rm Fe}=1/16$. In the case of $d=9$ kpc, this gives $M_{\rm Fe,in}$ between $0.15 M_{\odot}$ and $3.8 M_{\odot}$.
}

\subsection{An explosion model and illustrative simulation}

The current data do not allow us to unambiguously distinguish between the possible scenarios for the nature of such an object, so below we briefly discuss which of them appears to be more plausible.

First of all, all the observational properties point towards a scenario in which the newly found object is located in low density environment, suppressing efficiency of relativistic particles acceleration. 

There are several possibilities that might be a reason for such an environment of a (massive) progenitor star. First, the object might be located out the plane of the Galactic disc and at large Galactocentric distance. Indeed, if we take d$\sim 6$ kpc, corresponding to the further edge of the Perseus arm in this direction, we obtain the estimated energy of the explosion $E_{\rm tot}\sim1\times10^{50}$ erg, the age of 17 000 yrs and the size of 32 pc. 

{If the distance of 9 kpc would be assumed, one gets $E_{\rm tot}\sim(2-3)\times10^{50}$ erg, the age of 25 000 yrs and the size $\sim50$ pc. For this distance the density estimates from the spectral shape and normalisation become closer to each other and would correspond to the environment density $n_{e,0}\approx0.01$ cm$^{-3}$. At this distance, the object would be located 300 pc away from the Galactic plane in vicinity to the Outer Arm, which can naturally explain the required properties of the environment. In this case the parameters of explosion would be close to the canonical values for { core collapse and thermonuclear supernovae}.}

{The electron temperature inferred from the X-ray analysis most likely corresponds to the lower limit on the actual mean temperature of the gas, which would also include {the} temperature of ions. The latter indeed, is expected to exceed the electron temperature for the collisionless shocks \citep[e.g.][]{2007ApJ...654L..69G}, and, for the gas temperatures above 0.3 keV, the electron-ion temperature equilibration time scale might exceed $\tau _{ei}\sim 10^{11}$ cm$^{-3}$ s \citep[e.g.][]{1999LNP...520..189L}. This would then imply, that the observed X-ray emission might either reflect the non-equilibrium ionisation state of temperature equilibrated plasma, or, instead, correspond to equilibrium ionisation of the gas with the evolving temperature.}

{The explosion energy from the Eq.\ref{eq:etot} needs then to be adjusted upward by a factor depending on the Mach number of the shock and its effective collisional "age". This should bring the explosion energy estimate even closer to the canonical value of $10^{51}$ erg.}

\begin{figure}
\centering
\includegraphics[angle=0,bb=30 170 580 700,width=0.9\columnwidth]{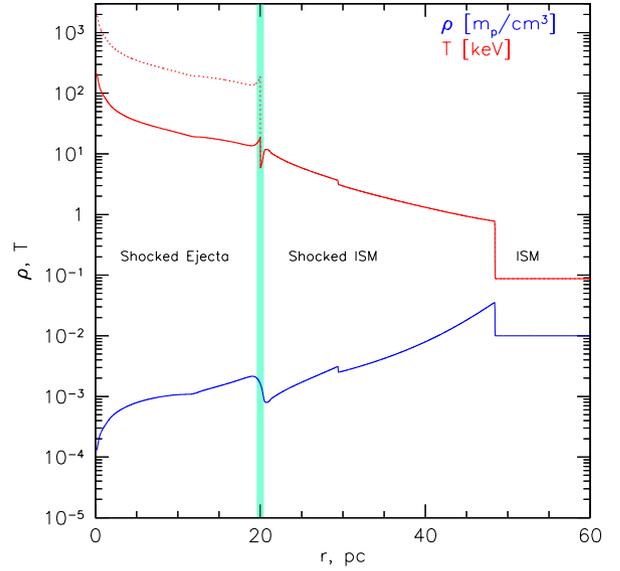}
\caption{{Simulated density and temperature profiles in a fiducial {scenario of a SN Ia explosion in a hot and low density ISM. The initial ISM density and temperature are $\rho=10^{-2}\,{m_p\rm cm^{-3}}$ and $T=10^6\,{\rm K}\approx 0.086\,{\rm keV}$, respectively.} The vertical line separates shocked ejecta from the shocked ISM. The solid and dotted lines show the {derived temperature of the ejecta assuming $\mu\sim2$ and $\mu\sim20$, respectively (in the ISM with the Solar abundance of heavy elements, $\mu\approx 0.62$).} The low temperature solution corresponds to fully ionized gas (e.g., pure Fe), while the hotter solution {formally} assumes that the gas is neutral. {Note that these simulations do not include radiative losses or effects of ionization,  cosmic ray acceleration, etc.  The two solutions for the temperature are intended only for illustration of the possible temperature range in the non-radiative case. In reality, the temperature of the ejecta can be much lower due to these effects. In the fiducial run, the SNR age is $\sim$25000~yr, the forward shock has a Mach number $M=5.4$. The radius of the forward shock is $\sim 48\,$pc, implying the distance to the SNR $\sim 9\,$kpc. }}}
\label{fig:prof_1d}
\end{figure}

{In order to check the consistency of this picture, we have performed a set of 1D simulations of {an SNR} in the uniform medium with the gas density $\rho=10^{-2} m_p\,{\rm cm^{-3}}$, temperature $kT=10^6$ K, total energy $E_{tot}=10^{51}$ erg and the ejecta mass 1.4 $M_\odot$. {Initially, the ejecta have an exponential density distribution suitable for type Ia supernovae \citep[see, e.g.][]{1998ApJ...497..807D}.}
The resulting radial profiles of the gas density and temperature at the moment 24400~yr after the explosion are shown in Figure \ref{fig:prof_1d}, corresponding to the {forward shock radius} radius 48 pc. The Mach number of the shock is $M=5.4$, the post-shock temperature $kT_2\sim$0.8 keV, and the ionisation parameter $\tau\sim10^{10}\,{\rm cm^{-3}~s}$ all broadly consistent with the inferred properties of G121.1-1.9. {As a caveat, we note that these 1D simulations ignore radiative losses and other processes that can influence the gas effective equation of state or mixing near the contact discontinuity.}}  

\begin{figure}
\centering
\includegraphics[angle=0,bb=10 160 590 700,width=0.9\columnwidth]{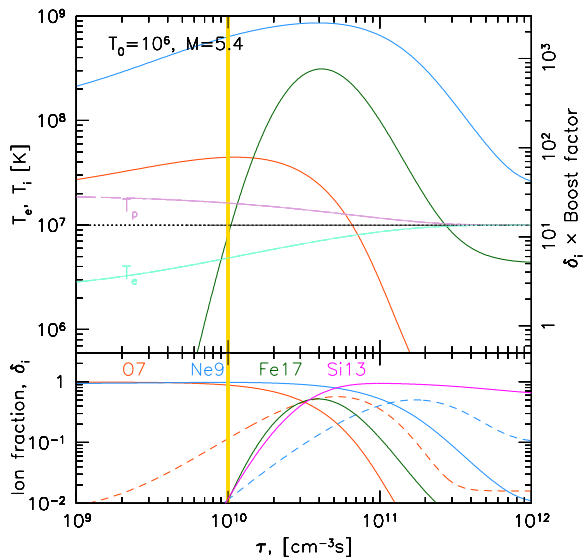}
\caption{{Evolution of the ionization balance and lines emissivities of the gas compressed by a plane shock with the Mach number $M=5.4$. The initial gas temperature upstream of the shock is $T_0=10^6\,{\rm K}$; the gas is initially in a collisional equilibrium. The electron temperature just behind the shock is assumed to change adiabatically, according to the density compression factor, i.e. the initial electron downstream temperature is $T_{e}=T_{0} C^{2/3}$, where $C$ is the density compression factor ($C=3.56$ in the fiducial model). The initial temperature of ions is based on the assumption that ions and electrons combined obey the Rankine-Hugoniot jump condition for $M=5.4$. The subsequent electron-ion temperature equilibration is mediated by Coulomb collisions (the electron and ion temperatures are shown in the top panel with a pale green and violet lines, respectively.) The bottom panel shows the evolution of the ion fractions for He-like ions O~VII (red), Ne~IX (blue), { and Si~XIII (magenta), as well as Ne-like Fe~XVII (green).} The dashed lines show the ion fractions for H-like O~VIII (red) and Ne~X (blue). The colored curves in the top panel show {qualitatively the time evolution of the emissivity} of the same He-like ions {of oxygen and neon, and the Ne-like ion of iron} compared to the pre-shock hot ISM. {The values of the boost factor (multiplied by the ion fractions, i.e. $\eta(\tau)\times \delta_i$),} are indicated by the vertical axis on the right side of the plot. A more rapid electron-ion temperature equilibration would shift all the curves to the left.}     
}
\label{fig:nei}
\end{figure}

{The small value of $\tau\sim 10^{10}\,{\rm cm^{-3}s}$ in our fiducial model implies that the gas downstream of the shock may be far from the equilibrium, both in terms of the electron and ion temperatures and the ionization balance. We illustrate potential departures from these equilibria in Fig.~\ref{fig:nei}. It shows the evolution of the ionization balance as a function of $\tau$ (bottom panel) and the changes in the electron and ion temperatures (top panel). Here we assume that (i) the mean plasma temperature changes at the shock front in accordance with the Rankine-Hugoniot jump condition for $M=5.4$, while the temperature of electrons changes adiabatically, i.e. $T_{e}=T_{0} C^{2/3}$, where $C$ is the density compression factor ($C\approx 3.6$) and (ii) subsequent temperature equilibration is mediated by Coulomb collisions. In addition, the top panel (right vertical axis) shows the impact of the temperature and the ionization state on the emissivity of the major lines. This was done by calculating the following quantity for a few transitions
\begin{eqnarray}
\eta(\tau)=\delta_i \left ( \frac{T_0}{T_e(\tau)}\right )^{1/2} e^{-E_i/kT_e(\tau)+E_i/kT_0} ,
\end{eqnarray}
where $\delta_i$ is the fraction of a given ion and $E_{i}$ is the energy of the most important transition. The colour lines in the upper panel show $\eta(\tau)$ for O~VII, Ne~IX and Fe~XVII. The yellow vertical line shows the characteristic value of the ionization age in the fiducial simulations.
One can see that the Coulomb collision do not guarantee the temperature equilibration and in this case, the ionization balance does not change much in the shocked ISM for $\tau\sim 10^{10}\,{\rm cm^{-3}s}$. The line emissivities for oxygen and, especially for neon, can be boosted by a large factor, potentially explaining the overabundance of neon in the outer layers of the SNR. This qualitative conclusion is broadly consistent with the results of fitting the outer shell spectrum with the \texttt{vrnei} model (see Table 1). In principle, changing the upstream gas temperature, the level of electron heating at the shock and the rate of temperatures equilibration (on top of Coulomb scatterings) provides enough flexibility in shifting the curves (mostly to the left) and offering a possibility of constraining these properties once deeper X-ray observations are available.

The spectrum of the inner region is likely contaminated by the contribution of outer layers (see Section \ref{ss:xrayspec}) and might have a much more complicated time history \citep[cf. the dedicated modelling for Ia supernovae by][]{2003ApJ...593..358B,2005ApJ...624..198B}. We defer the modelling of its spectrum and comparison with the data for future studies.
}

%------------------------
%-------------------------

%------------------------
%-------------------------

%------------------------
\begin{figure}
\centering
\includegraphics[angle=0,bb=10 160 590 700,width=0.9\columnwidth]{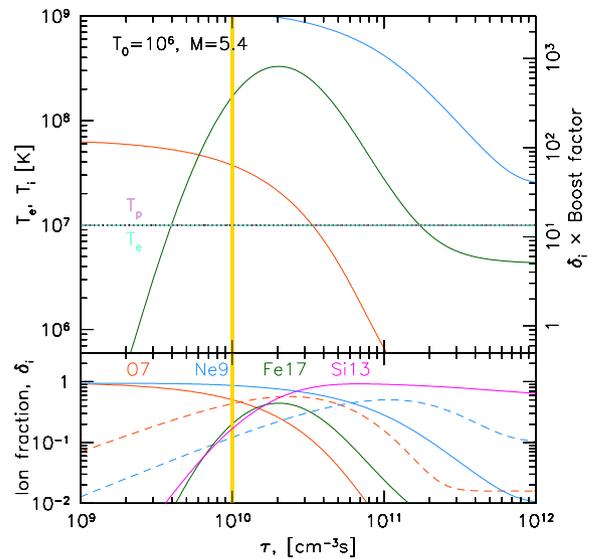}
\caption{ Same as Fig.\ref{fig:nei} but for the case of instantaneous equilibriation of the electron and ion temperatures. The appearance of higher ionization states, in particular Fe~XVII and Si~XIII, and onset of the corresponding line emission is already noticeable before $\tau=10^{10}$ cm$^{-3}$s.    
}
\label{fig:nei_fast}
\end{figure}
%-------------------------

\subsection{Possible progenitors and astrophysical context}

{ 
G121.1-1.9 in some regards looks like a sibling of G116.6-26.1 (\citetalias{2021MNRAS.507..971C}): 
both are (i) round-shaped, (ii) strongly decelerated, (iii) of extremely low radio surface brightness, and (iv) reside 
in the external Galactic sector $90^{\circ} < l < 270^{\circ}$. 

It is not surprising, that such objects should be among the first ones found in a blind all-sky X-ray survey, since 1) the objects in more typical environments have already been found by means of the radio observations, 2) the unperturbed spherical shape makes it much easier to identify them against spatially inhomogeneous fore- and background emission, 3) the life-time of such objects is prolonged thanks to inefficient cooling, so that they stay observable in X-rays for longer periods of time; 4) finally, the objects located in the inner parts of the Galaxy and in the disc suffer from the confusion due to high density of the objects.

Unlike G116.6-26.1 that most likely belongs to the halo, G121.1-1.9 has low Galactic latitude and at first glance lies in the Galactic disc. 
However, hydrodynamic simulations of the SNR expansion in the scenario of SN~Ia in the hot rarefied 
medium combined with the X-ray data indicate the large distance, of $\sim 9$ kpc, in which case 
the Galactic height is of $\sim300$\,pc in line with SN~Ia to be a likely progenitor. 

}

{A fact that we observe an excess of Ne and Fe emission might also be suggestive of the Type Ia scenario, since no oxygen-to-iron overabundance is produced in core collapse supernova, while the apparent overabundance of neon might be an indication of oxygen depletion or be a result of inaccuracies in the non-equilibrium emission modelling. Comprehensive models for the X-ray emission from SN~Ia remnants have been constructed  \citep[][]{1984ApJ...287..282H,2003ApJ...593..358B, 2005ApJ...624..198B}, but the current level of the X-ray data does not allow us to make direct comparisons. Somewhat similar morphology of the X-ray emission was discussed in the context of the supernova remnants RCW 86 \citep[][]{2011ApJ...741...96W}.
{For this SNR in particular, the importance of non-equillibrium ionisation in the post-shock gas has been considered by    \cite[][]{2022arXiv220107607S}, including a possible effect of the accelerated cosmic rays.}

Another object that shares similar properties might be G272.2-3.2, first discovered in the $ROSAT$ data \citep[][]{1994A&A...286L..35G} and later suggested to be a remnant of a type Ia supernova \citep[][]{2013A&A...552A..52S}. Both these objects, however appear to be sources of radio and even gamma-ray emission and feature the second, harder, component in the X-ray spectrum. } {Another difference might be connected to the fact in the distant and old SNR scenario, considered by us above, the reverse shock has already went through the entire ejecta (see Fig. \ref{fig:prof_1d}), which might not be the case for more nearby and younger objects (embedded in the ISM with the same density), so that we observe shock heating of the ejecta at present time.  }
{ There is interesting similarity between X-ray proprieties of G121.1-1.9 and the class of mixed-morphology SNRs in general \citep[][]{1998ApJ...503L.167R}, to which G272.2-3.2 also belongs. For some of the mixed-morphology SNRs, e.g. G65.2+5.7, soft X-ray emission from the outer shell has been observed \citep[][]{2004ApJ...615..275S}, making possible similarity even stronger. However, bright shell-like radio emission is always associated with these objects, and they are believed to be associated to denser-than-average environments. This is opposite to what we observe for G121.1-1.9, as no signatures of radio emission can be revealed in the currently available radio maps. 
}

{The newly discovered object might be a Galactic member of the class of iron rich supernova remnants discovered in Large Magellanic Cloud \citep[][]{2006ApJ...652.1259B,2014A&A...561A..76M,2016A&A...585A.162M,2016A&A...586A...4K} and later found in Small Magellanic Cloud as well \citep[][]{2019A&A...631A.127M}. If confirmed, G121.1-1.9 might offer an opportunity to study the enrichment processes by SN Ia in the evolved phase in our own Galaxy.}

{In this regard, it is interesting to compare the masses of iron derived from the spectral fitting results in Section \ref{s:discussion} (given by Eq. \ref{eq:miron}).
For $d=9$ kpc, the estimate gives $M_{\rm Fe,in}$ between $0.15 M_{\odot}$ and $3.8 M_{\odot}$ within $R=23.5$ pc. On {the} one hand, the actual mass of iron could be even higher if there is fraction of the gas that is not contributing to the observed X-ray emission (e.g. it is too cold or too over-ionized). On the other hand, clumpiness of the emitting medium would mean that smaller mass could be responsible for the same level of emission. Given all the uncertainties, we conclude that the estimated iron mass leaves open possibility for both core collapse and thermonuclear SNR scenarios.}

{It is important to notice, however, that direct physical interpretation of the spectral properties of the observed emission is challenging due to very limited count statistics hindering spatially resolved spectral analysis and application of the de-projection techniques. In particular, for the central part of the object several effects might be in play, e.g. strong overabundance of heavy elements (pure ejecta, or ejecta mixed with the shocked ISM), non-equilibrium ionization, inequality between ion and electron temperatures, thermal conduction and inhomogeneity of the emitting gas. Clearly, better X-ray data should allow clarifying some of these issues and we defer their discussion for future  detailed studies. 
}

{It is natural to ask how probable it would be for a Ia supernova to explode in the hot and tenuous phase of the Galactic disc.}
{ 
Based of the double-exponential stellar density distribution in the thick disc with radial and vertical scales $h_r = 3$ kpc and $h_z = 0.9$ kpc, respectively,  \citep{2017ApJ...849..115L} we expect to find in the external Galactic sector $90^{\circ} < l < 270^{\circ}$ $\sim 5.5$\% of all SNRs residing in the thick disc.
Earlier  \citepalias{2021MNRAS.507..971C}, the number of SNRs with the age $< 10^5$ yr related to SNe~Ia in the thick disc was 
estimated to be of the order of $10^2$, which implies that the expected number of SNRs resembling  G121.1-1.9 and  G116.6-26.1 in the 
external Galactic sector is $\sim 5$. We therefore are surprisingly close to the detection of all similar SNRs in the external part of Milky Way.
}

 { Finally, we consider a few other frequently discussed scenarios that might lead to circular X-ray-emitting structures.
In the scenario of a nova, the object with this temperature should have a characteristic size of 0.5 pc and hence be located at very small distance, $d\sim100$ pc. This is not consistent with the inferred absorbing column density and also no binary systems with white dwarfs are known in the center of the object. A scenario with a remnant of a failed supernova \citep[][]{1980Ap&SS..69..115N,2021PASJ...73L...6T} might in principle explain the observed temperature of the X-ray emitting gas, and even the absence of bright radio emission. However, typical terminal sizes for such objects would be $\sim 5$ pc, locating G121.1-1.9 at $d\sim1$ kpc. At this distance, the estimated energy according to Eq.\ref{eq:etot} would be $(1.5-3)\times 10^{48}$ erg, which is substantially below typical parameters considered for failed supernova \citep[e.g.][]{2021PASJ...73L...6T}. Given that the estimate by Eq.\ref{eq:etot} likely corresponds to the lower limit on the explosion energy, the failed supernova scenario appears less favourable compared to the regular supernova scenario.}

\subsection{Radio properties}
{
The upper limit on the radio flux from G121.1-1.9 discussed above provides rather stringent constraints on a possible SNR model. Indeed the minimal radio emission flux is expected in the case of the adiabatic compression of the ambient magnetic field and radio emitting electrons by a forward shock suggested by  \citet{1962MNRAS.124..125V}. The model may successfully explain the radio emission of extended sources like the Cygnus Loop \citep[see e.g.][]{2020ApJ...894..108R} and G116.6-26.1 \citepalias[][]{2022MNRAS.513L..83C}.
The adiabatic compression model allows to derive the synchrotron radio emissivity behind the SNR forward shock once the Galactic synchrotron emissivity distribution obtained from radio surveys  \citep[e.g.][]{2008MNRAS.388..247D,2017MNRAS.465.3163S,2018MNRAS.479.4041S,2022MNRAS.509.4923I} is available. The diffuse Galactic synchrotron radio emissivity can be estimated from the measurements of the free–free absorption of the radiation by intervening HII regions along the line of sight. From the Murchison Wide-field Array observations at 88 MHz of the region  250
< l < 355 degrees, and |b| < 2 degrees,  \citealt[][]{2017MNRAS.465.3163S} derived the synchrotron emissivities $\varepsilon_r$ between 0.39 and 1.45 K pc$^{-1}$ depending on the distance from the Galactic centre (with a mean of 0.77 K pc$^{-1}$ and a variance
of 0.14 K pc$^{-1}$). Then one can estimate the minimal excess of the synchrotron surface brightness  
from G121.1-1.9 as $\varepsilon_r \times R \times f_r$ where $f_r$ is the adiabatic boost factor of the volume emissivity of van der Laan's adiabatic compression model. 

The dependence of the boost factor $f_r$ on the compression factor can be seen in Fig. 5 from \citetalias{2022MNRAS.513L..83C}. Using as a representative the data from fourth Galactic quadrant and extrapolating $\varepsilon_r$ presented by \citet[][]{2017MNRAS.465.3163S} to 408 MHz with the spectral indexes given by \citealt{2022MNRAS.509.4923I} one can see that the excess is above $\sim$ 1 K for the parameter space domain which includes the SNR radii above 10 pc and the forward shock compression factor larger then 2. For the fiducial model of SNR of the age $\sim$25000~years with the radius $\sim$50~pc, and the compression factor of 3.56 illustrated in Fig.~\ref{fig:prof_1d} the estimated $f_r >$ 20. This imply rather low values of $\varepsilon_r$ to meet the derived upper limit of radio emission. 

For SNR propagating in the wind produced of the progenitor star or in the larger scale collective wind cavern produced by a nearby star association the value $\varepsilon_r$ can be substantially reduced due to the modulation effect of the wind on the low energy cosmic rays. Another possibility to understand the reduced local ambient synchrotron emissivity in the vicinity of G121.1-1.9 is to assume that it resides in the region filled with the hot interstellar gas where magnetic field could be low as it was suggested by \citealt{2017MNRAS.464L.105E}. 

The core collapse supernovae at some stage are propagating in the fast supersonic progenitor wind. The supersonic winds may have a low magnetic field and density before the wind termination shock and they can prevent the background radio emitting cosmic ray electrons to reach the SN shock. However, the radius of the termination shock of the supersonic wind of a massive star is typically well below 10 pc \citep[see e.g.] []{2013A&A...559A..69G}. Therefore, in the scenario with the core-collapse SNR, the effect described above  may help to reconcile the van der Laan's model with the radio data on G121.1-1.9 only for a rather nearby SNR. 

Bright radio emission observed from many young SNRs is associated with the acceleration of GeV range relativistic electrons by supernova blast waves \citep[see e.g. ][]{Helder12}. Recently, radio detection of an unusual SNR candidate J0624–6948 which represents a ring of a very good circularity with a low-surface brightness and rather flat radio spectral index  was reported by \citet{2022MNRAS.512..265F}. The estimated radio surface brightness $\Sigma $ of 1.6 $\times$ 10$^{-22}$ W m$^{-2}$ Hz$^{-1}$ sr$^{-1}$ at 1 GHz for the measured angular diameter of 196 arcsec would place the SNR candidate source J0624–6948  to  the bottom area of the $\Sigma-D$
diagram either to left or right corner depending on the assumed distance to the source and it is likely that the ambient matter is very tenuous. The estimated radio spectral index is generally consistent with that observed in a number of young SNRs with efficient particle acceleration by the blast wave. The radio flux density decreases with time in young SNRs at the Sedov phase or later because of the adiabatic cooling of relativistic particles and magnetic field decreasing in the expanding SNR \citep[see e.g.][]{Dubner15}. For old and evolved SNRs, the "minimal" van der Laan's adiabatic compression model, which we applied above to G121.1-1.9, is likely appropriate.}

{
Once the association of G121.1-1.9 with an SNR is confirmed, then one can use it as a probe of the local variations of cosmic ray electron fluxes and Galactic magnetic fields in addition to the technique based on the free–free absorption of the radiation by intervening HII regions along the line of sight.} 

\subsection{What is needed to remove ambiguities?}

{
More sensitive X-ray observations will allow to clarify the picture by, first of all, better characterisation of the morphology of X-ray emission, namely, by checking whether strong clumpiness of the X-ray emitting gas is observed. Second, given the inferred geometry allows conducting de-projection analysis to be performed in a robust manner, spectral analysis of the de-projected spectra might be performed. This should allow testing the models for the observed spectral variations across the object, namely disentangling scenarios in which brightening of the emission in iron lines is a result of time-dependent ionization effects or a result of chemical enrichment by the ejecta medium. { Detection of the line emission from high ionization states of silicon, sulphur etc from the inner part of object might be a clear indication of the progressive inward ionization against the iron enrichment scenario. Sensitive observations of the central part with the \textit{XMM-Newton} observatory would be best suited for this purpose \footnote{ A set of observations of G121.1-1.9 has been performed in July-August 2022 (PI Norbert Schartel), but very high and flaring particle background precludes robust measurement of the spectrum above 1.5 keV, although confirming the spectral shape of the emission observed from the central part by \textit{SRG}/eROSITA in the 0.5-1.3 keV band}. }

Finally, improved statistics of the data and improvement in their description by physically-motivated models should allow constraining the interstellar absorption, and hence better constrain the distance of the object. 

} 

{Clearly, detection of the counterparts at other wavelengths is crucial to pin down the nature of the object. Dedicated radio observations are needed to detect or further constrain intensity of radio emission, allowing to locate the object on the $\Sigma-D$ diagram and characterise properties of the object's environment \citep[as exemplified by the LOFAR detection of radio emission from G116.6-26.1 ][]{2022MNRAS.513L..83C}. Detection and characterisation of optical filaments coinciding with the object should also allow in depth comparison with other radio faint SNRs \citep[e.g.][]{2013A&A...549A.107F,2017A&A...597A.116K,2020ApJ...888...90R,2021A&A...655A..10R,2022MNRAS.tmp.1558P}.}

%-------------------------
\section{Conclusions}
\label{s:conclusion}
%-------------------------

{We report the discovery of diffuse X-ray source SRGe~J003602.3+605421=G121.1-1.9 in the course of \textit{SRG}/eROSITA all-sky survey. The object is located at (l,b)=(121.1$^\circ$,-1.9$^\circ$) in Galactic coordinates, it is $\approx36$ arcmin in angular diameter, { has nearly circular shape} and has soft X-ray spectrum dominated by emission lines of oxygen, neon and iron characteristic to the temperatures of 0.2-0.6 keV. Contrary to the bulk majority of the known supernova remnants located in the Galactic disc, the object lacks clear counterparts in radio, infrared and H$\alpha$ images, probably indicating high temperature and low density of the object's environment, akin the recently discovered candidate SN~Ia remnant {G116.6-26.1} in the Galactic halo \citep[][]{2021MNRAS.507..971C, 2022MNRAS.513L..83C}. {The lack of radio is not surprising since otherwise this SNR would have already be found in sensitive radio surveys. Interestingly, the two newly discovered objects (G121.1-1.9 and G116.6-26.1) have a very similar appearance and could become founding members of apparently not very numerous SNR class.}

Clear variations in spectral shape of the X-ray emission across the object are detected, with the inner (r$\lesssim9$ arcmin) and outer part (r$\gtrsim9$ arcmin) being dominated by emission lines of iron and oxygen/neon respectively. The non-equilibrium plasma emission model with the initial temperature 0.1 keV, final temperature 0.5 keV, and the ionisation parameter  $2\times10^{10}$ cm$^{-3}$~s is capable of describing emission from the outer part, while emission from the inner part appears to be more complex (partially due to contribution of the projected outer shell emission), but shows signatures of substantial relative overabundance of iron.  

The X-ray absorbing column density derived from the spectral fitting,  $N_H\approx(4-6)\times10^{21}$ cm$^{-2}$, locates the object at the distance beyond 1.5 kpc, implying the age of the supernova $\sim(5-30)\times1000$ yrs. A illustrative model of an SN~Ia explosion with energy $10^{51}$ erg in the environment with the gas temperature $kT\sim10^6$ K and number density $10^{-2}$ cm$^{-3}$ at the distance of 9 kpc is able to reproduce main observational properties of the object.

 More sensitive X-ray, radio and optical observations are needed to clarify the nature of the object, which might be a rare example of an old SN Ia remnant in our Galaxy bearing signatures of the ejecta enrichment in the central part and non-equillibrium processes in the outer shell allowing us to probe the thermal and non-thermal properties of the hot phase of the ISM above the cold disc.
 
 }

\appendix

\section{Atlas of the X-ray, H$\alpha$ and radio emission from the region of G121.1-1.9}
\label{s:atlas}

Here we present maps of H$\alpha$ and radio emissions covering 1.4$\times$1.4 region around G121.1-1.9 and the map of its X-ray surface brightness, indicating its position, size and morphology. All images are on the linear scale. No associated emission in other bands can be noticed.

%------------------------
\begin{figure*}
\centering

\includegraphics[angle=0,width=0.62\columnwidth, bb = 50 160 550 640]{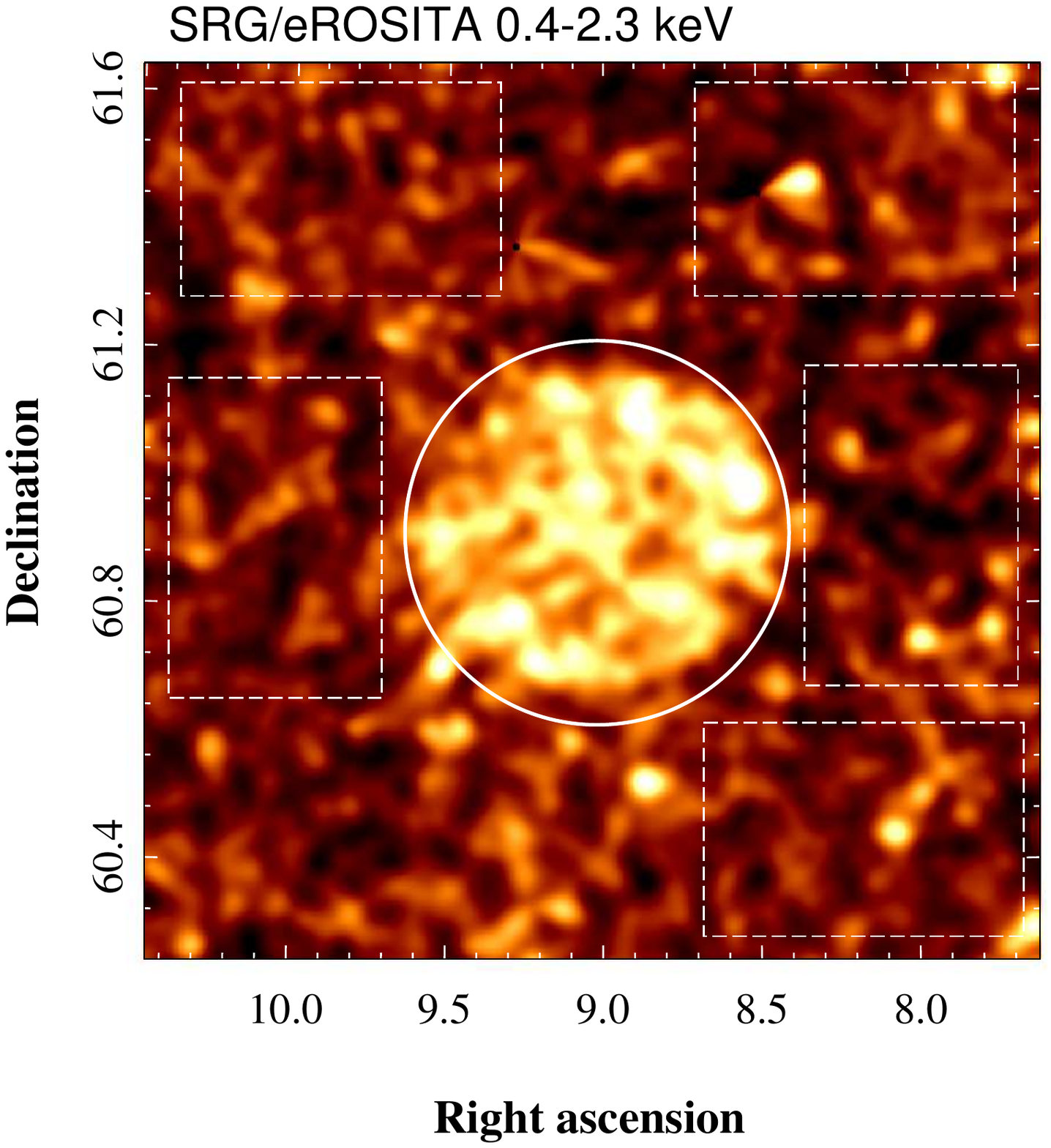}
\includegraphics[angle=0,width=0.62\columnwidth, bb = 50 160 550 640]{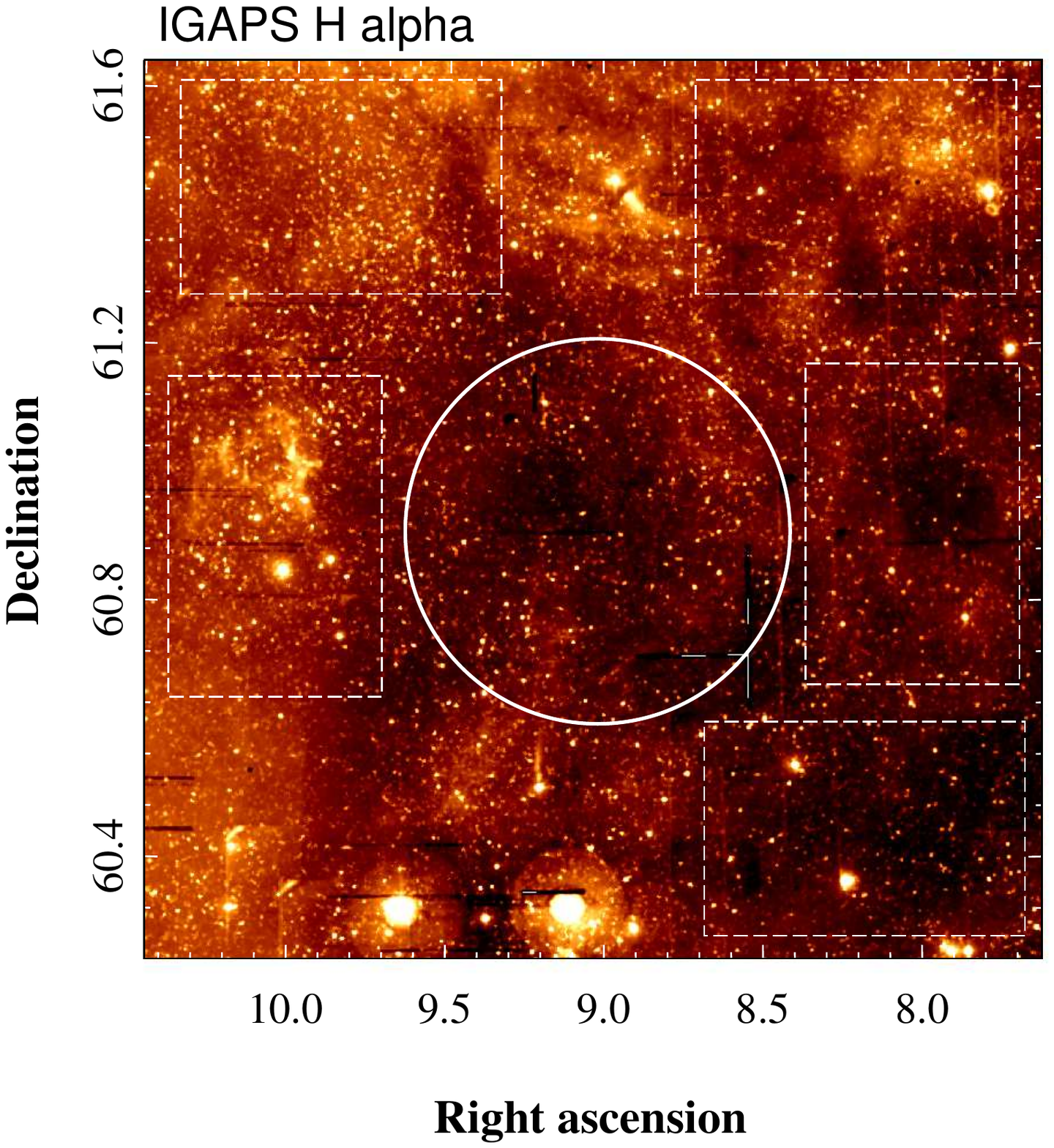}
\includegraphics[angle=0,width=0.62\columnwidth, bb = 50 160 550 640]{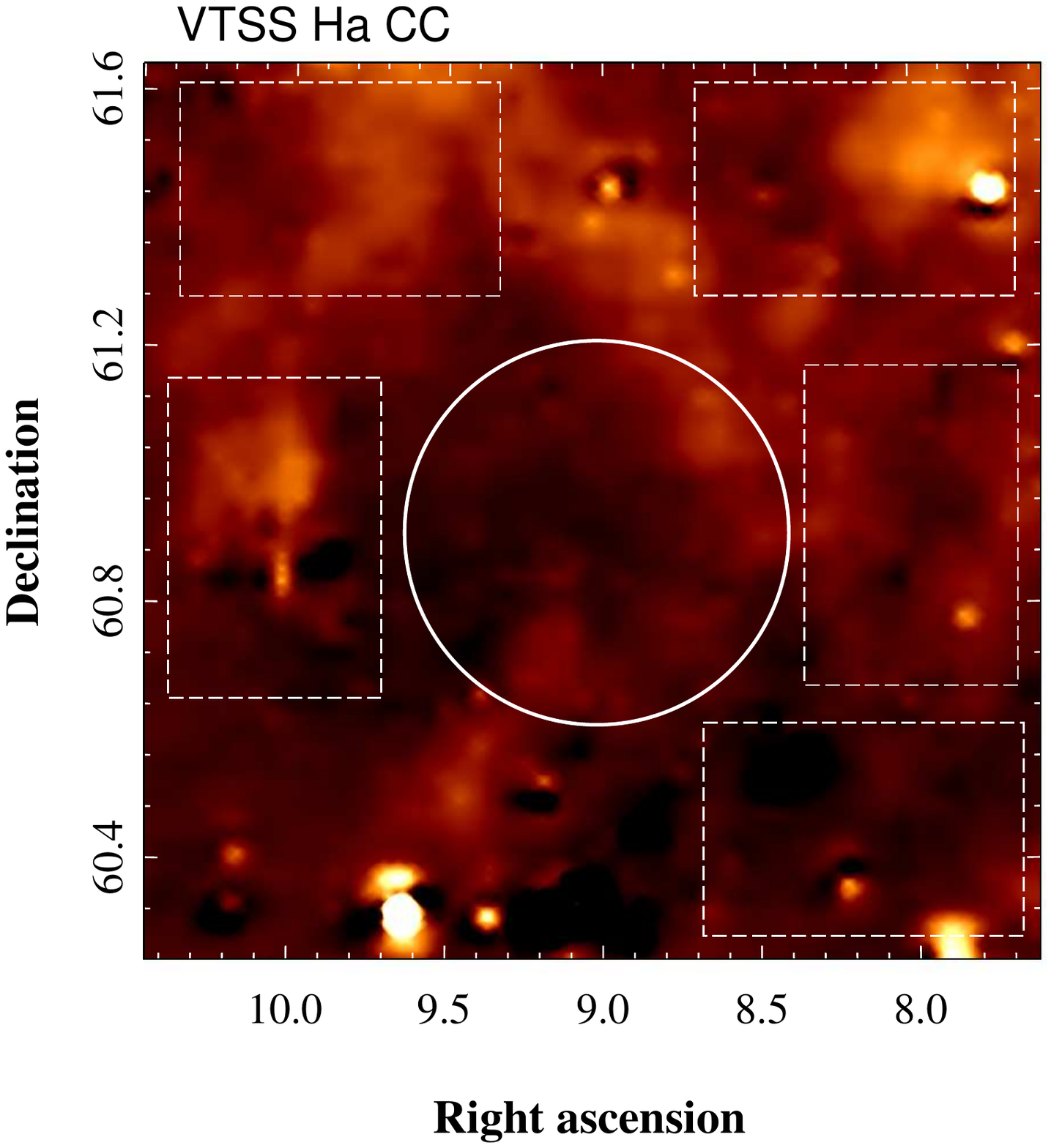}

\includegraphics[angle=0,width=0.62\columnwidth, bb = 50 160 550 660]{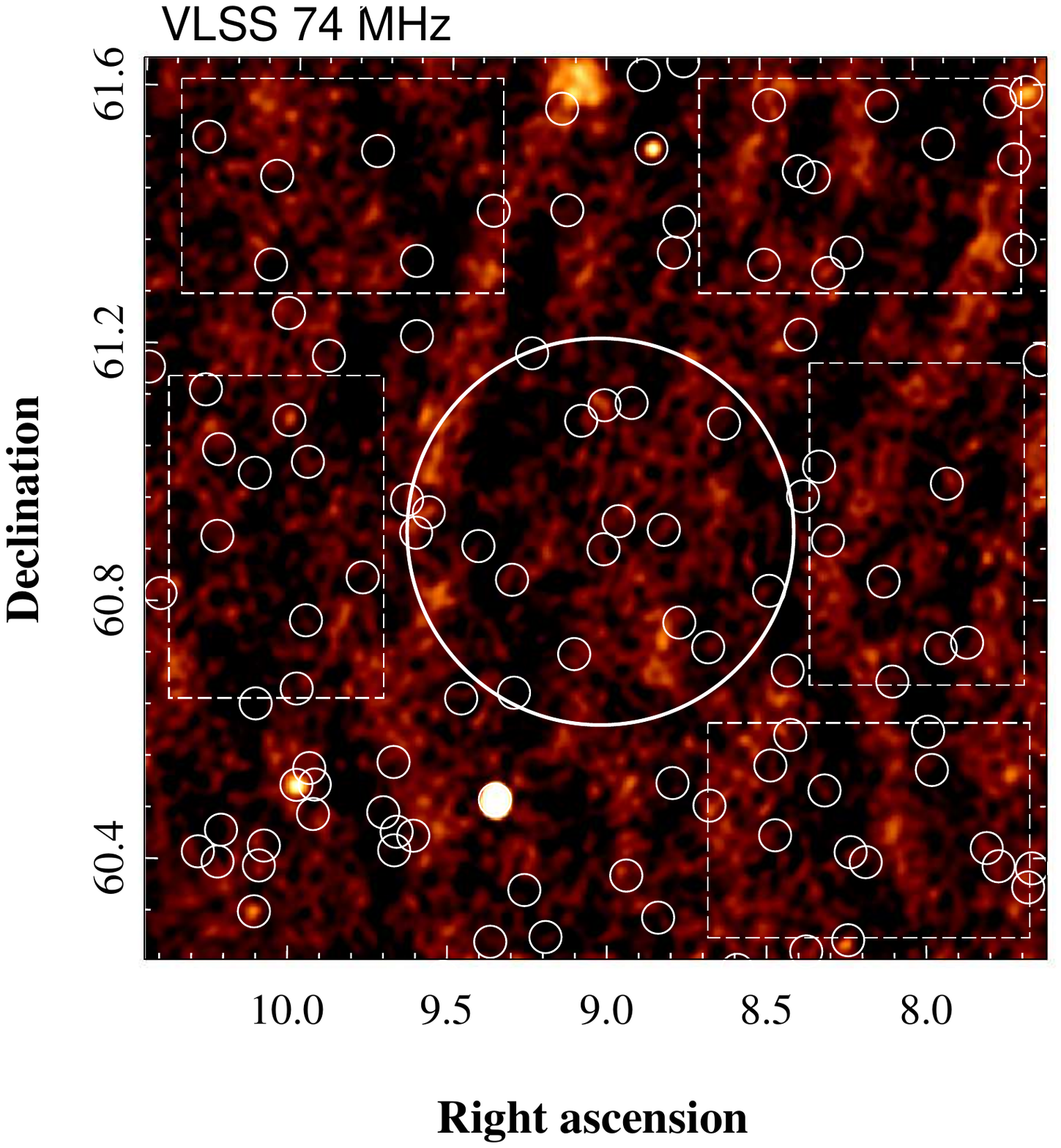}
\includegraphics[angle=0,width=0.62\columnwidth, bb = 50 160 550 660]{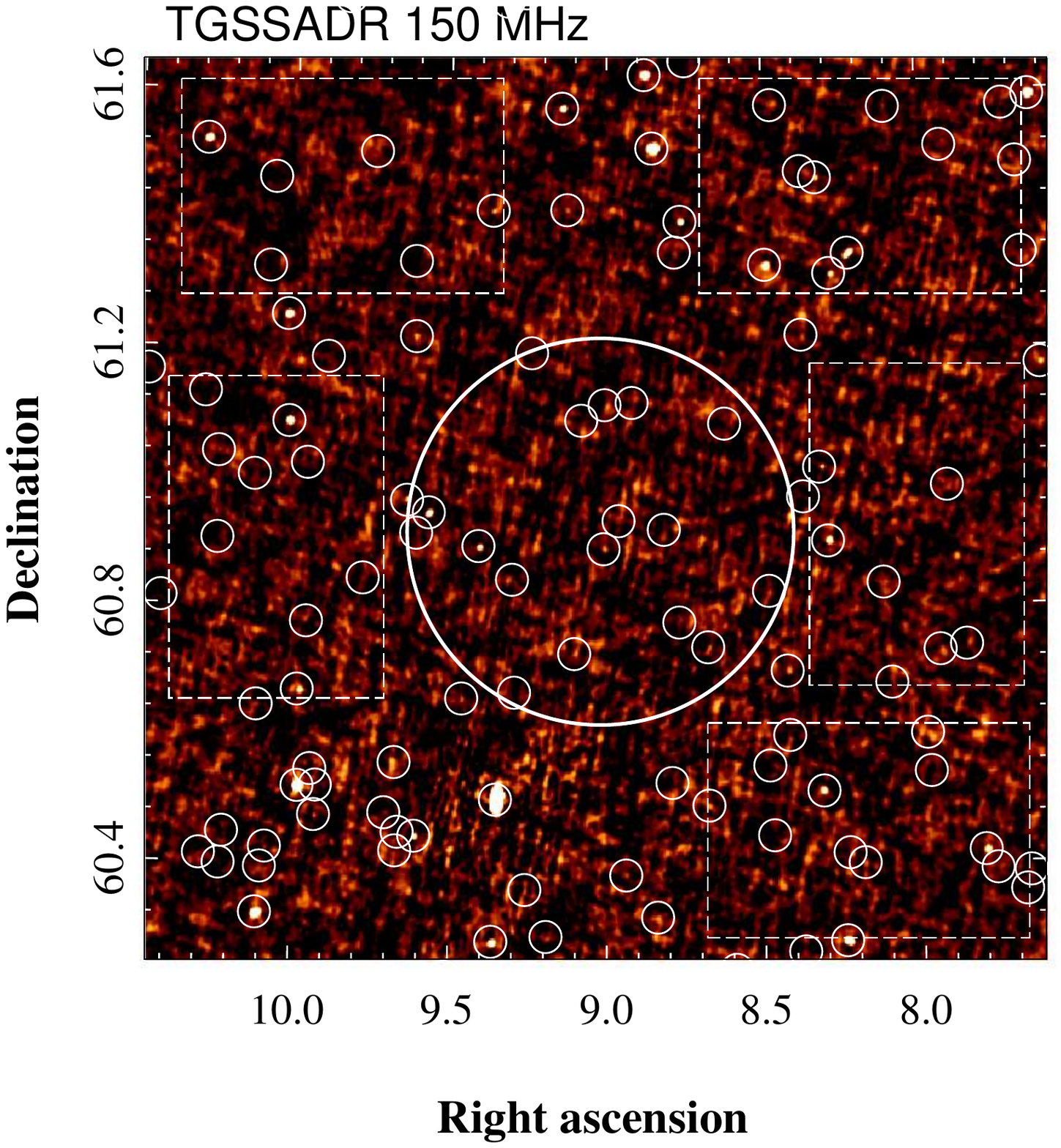}
\includegraphics[angle=0,width=0.62\columnwidth, bb = 50 160 550 660]{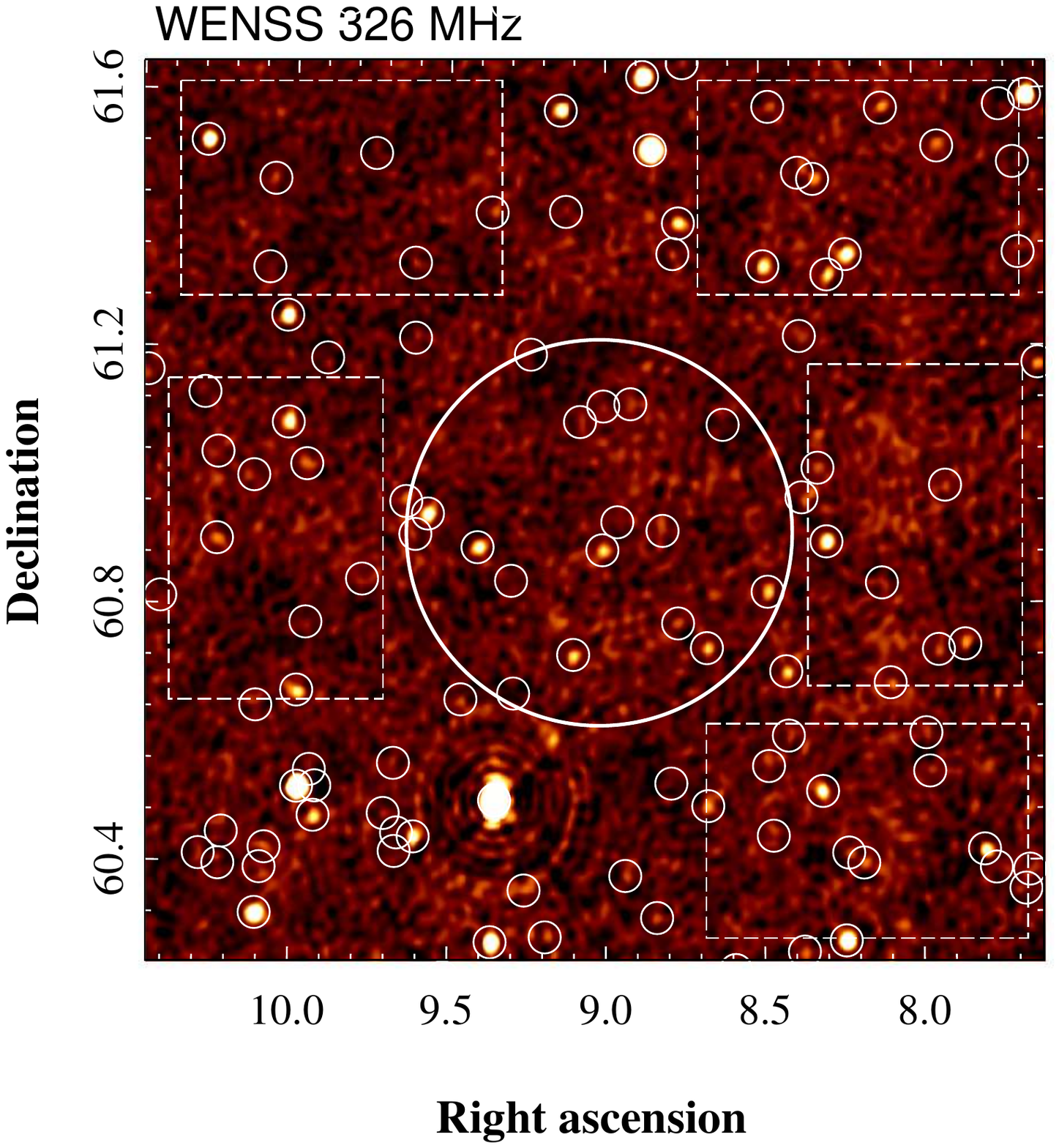}
\includegraphics[angle=0,width=0.62\columnwidth, bb = 50 160 550 660]{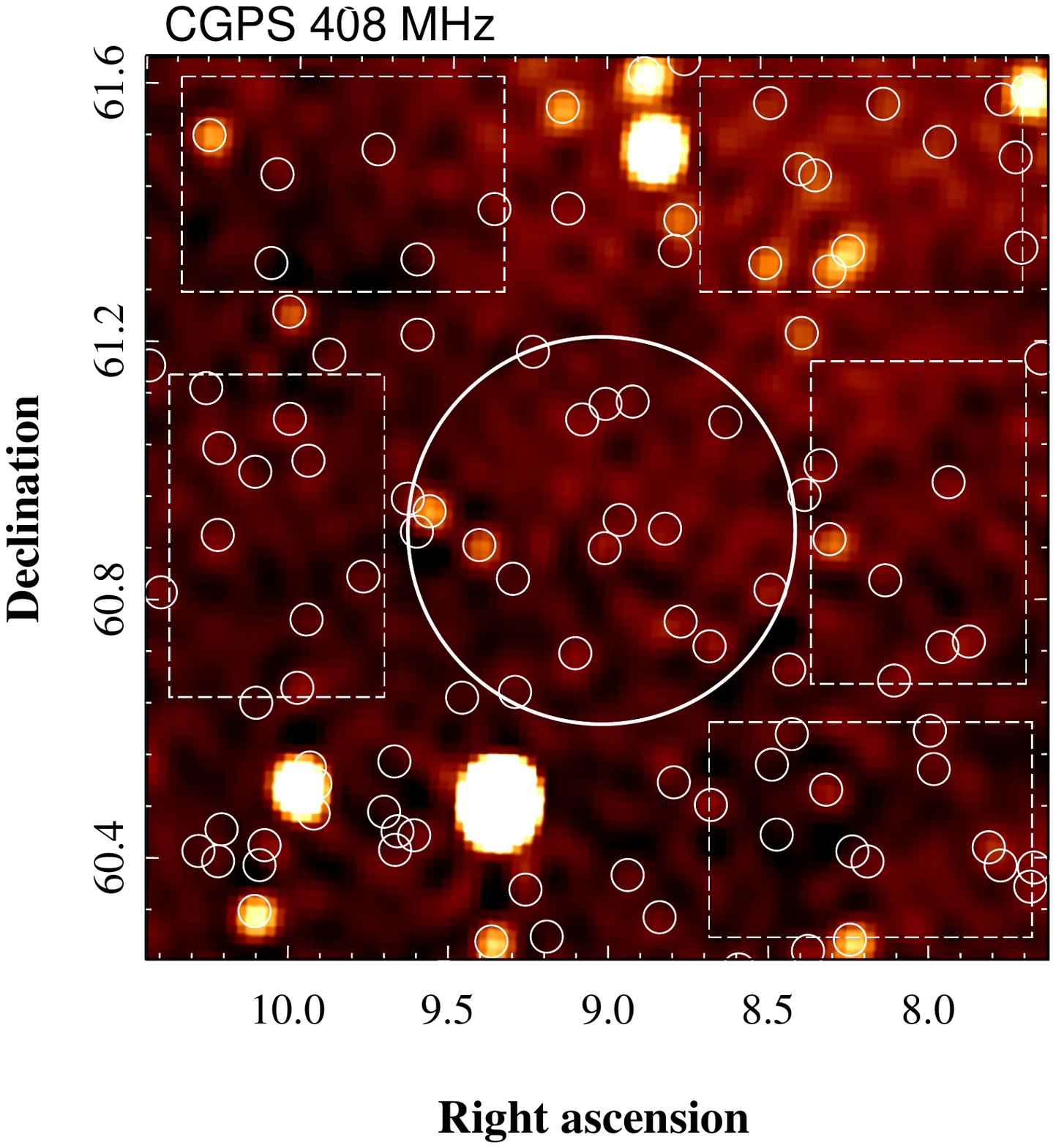}
\includegraphics[angle=0,width=0.62\columnwidth, bb = 50 160 550 660]{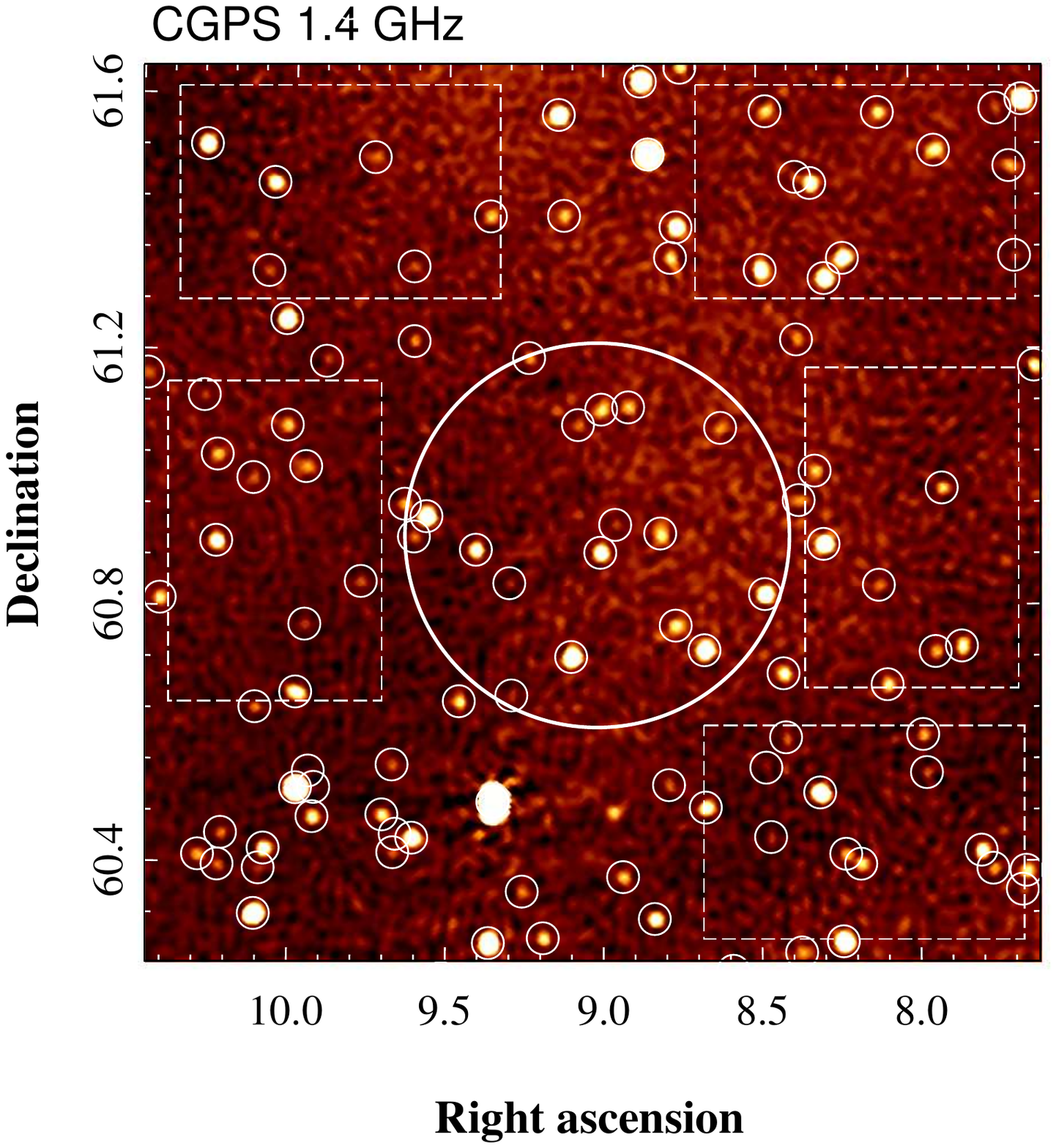}
\includegraphics[angle=0,width=0.62\columnwidth, bb = 50 160 550 660]{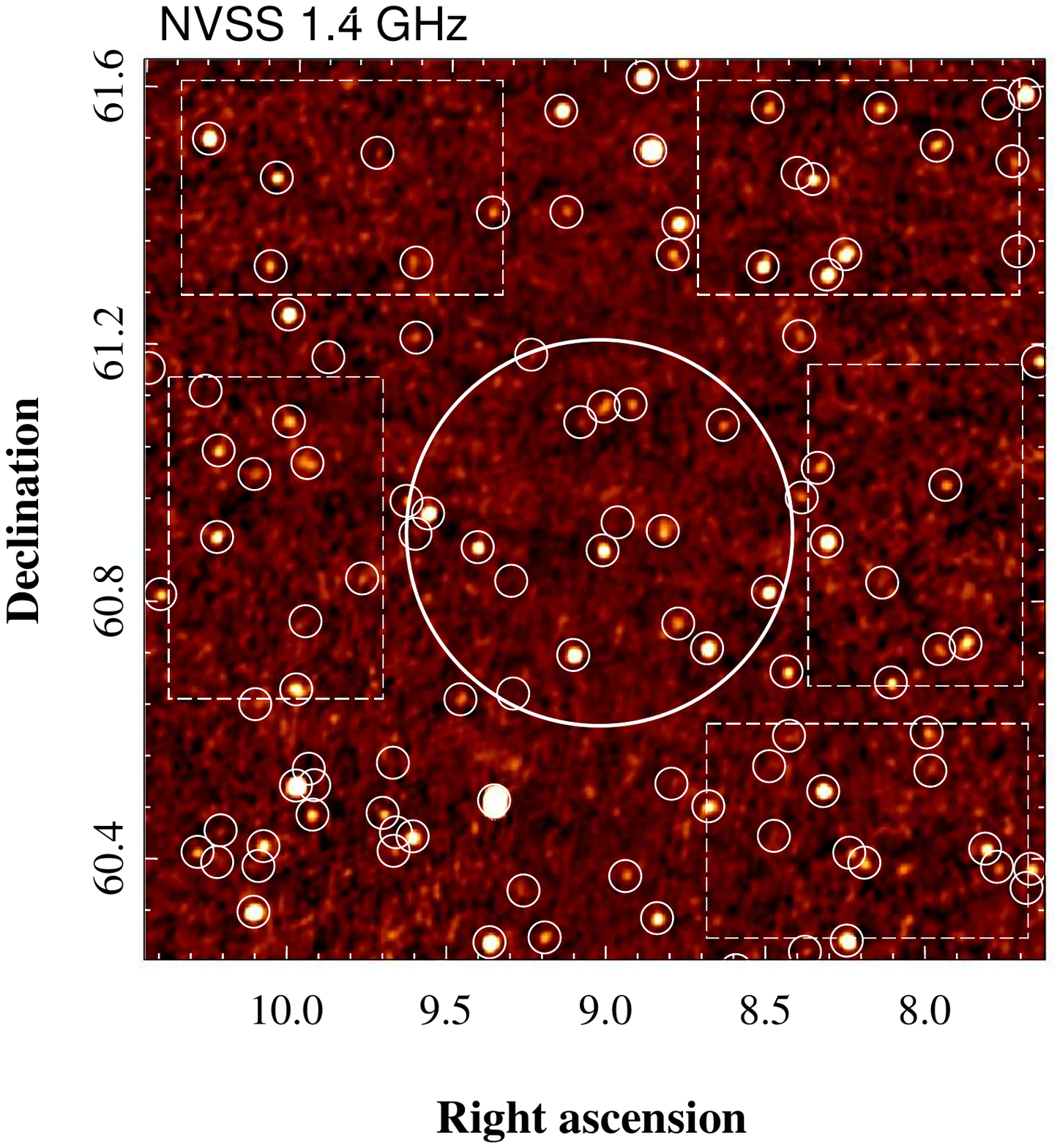}
\includegraphics[angle=0,width=0.62\columnwidth, bb = 50 160 550 660]{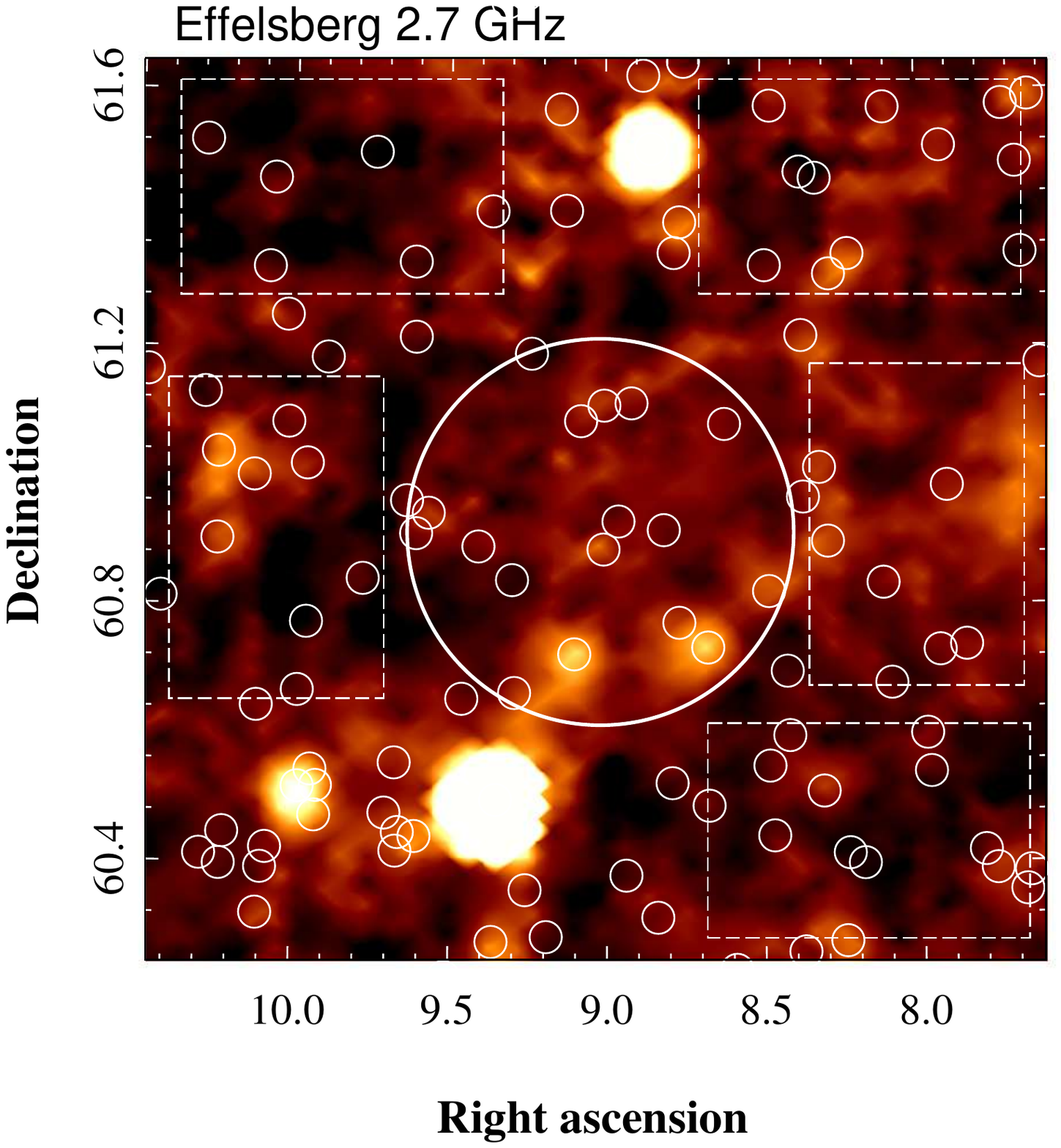}
\includegraphics[angle=0,width=0.62\columnwidth, bb = 50 160 550 660]{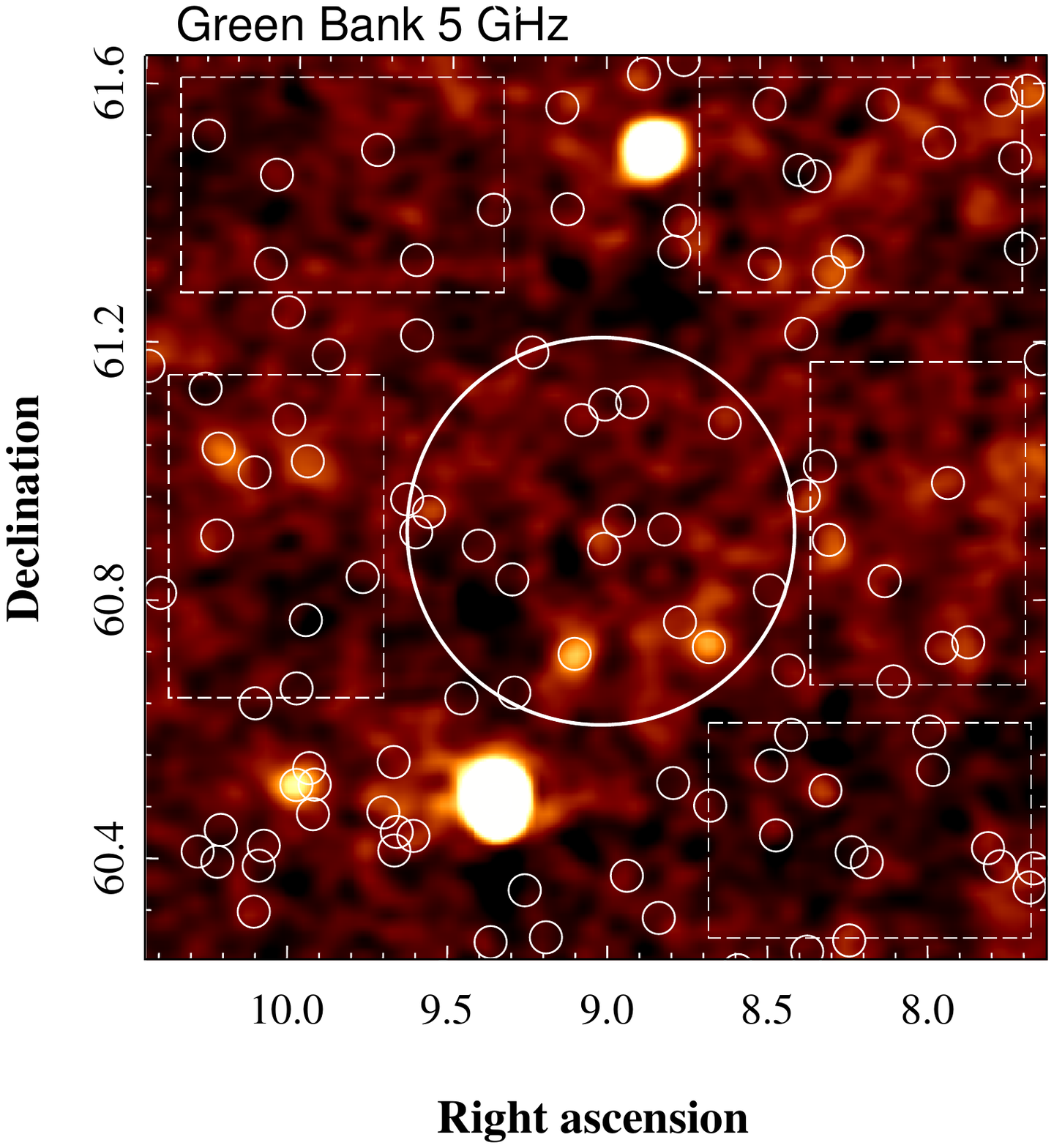}
\includegraphics[angle=0,width=0.62\columnwidth, bb = 50 160 550 660]{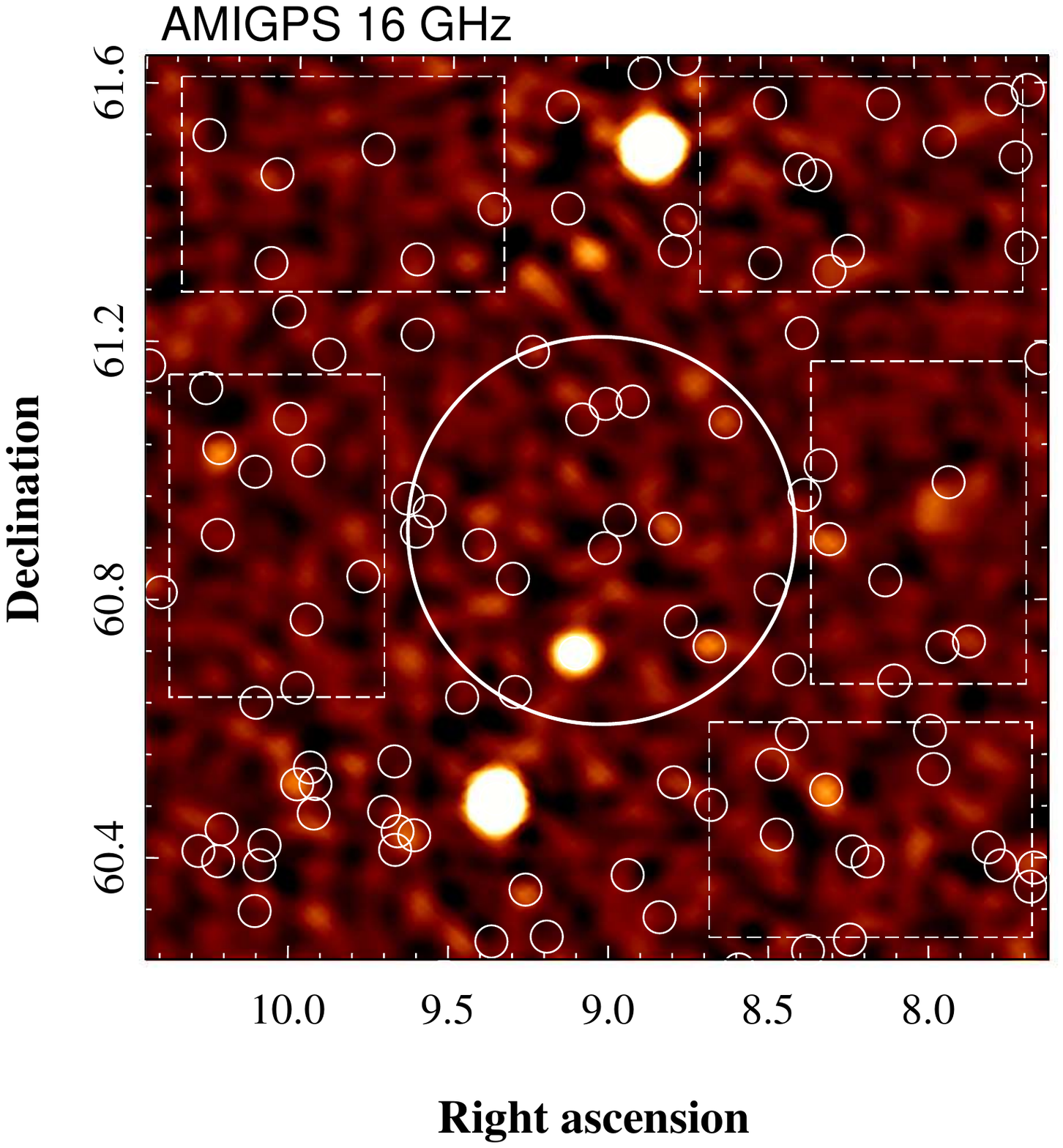}
\caption{
An atlas of $1.4\times1.4$ deg$^2$ images centred on the position of the newly discovered SNR candidate G121.1-1.9, starting from the background-subtracted vignetting corrected X-ray image obtained by $SRG$/eROSITA in the 0.4-2.3 keV band (the image was smoothed with $\sigma=1$~arcmin gaussian kernel after masking of the point sources down to the flux limit of $3\times10^{-14}$ erg s$^{-1}$ cm$^{-2}$ in 0.5-2 keV band). Next two figures in the first row show H$\alpha$ images constructed from the IGAPS and VTSS data, while the images on the following panels show radio emission at 74~Mhz (VLSS), 150~Mhz (TGSSADR), 326~MHz (WENSS), 408~MHz (CGPS), 1.4 GHz (CGPS and NVSS), 2.7 GHz (Effelsberg), 5~GHz (Green Bank), and 16 GHz (AMIGPS). The small white circles indicate point sources detected at 14 GHz in the CGPS data, which are excluded for the radio flux estimation from the "source" (big white circle, R=18 ~arcmin) and "background" (white dashed boxes) regions. {All images are on the linear scale and show no signatures of excess emission from the G121.1-1.9 region with the only exception of the X-ray image.} }  
\label{fig:atlas}
\end{figure*}
%-------------------------

\section{Cutouts from the PPV cube of HI emission}
%------------------------
\label{s:ppv}

\begin{figure*}
\centering
\includegraphics[angle=0,width=0.65\columnwidth, bb = 70 10 560 500]{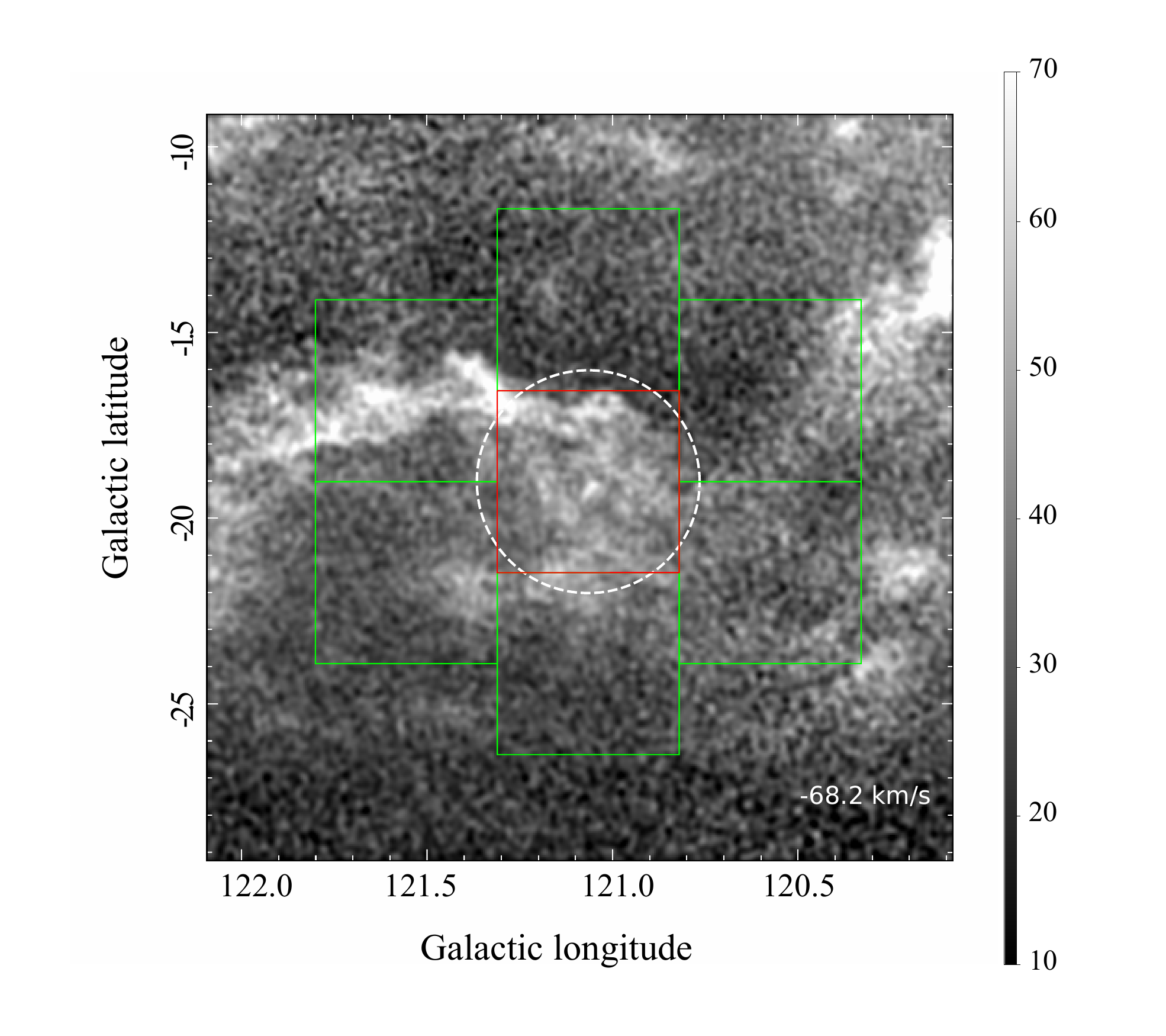}
\includegraphics[angle=0,width=0.65\columnwidth, bb = 70 10 560 500]{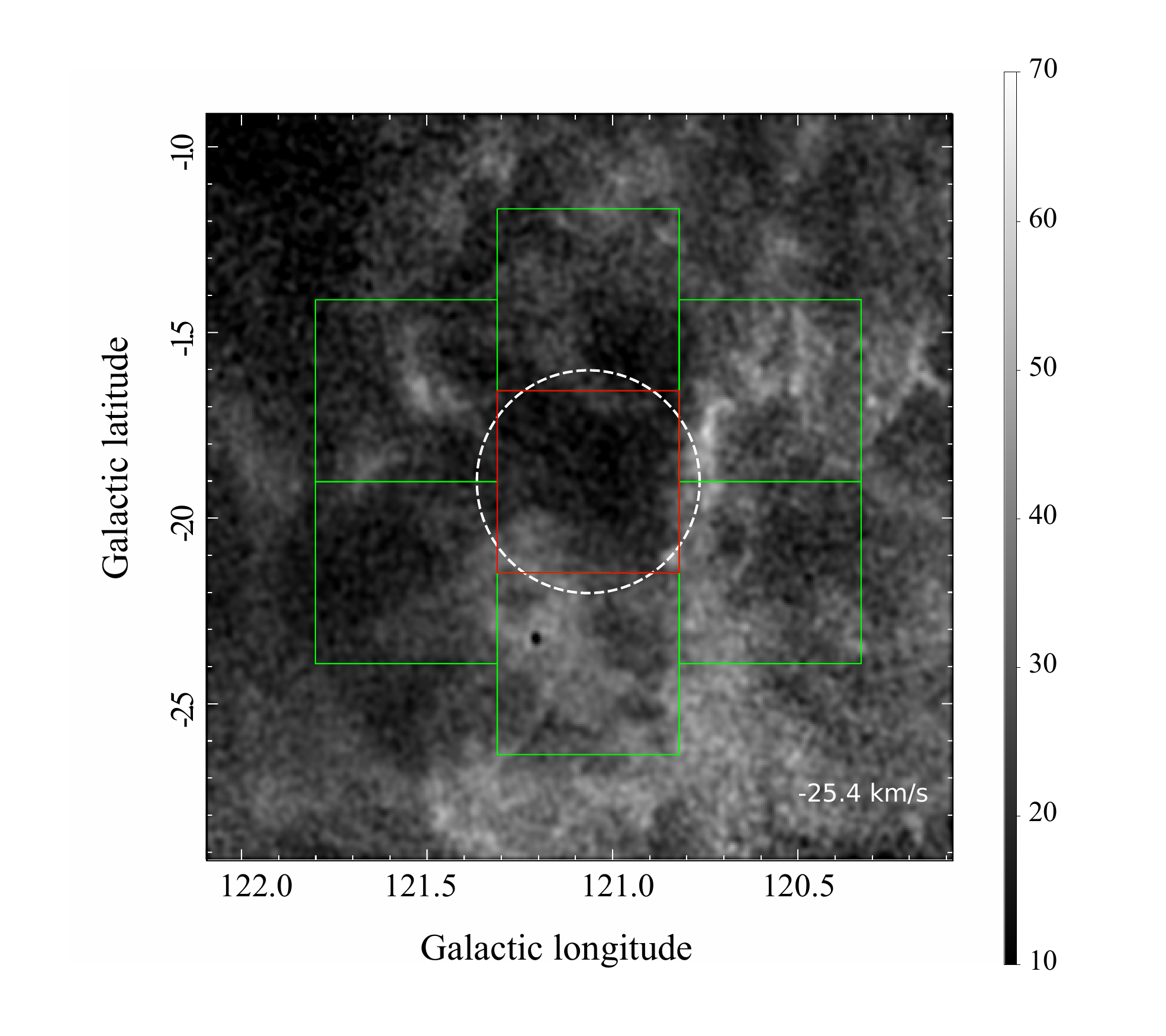}
\includegraphics[angle=0,width=0.65\columnwidth, bb = 70 10 560 500]{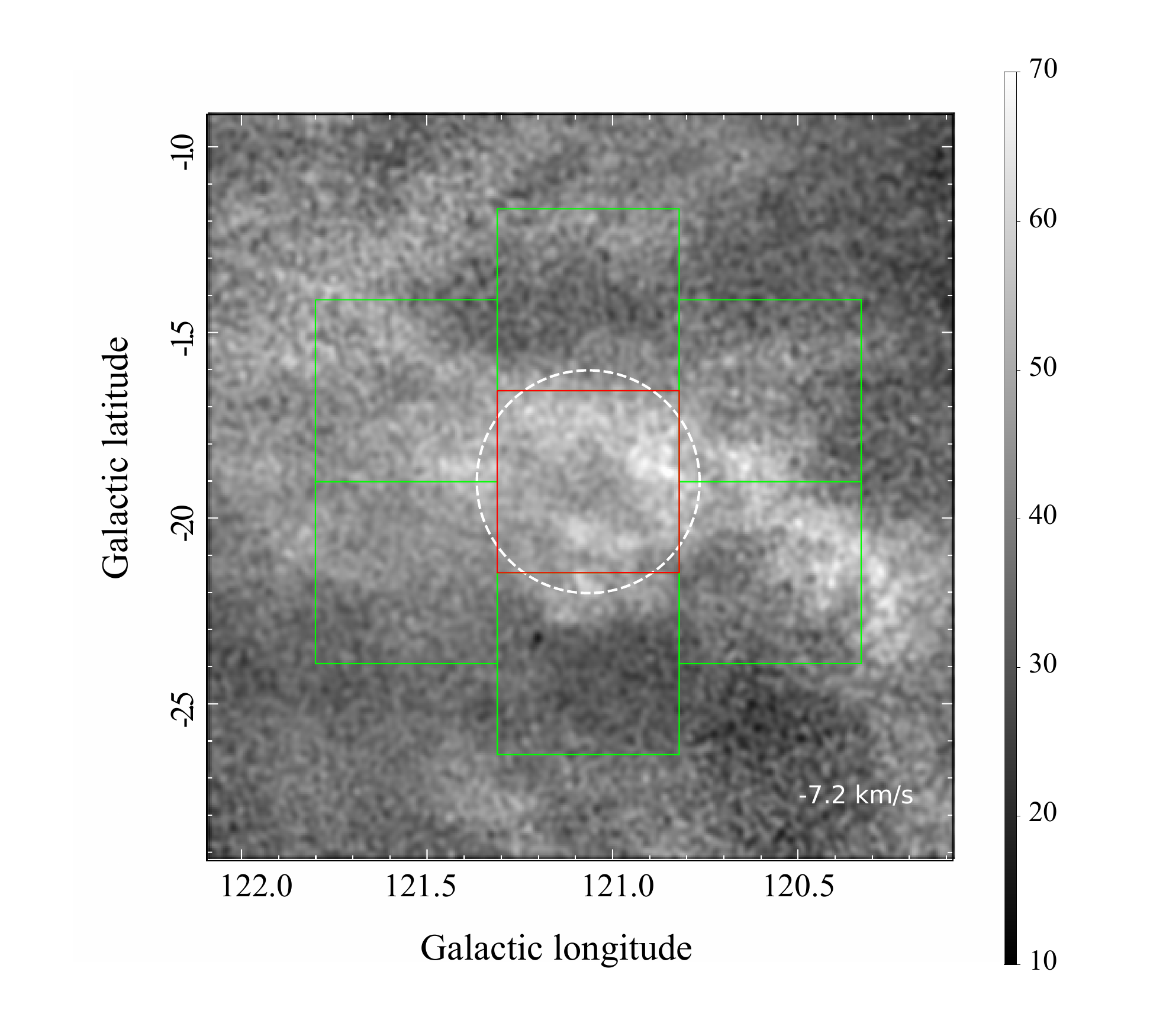}
\caption{Cutouts from the HI emission cubes at $V_{los}$=-68.2 km/s (left), -25.4 km/s (center), -7.2 km/s (right). The white circle shows R=18' extend of G121.1-1.9, while the 30'x30' boxes show regions used for the velocity profile extraction, with the red box being the source region, and the green boxes are background regions. The images and the box regions are aligned with Galactic coordinates, according to the orientation of the original data cubes.}
\label{fig:hiimages}
\end{figure*}

In this Section we demonstrate cutouts (in Galactic coordinates) from the position-position-velocity cubes of the HI 21 cm emission from the CGPS survey \citep[][]{2003AJ....125.3145T}. Namely, Figure \ref{fig:hiimages} shows slices corresponding to $V_{\rm los}$=-68.2 km/s (left), -25.4 km/s (center), -7.2 km/s (right), where emission from the region of G121.1-1.9 (red central box) significantly differs from the average profile extracted from the surrounding background regions (green boxes).

\section{Degeneracy in the parameters of the NEI plasma emission model}
\label{s:deg}
In this section we show degeneracies between the parameters of the absorbed non-equilibrium plasma emission models, as constrained by the currently available X-ray data for the inner (0'-9') and outer (9'-18') regions of the object.

\begin{figure*}
\centering
\includegraphics[angle=0, width=0.66\columnwidth]{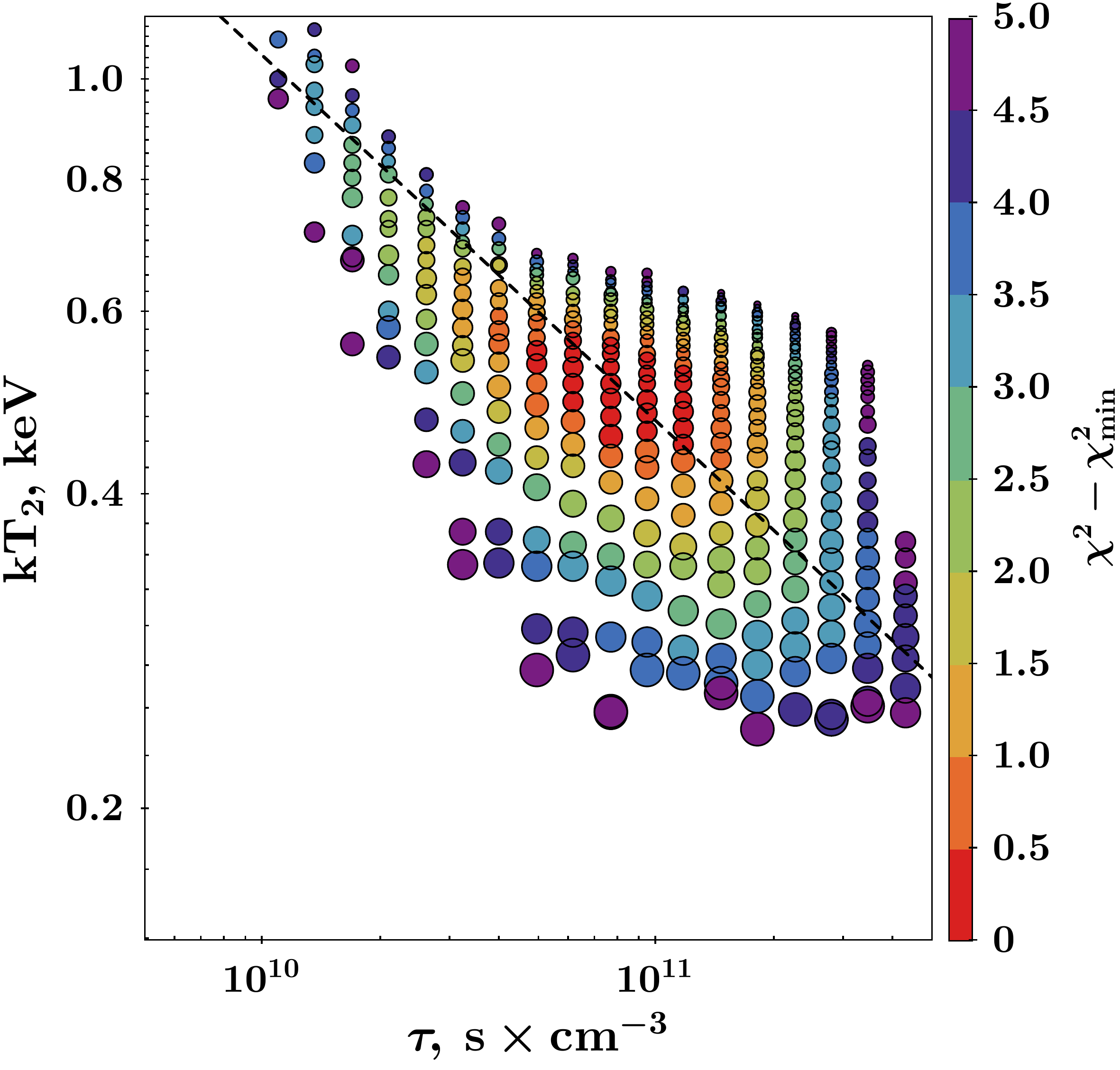}
\includegraphics[angle=0, width=0.66\columnwidth]{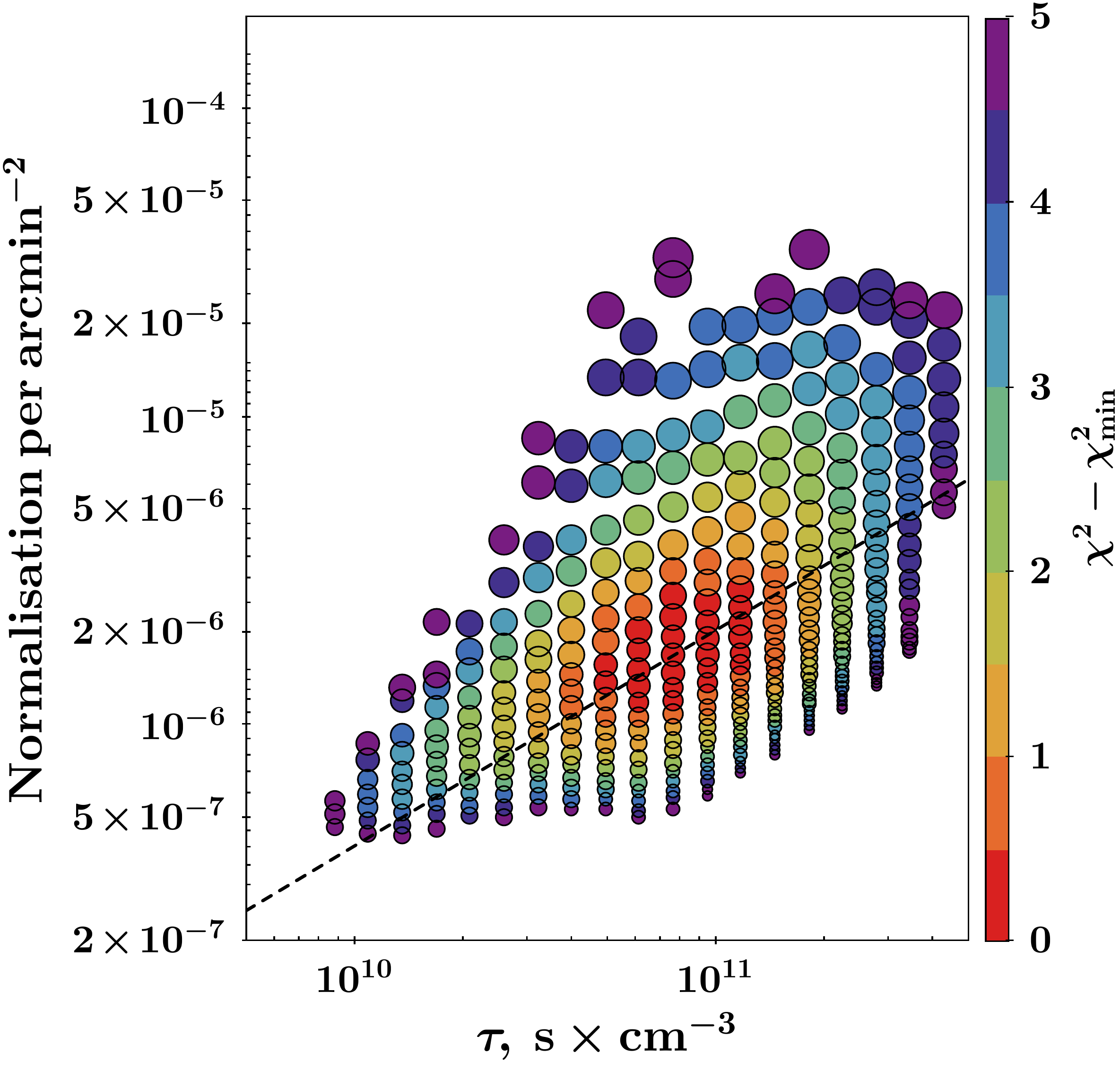}
\includegraphics[angle=0, width=0.66\columnwidth]{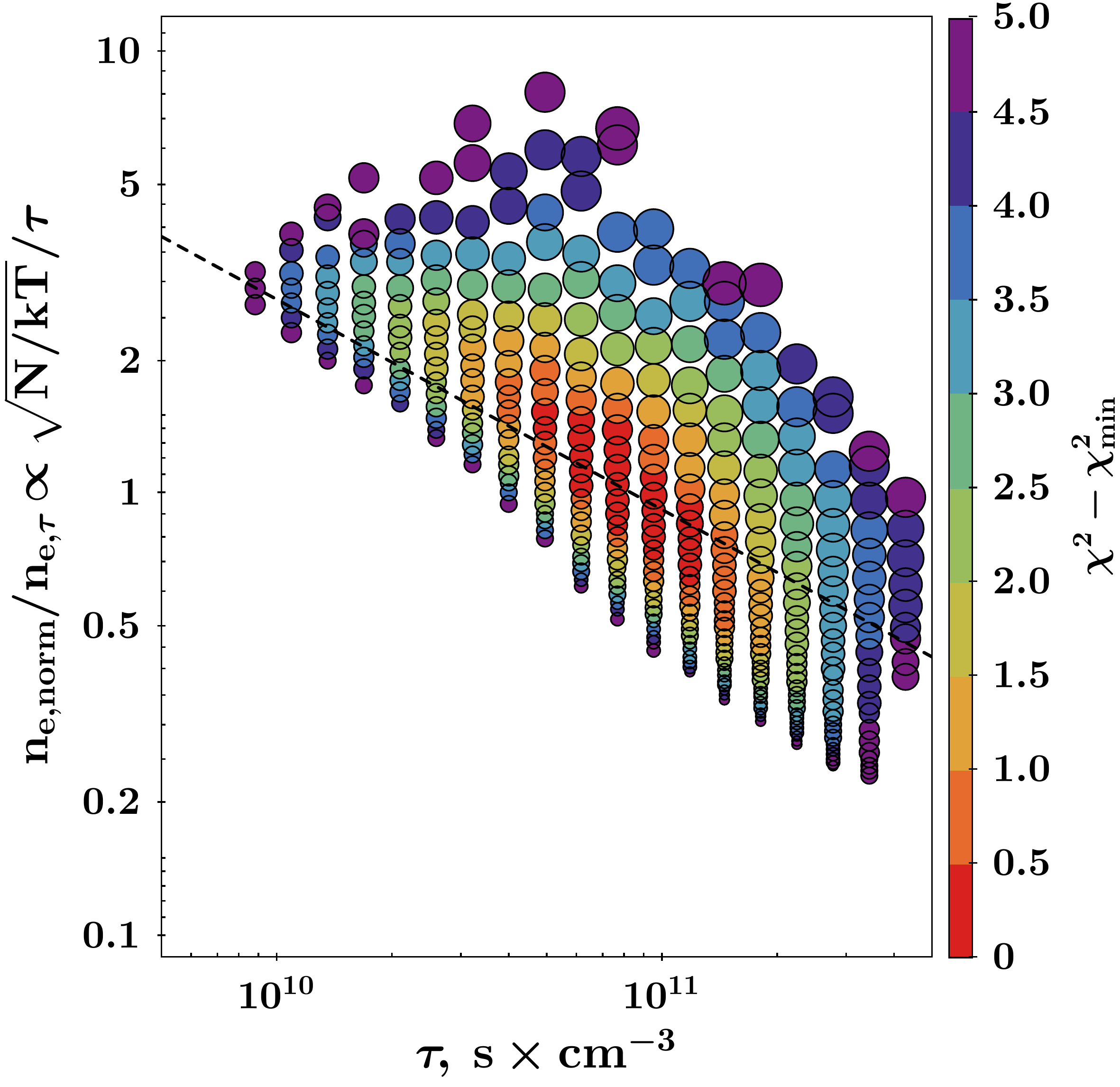}
\caption{Degeneracy between the fit parameters for the inner 9' region.The right panel shows dependence of the boost factor on $\tau$. The color of the symbol shows the excess in the $\chi^2$ statistic over the minimum value, while the size is proportional to the value absorbing column density value. The dashed lines show simple powerlaw approximations for the ridge of the distribution.}
\label{fig:deg_r9}
\end{figure*}

\begin{figure*}
\centering
\includegraphics[angle=0, width=0.66\columnwidth]{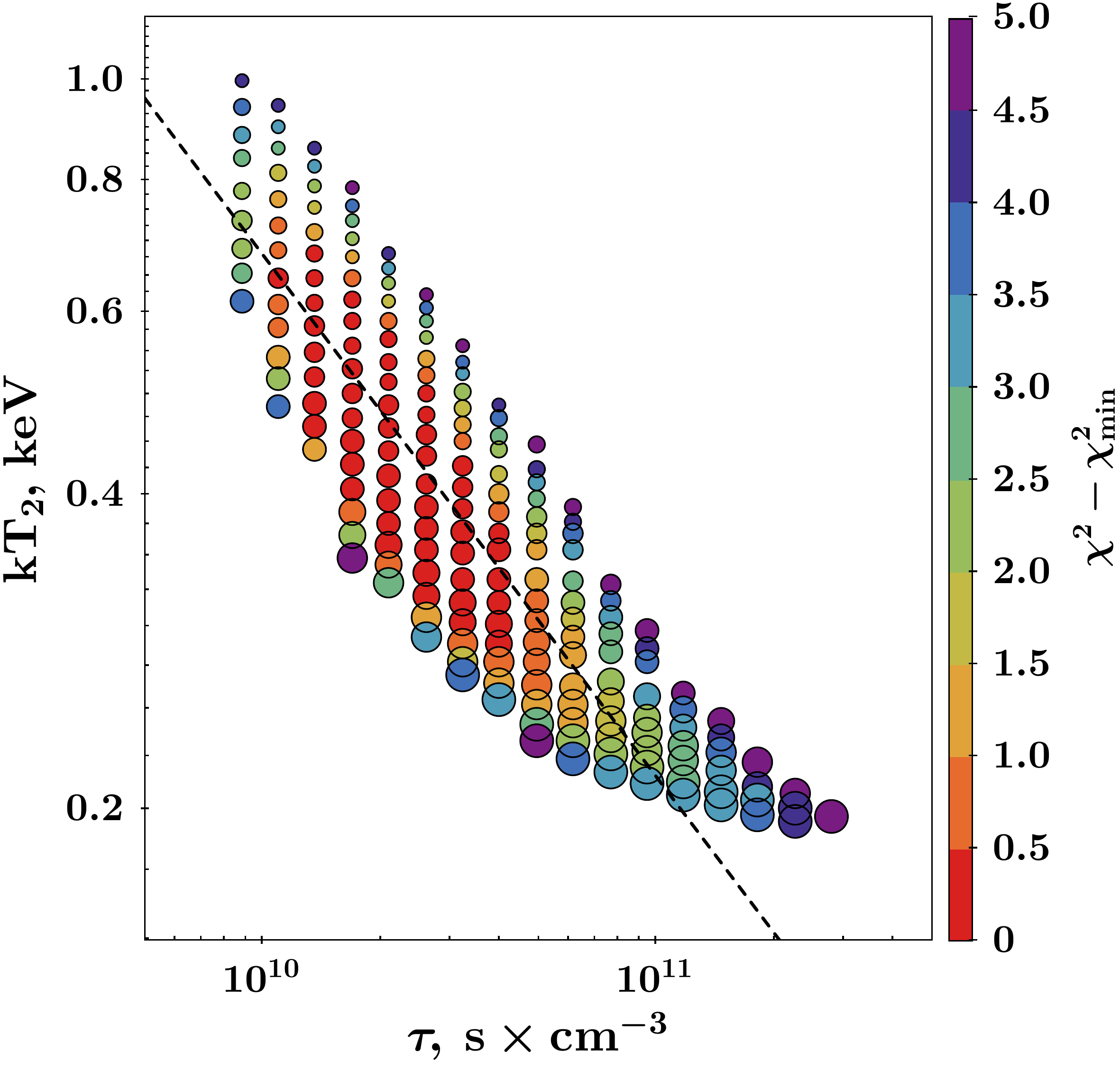}
\includegraphics[angle=0, width=0.66\columnwidth]{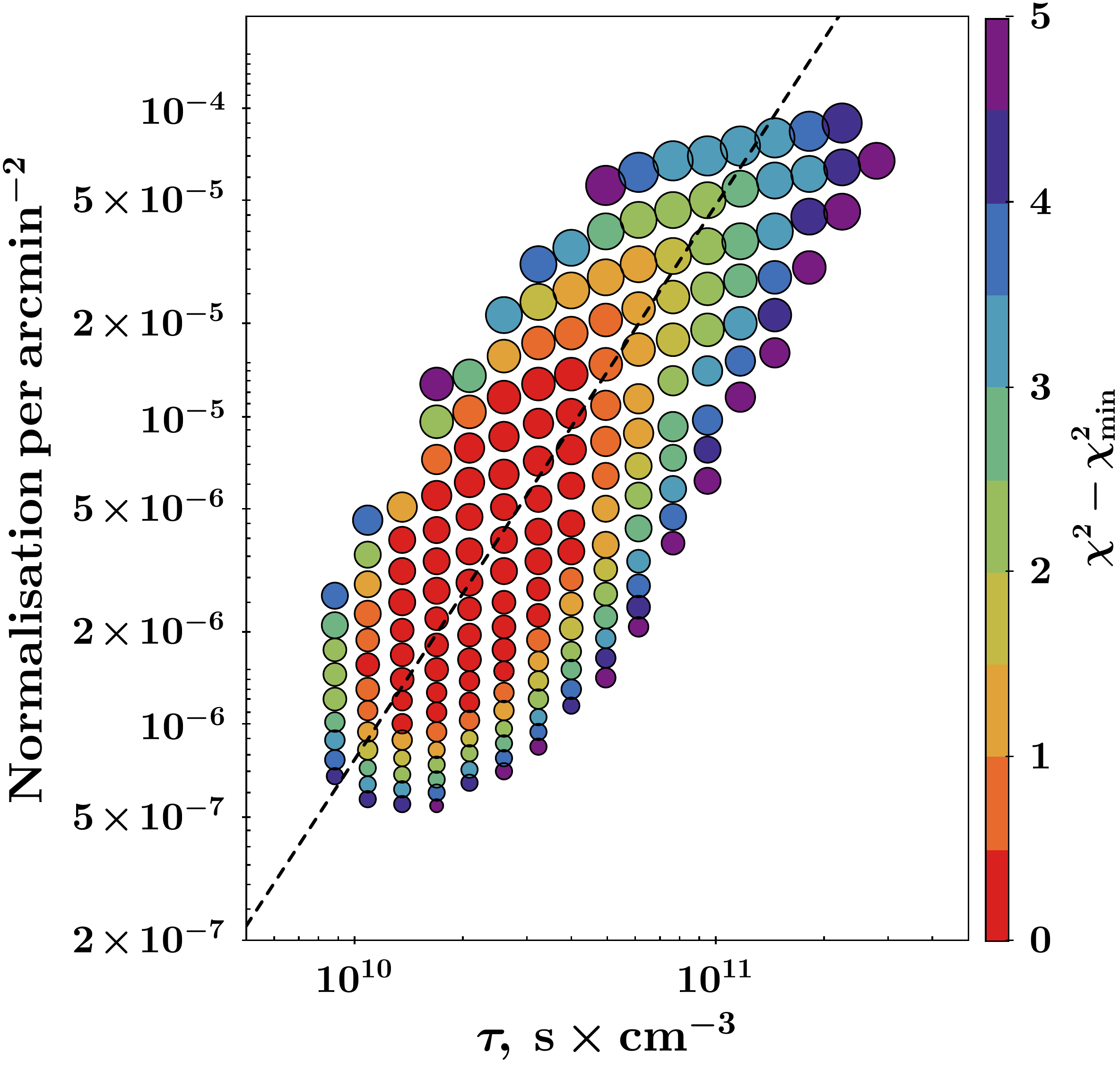}
\includegraphics[angle=0, width=0.66\columnwidth]{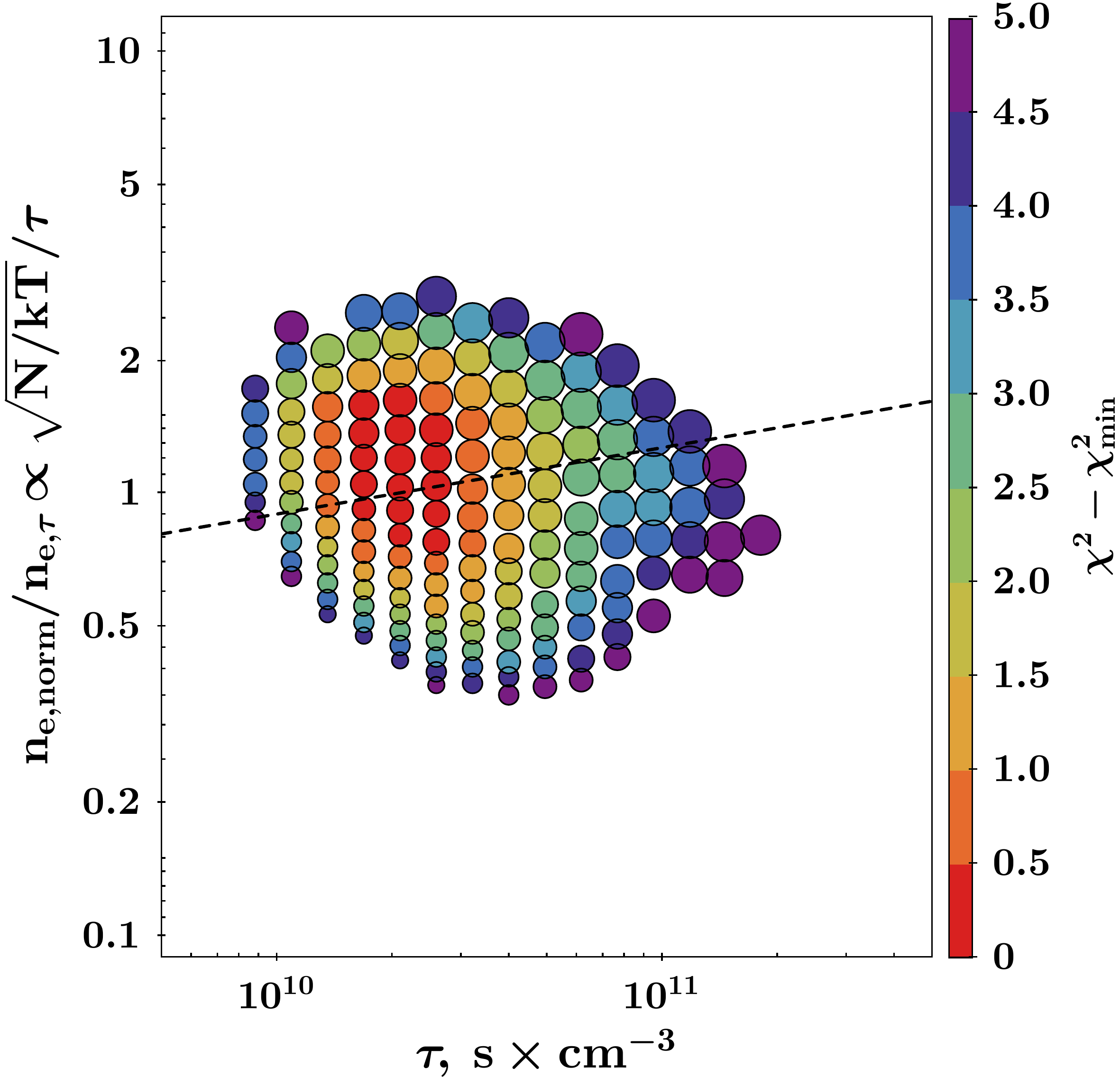}
\caption{The same as Figure \ref{fig:deg_r9} but for the 9'-18' annuli. }
\label{fig:deg_r18}
\end{figure*}

%%%%%%%%%%%%%%%%%%%%%%%%%%%%%%%%%%%%%%%%%%%%%%%%%%
%\newpage
\section*{Acknowledgements}

 IK acknowledges support by the COMPLEX project from the European Research Council (ERC) under
the European Union’s Horizon 2020 research and innovation program grant agreement ERC-2019-AdG 882679. Modelling of radio emission at the Joint Supercomputer Center JSCC RAS and at the ``Tornado'' subsystem of the St.~Petersburg Polytechnic University supercomputing center by A.M.B. was supported by the RSF grant 21-72-20020. { We thank Alexander Nezhin for testing some of the ionization calculations. We are grateful to the anonymous referee for the careful reading and constructive suggestions.}

This work is based on observations with the eROSITA telescope onboard \textit{SRG} space observatory. The \textit{SRG} observatory was built by Roskosmos in the interests of the Russian Academy of Sciences represented by its Space Research Institute (IKI) in the framework of the Russian Federal Space Program, with the participation of the Deutsches Zentrum für Luft- und Raumfahrt (DLR). The eROSITA X-ray telescope was built by a consortium of German Institutes led by MPE, and supported by DLR. The SRG spacecraft was designed, built, launched, and is operated by the Lavochkin Association and its subcontractors. The science data are downlinked via the Deep Space Network Antennae in Bear Lakes, Ussurijsk, and Baikonur, funded by Roskosmos. The eROSITA data used in this work were converted to calibrated event lists using the eSASS software system developed by the German eROSITA Consortium and analysed using proprietary data reduction software developed by the Russian eROSITA Consortium.

This research made use of Montage. It is funded by the National Science Foundation under Grant Number ACI-1440620, and was previously funded by the National Aeronautics and Space Administration's Earth Science Technology Office, Computation Technologies Project, under Cooperative Agreement Number NCC5-626 between NASA and the California Institute of Technology. Catalogue manipulations have been performed using the \texttt{TOPCAT/STILTS} software \citep{2005ASPC..347...29T}.
This study has used part of an image
obtained by the Virginia Tech Spectral-Line Survey, which is supported by
the National Science Foundation.
We acknowledge the use of data provided by the Centre d'Analyse de Données Etendues (CADE), a service of IRAP-UPS/CNRS (http:\/\/cade.irap.omp.eu). This research has made use of the SIMBAD database, operated at CDS, Strasbourg, France.

This work has made use of data from the European Space Agency (ESA) mission
{\it Gaia} (\url{https://www.cosmos.esa.int/gaia}), processed by the {\it Gaia}
Data Processing and Analysis Consortium (DPAC,
\url{https://www.cosmos.esa.int/web/gaia/dpac/consortium}). Funding for the DPAC
has been provided by national institutions, in particular the institutions
participating in the {\it Gaia} Multilateral Agreement.

%%%%%%%%%%%%%%%%%%%%%%%%%%%%%%%%%%%%%%%%%%%%%%%%%%
\section*{Data availability}

X-ray data analysed in this article were used by permission of the Russian SRG/eROSITA consortium. The data will become publicly available as a part of the corresponding SRG/eROSITA data release along with the appropriate calibration information. 
All other used data are publicly available and were can be accessed at the corresponding public archive servers. 


%%\clearpage
\bibliographystyle{mnras}
\bibliography{references} % if your bibtex file is called example.bib


%%%%%%%%%%%%%%%%%%%%%%%%%%%%%%%%%%%%%%%%%%%%%%%%%%

%%%%%%%%%%%%%%%%% APPENDICES %%%%%%%%%%%%%%%%%%%%%

%\appendix


%%%%%%%%%%%%%%%%%%%%%%%%%%%%%%%%%%%%%%%%%%%%%%%%%%

% Don't change these lines
\bsp	% typesetting comment
\label{lastpage}
\end{document}